\documentclass[amsfonts,11pt]{article}
\usepackage{amsmath,amssymb,graphicx,makeidx}

\begin{document}


\tableofcontents
\newpage

\section{Introduction}\label{Introduction}

In standard quantum mechanics, in the Schr\"{o}dinger or Heisenberg formulation, 
phase space and its methods - intensely studied in classical mechanics -
play no role. In the quantization
procedure the classical variables $q$ (space) and $p$ (momentum) are promoted to 
operators $Q$ and $P$ satisfying the commutation relation $[Q,P]=i$, where
throughout the article $\hbar=1$. These act on a Hilbert space of state vectors 
in the well-known way. Formally there is a strong resemblance between 
classical dynamics formulated in terms of Poisson brackets and quantum dynamics
formulated with commutators (Heisenberg picture). But the classical
Heisenberg equations of motion are covariant under canonical transformations,
a property related to the symplectic structure of the underlying phase space
manifold, i.e. a geometric structure. On the other hand 
the canonical quantization procedure works only in one special 
coordinate system, namely Cartesian coordinates, as was pointed out by 
Dirac\cite{Dirac}\footnote{For an example that one can obtain wrong results
in other coordinate systems see e.g. \cite{Shankar}}.
This certainly is a dilemma, 
since quantization should not depend on something so
arbitrary as the choice of coordinates. And moreover Cartesian coordinates
exist only in flat space. 

This was fully recognized and many efforts were made to reformulate quantum
mechanics in phase space. This led e.g. to the so-called deformation
quantization and to geometric quantization 
schemes, i.e. quantizations which
respect the free choice of coordinates and allow 
canonical transformations. A short introduction to the general ideas 
of these quantizations will 
subsequently be given since some of their ideas are borrowed in the present 
work.\\

{\bf Deformation Quantization}\label{IntroDefQuant}\\

The oldest formulation of quantum mechanics in phase space is due to Wigner
\cite{Wigner1}\cite{Wigner2}, Moyal \cite{Moyal} and others
\cite{Baker}\cite{Fairlie}\cite{Bayen}\cite{Curtright1}\cite{Curtright2}\cite{Zachos}.

Wigner's idea was to introduce a quasi-probability function\footnote{Here
``quasi-probability" refers to the fact that these functions are 
not positive everywhere for general $\psi$.} on phase space 
(here 2-dimensional for sake of simplicity)

\begin{equation*}
F(p,q)={\textstyle\frac{1}{2\pi}}{
\textstyle\int}\psi^*(q-x/2)e^{-ixp}\psi(q+x/2)dx
\end{equation*}

\noindent which has the properties
$\int F(p,q) dp=|\psi(q)|^2$ and $\int F(p,q)dq=|\tilde{\psi}(p)|^2$ giving
the right quantum mechanical expectations for the state $|\psi\rangle$ in the
$q$- and $p$-representation respectively. 

Moyal \cite{Moyal} showed that when quantum mechanics is expressed in 
phase space
in this way quantum dynamics (and therefore the analog of the classical
Poisson bracket) is governed by the Moyal sine-bracket, which is the 
quantum deformed version of the Poisson bracket and the Moyal
equations of motion - analog to the Liouville equations and
with $H(p,q)$ the classical Hamiltonian - become:

\begin{equation*}
\partial_tF(p,q,t)=2\sin\{{\textstyle \frac{1}{2}}
[\partial_{p_F}\partial_{q_H}-\partial_{p_H}
\partial_{q_F}]\}H(p,q)F(p,q,t)
\end{equation*}

\noindent In the same paper Moyal also showed the equivalence of this 
statistical 
approach to quantum mechanics to the usual Schr\"{o}dinger formulation
(the Wigner function is a special representation of the density matrix in 
the so-called Weyl-correspondence, see e.g. \cite{Zachos}). The more modern
notation uses the $\star$-product, defined as $\star:=\exp\{\frac{i}{2}(
\overleftarrow{\partial}_x\overrightarrow{\partial}_p-
\overleftarrow{\partial}_p\overrightarrow{\partial}_x)\}$, 
where the arrows indicate the side this operator acts on, 
and the Moyal equations
become $i\partial_tF=H\star F-F\star H$. Since the Schr\"{o}dinger and this
phase space formulation are equivalent it is not surprising that things
like Heisenberg uncertainty relation etc. emerge naturally (\cite{Zachos}), but
on the other hand this means that it is not surprising either that canonical
transformations do not in general leave the $\star$-product 
invariant\footnote{Only linear canonical transformations do, \cite{Zachos}}.

An interesting point is the connection of the Wigner functions (or more precisely
of their Fourier transform) to coherent states. This will be discussed later 
after having introduced the concept of coherent states.\\

{\bf Geometric Quantization}\label{IntroGeomQuant}\\

Geometric quantization tries - as the name already implies - to formulate
the quantization procedure in such a way that it is fully geometric in 
character. The phase space $M$ is taken as the starting point and functions
$\psi(p,q)$, $(p,q)\in M$, are introduced which are supposed to be 
square integrable. The basic quantum kinematical operators $Q$ and $P$ are
then represented with linear differential operators according to some rule, 
e.g. $Q=i\partial_p+q/2$, $P=-i\partial_q+p/2$, the only requirement being that
they fulfill the Heisenberg commutation relation $[Q,P]=i$. 
But there is a problem since
these representations are reducible\footnote{For the example given above
the operators $\tilde{Q}=i\partial_p-q/2$ and $\tilde{P}=-i\partial_q-p/2$
commute with $Q$ and $P$ and are certainly not multiples of the identity.} and
this very fact - although non-critical from a mathematical point of 
view - simply leads to inconsistencies with physics. So the whole process
up to this point was eventually called pre-quantization.

The idea to remedy the situation is ``polarization" and initially it came
from the fact that the Schr\"{o}dinger
wave functions depend on $q$ only (and the operators $P=-i\partial_q$ and 
$Q=q$ are irreducibly represented). A real polarization $\partial_p\psi
(p,q)=0$ would lead to functions which depend on $q$ only but not to 
square integrable functions of phase space. So complex linear polarizations, 
say
$[\partial_q-i(\partial_p+p)]\psi(p,q)=0$ just to give an example, 
were imposed to
reduce the space of admissible functions since $L^2(M)$ was too big to lead to
irreducible representations of the basic kinematical operators $Q$ and $P$. 
Finally, by one way or another, schemes to deal with kinematics could be 
established which were also consistent with the kinematical versions 
of (standard)
quantum mechanics\footnote{For further reading see e.g. \cite{Sniatycki}}. 
The breakdown of the theory came with dynamics. The idea
was to copy the classical Poisson bracket and establish an operator formalism
which respects the same relations (thus conserving the geometric character
of the classical world). Suffice it to say that the rules derived in this
way simply led to wrong answers except for linear Hamiltonians (and by chance
for the harmonic oscillator) and the whole program was largely abandoned.  

The important connection for the present work is the idea of polarization. 
As will be 
shown some coherent states already have a linear complex polarization 
condition built into them by their very definition (and they automatically
form a subset
of $L^2(M)$ which leads to irreducible representations of the kinematical
operators). \\


{\bf Coherent States and their Path Integrals}\label{IntroCS}\\

Coherence is a very general concept and especially important in the field of
optics where it was studied for more than three centuries. It is thus not
surprising that the first coherent states, nowadays called canonical coherent
states, arose in quantum optics \cite{KlauderSudarshan}. 
The name ``coherent states" was chosen as an
analog to the classical coherent ensemble, i.e. an ensemble for which 
$\langle V^*(t_1)...V^*(t_r)V(t_{r+1})...V(t_{2r})\rangle
=Y^*(t_1)...Y^*(t_r)Y(t_{r+1})...Y(t_{2r})$, where $\langle...\rangle$ denotes
the ensemble average and $Y$ is a deterministic function. A (canonical)
coherent state $|\psi\rangle$ is a state for which similarly
$\langle\psi|a^\dagger(t_1)...a^\dagger(t_r
)$ $a
(t_{r+1})...a(t_{2r})|\psi\rangle
=z^*(t_1)...z^*(t_r)z(t_{r+1})...z(t_{2r})$, where $a$ and $a^\dagger$ are
the harmonic oscillator creation and annihilation operators and $z$ is a
complex valued function. Soon
the concept of coherent states gradually broadened.  
Since they will be extensively discussed in the following chapter it may 
suffice at this point to say that, generally speaking, 
they are by their very definition related
to a certain underlying phase space. Hence they are a third way to describe
quantum mechanics in such a way that phase space methods can be used 
(e.g. canonical 
transformations of the underlying phase space do not change the set of coherent
states associated with it). A quantization by path integrals based on coherent
states thus promises to be of a geometric nature once all difficulties with
well-definedness are overcome. Indeed, such a path integral involving a
metric regularization that leads to a
so-called Wiener measure has been developed by I. Daubechies and J.R. Klauder
and T. Paul, and this result is the basis to understand the present work which 
studies the existence of a similar path integral for even more general states.
These states are called weak coherent states and together with the coherent 
states they form the class of Klauder states.

\newpage
\section{Coherent States and the Coherent State Path Integral}\label{CS}

Initially, the term ``coherent states" was used in a
quite narrow sense, namely what are called ``canonical" coherent states today.
Gradually, the concept was extended, and ``generalized" coherent states or
``Overcomplete Families of States" were used in many applications. Klauder
was the first to lay open the fundamental properties they share, and they
are all subsumed in the sense of his very general definition as coherent 
states nowadays.

For a better understanding and to get a feeling for coherent states
this mathematical framework is not taken as a starting point 
for section (\ref{CSCCS}), but rather
the original canonical coherent states. Many 
interesting properties of coherent states can already be learned in the 
discussion of this example. This also makes it easier to understand the
general concept in the following section (\ref{CSCS}) which otherwise
might seem quite technical. Next, section (\ref{CSGroupCS}) specializes again
to so-called group-defined coherent states which will be of most interest 
in the subsequent work. The ultimate goal of the whole chapter is to establish
a well-defined coherent state path integral which will provide a 
quantization scheme fully geometric in nature. To this end
the standard coherent state path integral is presented first 
in section (\ref{CSsCSPI}) and 
its shortcomings are pointed out in section (\ref{CSMathPI}), followed by 
section (\ref{CSMathTools})
containing mathematical tools which prepare the ground for the understanding
of the final coherent state path integral with Wiener measure [section
(\ref{CSCSPI})].

\subsection{Canonical coherent states}\label{CSCCS}

Still the best known and most widely used coherent states are the canonical
coherent states. 

Let $a$, $a^\dagger$ be operators obeying the Heisenberg algebra, i.e. 
$[a,a^\dagger]=1$, and let $|0\rangle$ be the unique normalized state 
annihilated by $a$, i.e. $a|0\rangle=0$. Denote the eigenstates of the
number operator $N=a^\dagger a$ by 
$|n\rangle:=\frac{1}{\sqrt{n!}}(a^\dagger)^n|0\rangle$. 
Then the coherent states are defined as \cite{Klauder}

\begin{eqnarray}\label{CanonicalCS}
|z\rangle&:=&e^{za^\dagger-z^*a}|0\rangle=e^{-{\textstyle\frac{1}{2}}
|z|^2}e^{za^\dagger}
e^{-z^*a}|0\rangle\nonumber\\
&=&e^{-{\textstyle\frac{1}{2}}|z|^2}e^{za^\dagger}|0\rangle
=e^{-{\textstyle\frac{1}{2}}|z|^2}
\sum_{n=0}^\infty \frac{1}{\sqrt{n!}}z^n|n\rangle
\end{eqnarray}

\noindent The relation $a e^{za^\dagger}=e^{za^\dagger}e^{-za^\dagger}a 
e^{za^\dagger}=
e^{za^\dagger}(a+z)$ immediately follows and applying this to
the second but last equality in (\ref{CanonicalCS}) 
it is easily seen that $a|z\rangle=z|z\rangle$,
i.e. the canonical coherent states are the generalized eigenstates of the
annihilation (or lowering) operator $a$.
From Eq. (\ref{CanonicalCS}) the overlap is easily seen to be

\begin{equation}
\langle z_2|z_1\rangle=e^{-{\textstyle\frac{1}{2}}|z_1|^2
-{\textstyle\frac{1}{2}}|z_2|^2+z_2^*z_1}
\end{equation}

\noindent which is a jointly continuous, nowhere vanishing function 
of $z_1$ and $z_2$.

By the foregoing $\langle z|a|z\rangle=z\langle z|z\rangle=z$, 
so the label $z$ is the mean of the operator $a$ in the
coherent state $|z\rangle$.

Consequently each state $|z\rangle$ satisfies $\langle z|a^\dagger a|z\rangle
=|z|^2=|\langle z|a|z\rangle|^2$ and since the uncertainty relation for
$a$ and $a^\dagger$ simply is $\langle a^\dagger a\rangle\geq\langle a^\dagger 
\rangle\langle a\rangle$ every single canonical coherent state is a minimum
uncertainty state.

The vector-valued function $|z\rangle$ is continuous in $z$  
[i.e., if $z\rightarrow z^\prime$ then $\parallel\!\! |z\rangle
-|z^\prime\rangle \!\!\parallel \rightarrow 0$] and the complex-valued
function $\psi(z)=\langle z|\psi\rangle$ is bounded and continuous
in $z$ for all $|\psi\rangle\in\mbox{Hilbert space}$. Further $\psi(z)
=0\Leftrightarrow |\psi\rangle=0$ holds establishing a one-to-one relation
between functions $\psi(z)$ and vectors $|\psi\rangle$, and thus $\psi(z)$
can be taken as a representative of the vector $|\psi\rangle$.

With the help of the last expression in Eq. (\ref{CanonicalCS}) and the 
use of polar coordinates at an intermediate step it can be seen that
the canonical coherent states resolve unity in the form

\begin{eqnarray}
&&\pi^{-1}\int|z\rangle\langle z|d^2z:=\pi^{-1}\sum_{n,m}\frac{1}{\sqrt{n!m!}}
\int e^{-|z|^2}z^{*n}z^m|m\rangle\langle n|\,d[\Re (z)]d[\Im (z)]\nonumber\\
&&=\sum_n\frac{1}{n!}\int e^{-|z|^2}|z|^{2n}|n\rangle\langle n|\,d[|z|^2]
=\sum_n|n\rangle\langle n|=1\!\!1
\end{eqnarray}

\noindent Note that a resolution of unity guarantees the existence of a 
coherent
state representation of Hilbert space (it is a stronger property than
$\psi(z)=0\Leftrightarrow |\psi\rangle=0$ which was already sufficient for this
purpose). 

Since the canonical coherent states depend only on $z$ but not on $z^*$ 
the functions $\psi(z):=\langle z|\psi\rangle=e^{-|z|^2/2}f(z^*)$ 
in the coherent state representation are entire functions $f(z^*)$ apart 
from an overall factor. This shows that the canonical coherent states are
overcomplete (especially they can not all be orthogonal) since by analyticity
the vanishing of $\psi(z)$ on a characteristic set (e.g. any curve of nonzero
length in the complex plane) already implies that $\psi(z)\equiv 0$.

By analyticity one has $\partial_z(e^{|z|^2/2}\langle z|\psi\rangle
=e^{|z|^2/2}(\partial_z+z^*/2)\langle z|\psi\rangle=0$ and thus
$(\partial_z+z^*/2)\langle z|\psi\rangle=0$ which is a complex polarization
condition. Note that this is really a consequence of $a|0\rangle=0$ since
$(\partial_z+z^*/2)\langle z|\psi\rangle=\langle 0|(\partial_z+z^*/2)
e^{-{\textstyle\frac{1}{2}}|z|^2}e^{za^\dagger}|\psi\rangle
=\langle 0|-a^\dagger e^{-{\textstyle\frac{1}{2}}|z|^2}e^{za^\dagger}
|\psi\rangle=0$. 

A further consequence is that an operator $B$ is already determined by its
diagonal coherent state matrix elements $B(z):=\langle z|B|z\rangle
=\exp\{-|z|^2\}
$ $\sum_{n,m}1/\sqrt{n!m!}\cdot\langle n|B|m\rangle z^{*n}z^m$ since
every monomial $z^{*n}z^m$ can be isolated by its dependence on $\theta$ and
$r$ (with $z:=re^{i\theta}$) and analytically extended to $z^{*n}z^{\prime m}$
resulting in the general matrix element $\langle z|B|z^\prime\rangle$.
The diagonal matrix element $B(z)$ as a function of $z$ is 
called the upper symbol or in this
case of canonical coherent states the normal ordered symbol.

Additionally there are two other operator representations. The first is 
called diagonal representation and is given by $B=\pi^{-1}\int b(z)|z\rangle
\langle z|d^2z$. The function $b(z)$ is called the lower symbol or in the case 
of canonical
coherent states the anti-normal ordered symbol\footnote{Many authors use 
the terminology exactly the other way round and call the upper symbol the lower
one and vice versa. The reason is that the upper symbol (their lower symbol) is
involved in a lower bound in certain inequalities, 
the Berezin-Lieb inequalities, and
the lower symbol (their upper symbol) is involved in an upper bound in the same 
inequalities.}.

The second is a 
differential operator representation. The creation and annihilation operators
e.g. are represented by $(a^\dagger\leftrightarrow z^*)$ and $(a\leftrightarrow
z/2+\partial_{z^*})$ since $\langle z|a^\dagger|\psi\rangle=
z^*\langle z|\psi\rangle$ and $\langle z|a|\psi\rangle=(z/2+\partial_{z^*})
\langle z|\psi\rangle$. A general operator $B(a^\dagger, a)$ is consequently
represented by $B(z^*, z/2+\partial_{z^*})$. 

To show just a glimpse of the power of coherent states consider the so-called
Segal-Bargmann (or holomorphic) representation of Hilbert space. Let
$z^*=\zeta$ and let the
normalization factor be absorbed into the measure $d\mu(\zeta)=\pi^{-1}
\exp\{-|\zeta|^2\}d^2\zeta$. 
Then $\psi(\zeta):=\langle \zeta|\psi\rangle$ are entire
(holomorphic) functions with inner product $\langle\psi|\psi\rangle
=\int|\psi(\zeta)|^2d\mu(\zeta)$ on the Hilbert space. In this form the 
operators
$a^\dagger$ and $a$ are represented by $\zeta$ and $\partial_\zeta$ 
respectively. The eigenvalue equation for the number operator $N=a^\dagger a$
becomes $\zeta\partial_\zeta\psi(\zeta)=\lambda\psi(\zeta)$ with solutions
$\psi(\zeta)=c\zeta^\lambda$. But by analyticity $\lambda\in\mathbb{N}_0$ and
therefore the spectrum of the number operator is derived in just these few 
lines!

Another example of the outstanding properties of the canonical 
coherent states is the
stability under time evolution in a harmonic potential: $\exp\{-it\omega 
a^\dagger a\}|z\rangle=\exp\{-|z|^2/2\sum_n 
{\textstyle\frac{1}{\sqrt{n!}}} z^n\exp\{
-in\omega t\}|n\rangle=|\exp\{-i\omega t\}z\rangle$, and since all canonical
coherent states are minimum uncertainty states this property is preserved 
for all times.

For a wide range of applications of the canonical coherent states and the
general coherent states which will be defined in the following section, 
see e.g. \cite{Klauder}.

\subsection{General coherent states}\label{CSCS}

The definition of coherent states used here follows Klauder \cite{Klauder}.
Let $\mathfrak{L}$ be a topological space called the ``label space" and for 
$l\in\mathfrak{L}$ denote by $|l\rangle$ a vector in a Hilbert space 
$\mathfrak{H}$. For $|l\rangle$ to be called a coherent state two properties
are required:\\

{\bf 1) Continuity}: The states $|l\rangle$ are a strongly continuous 
vector-valued function of the label $l$.\\

{\bf 2) Resolution of unity}: There is a positive measure $\delta l$ on the label 
space $\mathfrak{L}$ such that the identity operator $1\!\!1$
on $\mathfrak{H}$ can upon integration over $\mathfrak{L}$ be represented as 

\begin{equation}\label{ResUnity}
1\!\!1=\int |l\rangle\langle l|\delta l
\end{equation}

Condition 1) says: If $l^\prime\rightarrow l$ in $\mathfrak{L}$ then 
$\parallel\!\! |l^\prime
\rangle-|l\rangle\!\!\parallel\rightarrow 0$ in $\mathfrak{H}$. Hence, the set 
$\{|l\rangle:l\in\mathfrak{L}\}$ is a continuous submanifold of $\mathfrak{H}$
if $\mathfrak{L}$ is connected.

Condition 2) is to be understood in the sense of weak convergence, 
i.e. arbitrary matrix
elements converge in $\mathbb{C}$. Condition 2) implies completeness of the 
set
$\{|l\rangle:l\in\mathfrak{L}\}$. Without loss of generality (w.l.o.g.) all
$|l\rangle\in\mathfrak{H}$ can be assumed to be normalized to 1 since the 
measure $\delta l$ on $\mathfrak{L}$ can be rescaled at every point to 
allow for this
choice. Unless stated otherwise this convention will be adopted throughout
the article.

To get a better feeling of what coherent states ``are" it is certainly 
instructive to see what they are not. They can not be a set of discrete
orthonormal vectors $\{|n\rangle: n\in\mathbb{N}\}$ which is not continuous
in the labels and they can not be a set of $\delta$-orthonormalized continuum
vectors $\{|x\rangle: x\in\mathbb{R}\}$ which are continuous in the labels but
do not form a continuous set 
(apart from the fact that they are not vectors in Hilbert
space, but so-called generalized vectors). In other terms, coherent states 
can not be the eigenvectors (in the usual and generalized sense) of any self
adjoint operator.

Also there is a difference in the resolution of unity afforded by the coherent
states to the usual one furnished by generalized eigenvectors of self-adjoint
operators (and there should be since the latter always exists. So it would not 
make sense to include the resolution of unity in the defining properties for 
coherent states if there wasn't more to it). The one-dimensional projection
operators $|l\rangle\langle l|$ in Eq. (\ref{ResUnity}) are not in general
mutually orthogonal. The resolution of unity gives rise to a functional
representation

\begin{equation}\label{FuncRepr}
\langle l|\phi\rangle=\int\langle l|l^\prime\rangle\langle l^\prime|\phi\rangle
\delta l^\prime
\end{equation}

For $\delta$-orthonormalized continuum
vectors an equation of this form would merely be an identity, but 
because of the continuity requirement for coherent
states the kernel $\langle l|l^\prime\rangle$ is 
a jointly continuous function of $l$ and $l^\prime$, 
nonzero for $l=l^\prime$ (for the
normalized coherent states the value at this point is 1) 
and therefore nonzero in an open
neighborhood due to continuity. This implies severe restrictions on the 
admissible functions. 
One speaks of a reproducing kernel and the subspace spanned by the coherent
states is itself a so-called reproducing kernel Hilbert space (see e.g. \cite
{Meschkowski}).

The coherent states are linearly dependent as is evident from
$|l^\prime\rangle=\int |l\rangle\langle l|l^\prime\rangle \delta l$. This is why 
they are often called an ``overcomplete" family of states.

\subsection{Group defined coherent states}\label{CSGroupCS}

In this work group related coherent states will play an important
role. These are 
defined in the following way \cite{Klauder}\cite{Perelomov}:

Let $G$ be a Lie group and $U(g)$, $g\in G$, a strongly continuous irreducible
unitary representation on a Hilbert space $\mathfrak{H}$. Let $|\psi_0\rangle$
be an arbitrary but fixed normalized vector in $\mathfrak{H}$ called the 
``fiducial"
vector, then the action of $U$ on $|\psi_0\rangle$ creates the coherent states
of the group $G$:

\begin{equation}
|g\rangle:=U(g)|\psi_0\rangle
\end{equation}

\noindent where $g$ runs over the whole group.

Assume for a moment that the group has $n$ infinitesimal self-adjoint 
generators $L_i$, 
$i\in\{1,...,n\}$, and let $l\in\mathfrak{L}$ be an element of an n-dimensional
``label" space (a topological space) such that the group is represented in 
group coordinates of the second kind as $U(l)=e^{il_1L_1}...e^{il_nL_n}$
which is supposed to be irreducible. Then we call the label space the 
phase space associated with the group. Under canonical transformations of the 
phase space variables such that $l_1=l_1(\bar{l_1},...,\bar{l_n})$, ... ,
$l_n=l_n(\bar{l_1},...,\bar{l_n})$ and with an arbitrary phase 
$\bar{F_0}(\bar{l_1},...,\bar{l_n})$ it follows that

\begin{equation}
|l_1...l_n\rangle\rightarrow\overline{|\bar{l_1}...\bar{l_n}\rangle}
:=e^{-i\bar{F_0}(\bar{l_1},...,\bar{l_n})}|l_1(\bar{l_1},...,\bar{l_n})
...l_n(\bar{l_1},...,\bar{l_n})\rangle
\end{equation}

\noindent Apart from a phase factor the set of coherent state vectors has 
remained the
same, the only difference being the way they are labeled. The 
``new" coherent states are still generated by the action of the group $G$ on 
the fiducial vector $|\psi_0\rangle$ but the group is generally no longer 
represented in canonical group coordinates \cite{KlauderQisG}. 

The overlap of two states $|g\rangle$ and $|g^\prime\rangle$, 
$\langle g|g^\prime\rangle=\langle\psi_0|U(g)^\dagger U(g^\prime)|\psi_0\rangle
=\langle\psi_0|U(g^{-1}*g^\prime)|\psi_0\rangle$, is bounded (by unity) and
jointly continuous in the group parameters. 

Completeness of these states follows immediately from the irreducibility of
$U$ \cite{Perelomov} but
to qualify as coherent states they must
admit a resolution of unity in the form $1\!\!1=\int |l\rangle\langle l|\delta l$.
Klauder has shown \cite{KlauderCRTI}\cite{KlauderCRTII} 
that the integration measure 
can be taken as the left-invariant group measure $dg$\footnote{For 
compact groups
left- and right-invariant measure always coincide. More generally left- and
right-invariant group measure coincide if the structure constants of the
group algebra satisfy $C_{ab}^b=0$.}. If the representation
$U$ of $G$ is square integrable, i.e. $\int |\langle\psi_0|U(g)|
\psi_0\rangle|^2dg<\infty$\footnote{This bound 
is automatic if the group volume is finite which is
always the case for compact groups, but it can impose restrictions for 
non-compact groups.}, then - with the normalization
$\int |\langle\psi_0|U(g)|\psi_0\rangle|^2dg=1$ - the invariance of the
measure along with Schur's Lemma provides a resolution of unity 

\begin{equation}
1\!\!1=\int_G|g\rangle\langle g|dg
\end{equation}

Next consider a general operator ${\cal H}$ on the Hilbert 
space $\mathfrak{H}$.
Associated with it are the upper symbol 

\begin{equation}\label{UpperSymbol}
H(g)=\langle g|{\cal H}|g\rangle
\end{equation}

\noindent which is the diagonal coherent state matrix element for given $g$, 
and (if it exists) the lower symbol 
implicitly defined as the diagonal representative of ${\cal H}$ 

\begin{equation}\label{LowerSymbol}
{\cal H}=\int h(g)|g\rangle\langle g|dg
\end{equation}

\noindent It was shown by Lieb (\cite{Klauder}, page 35) that the set of 
operators
admitting a diagonal representation, i.e. for which the lower symbol exists,
is identical to the set of operators uniquely specified by the diagonal
coherent state matrix elements alone, i.e. by the upper symbol.

If ${\cal H}$ is interpreted as the quantum Hamiltonian then each of the 
symbols is interpreted in various physical applications as the associated
classical Hamiltonian. For the upper symbol this association is known as
the ``weak correspondence principle" \cite{KlauderCRTII}\cite{KlauderPI}.

The symbols associated with an operator will be very important in the 
coherent state path integral formulation.

In chapter \ref{WCS} this work will deal almost entirely 
with affine coherent states 
which are special coherent states generated by the affine group, often 
called the
ax+b-group, which is a subgroup of SU(1,1). 
Nevertheless for illustration and for the 
sake of comparision it is useful to start with the most common coherent states,
the canonical coherent states, which are the coherent states of the 
Heisenberg-Weyl group. After that example a short overview of the spin 
coherent states, the coherent 
states of the SU(2) group, will follow before the affine coherent states 
are discussed. \\

{\bf Canonical coherent states - phase space formulation}:\\

The Heisenberg-Weyl group is generated by the self-adjoint canonical operators
$P$ and $Q$ satisfying the Heisenberg commutation relation $[Q,P]=i$ and
has exactly one irreducible unitary representation 
up to unitary equivalence which is e.g. given in canonical group coordinates of
the second kind by

\begin{equation}
U(p,q)=e^{-iqP}e^{ipQ}
\end{equation}

\noindent where $(p,q)\in\mathbb{R}\times\mathbb{R}$.

The canonical coherent states are generated by the action of this group on 
the harmonic oscillator ground state $|0;\omega\rangle$ which satisfies
$(\omega^{1/2}Q+i\omega^{-1/2}P
)$ $|
0;\omega\rangle=0$ 
($\omega\in\mathbb{R}$ is the oscillator frequency).  
The ground state will often be written as $|0\rangle$ with the 
$\omega$-dependence left implicit. As stated previously, the ground state is a 
minimum uncertainty state. The definition of the coherent states is

\begin{equation}\label{DefCCS}
|pq\rangle:=U(p,q)|0\rangle
\end{equation}

The canonical coherent states introduced in the beginning of this 
chapter in their complex characterization are the same states 
apart from an unimportant phase factor. To see this rewrite the 
annihilation operator as $a=2^{-1/2}(\omega^{1/2}Q+i\omega^{-1/2}P)$, 
the creation operator as $a^\dagger
=2^{-1/2}(\omega^{1/2}Q-i\omega^{-1/2}P)$ 
and the complex label $z$ as $z:=2^{-1/2}(\omega^{1/2}q+i\omega^{-1/2}p)$. 
Then
$U(z):=e^{za^\dagger-z^*a}=e^{i(pQ-qP)}=:\tilde{U}(p,q)$ and the group 
representation is given in canonical group coordinates of the first kind.

The overlap of two states defined by Eq. (\ref{DefCCS}) is

\begin{equation}\label{OverlapCCS}
\langle pq|p^\prime q^\prime\rangle=e^{i/2(p+p^\prime)(q-q^\prime)}
e^{-1/4[\omega^{-1}(p-p^\prime)^2+\omega(q-q^\prime)^2]}
\end{equation}

\noindent whereas the phase would be different if the group had been defined
with group coordinates of the first kind, namely 
$\langle pq|p^\prime q^\prime\rangle=\exp\{\frac{i}{2}(p^\prime q-q^\prime p)
\}$ $
\exp\{-\frac{1}{4}[\omega^{-1}(p-p^\prime)^2+\omega(q-q^\prime)^2]\}$.
The Weyl operator $U$ serves as a translation operator for $P$ 
and $Q$

\begin{eqnarray}\label{TranslationOperators}
U^\dagger (p,q)QU(p,q)&=&Q+q\nonumber\\
U^\dagger(p,q)PU(p,q)&=&P+p
\end{eqnarray}

\noindent showing that the coherent states $|pq\rangle$ are labeled by the 
means of $p$ and $q$ since $\langle 0|Q|0\rangle=\langle 0|P|0\rangle=0$ and so
$\langle pq|Q|pq\rangle=q$ and $\langle pq|P|pq\rangle=p$.

The resolution of unity reads

\begin{equation}
1\!\!1=(2\pi)^{-1}\int|pq\rangle\langle pq|dp\,dq
\end{equation}

\noindent Since the
group measure is both left- and right-invariant, this implies that all fiducial
vectors lead to square-integrable representations if just one vector 
does \cite{Klauder}. And clearly there is one such vector, namely the harmonic
oscillator ground state $|0\rangle$.

General bounded operators, or operators polynomial in $P$ and $Q$, are uniquely
determined by their diagonal coherent state matrix elements\footnote{This
is a direct consequence of analyticity most easily exhibited in the
complex representation of the canonical coherent states.}. In other words
the upper symbol $H$ associated with an operator ${\cal H}$ of this type 
uniquely determines the operator. At the same time the lower symbol $h$
exists for these operators and 
generally $H(p,q)-h(p,q)=O(\hbar)$, which is unfortunately
not easy to see here due to the suppression of $\hbar$ in this work.

The coherent states of the Heisenberg-Weyl group in general, i.e. with an
arbitrary fiducial vector, do not (in general) have the property of 
analyticity\footnote{This is most easily seen in the complex representation of 
the canonical coherent states.}
but otherwise share all essential properties of the canonical coherent states
such as overcompleteness, resolution of unity, universal representation of
the operators $P$ and $Q$ by differential operators, etc.

The upper symbol uniquely determines an operator polynomial in $P$ and $Q$ and
this is even true for a general operator if the coherent state overlap
(reproducing kernel) nowhere vanishes, which happens for uncountably many 
fiducial vectors \cite{KlauderCRTIII}. 
The existence of the lower symbol is guaranteed
for these operators.

An interesting property of the Heisenberg-Weyl group is the relation it
establishes between its coherent states and the Wigner/Moyal phase space
formulation of quantum mechanics. 

For any states $|\psi_1\rangle$ and $|\phi_1\rangle$ 
the $x$-representation
of $\langle \psi_1|U(p,q)|\phi_1\rangle=\int\psi_1^*(x+q/2)e^{ipx}\phi_1
(x-q/2)dx$ is the Fourier transform of a general Wigner 
function \cite{Moyal}\cite{Curtright2}. Also, for arbitrary  
$|\psi_2\rangle$ and $|\phi_2\rangle$,

\begin{equation}
(2\pi)^{-1}\int\langle\psi_1|U(p,q)|\phi_1\rangle\langle\psi_2|U(p,q)|\phi_2\rangle^*
dp\,dq =\langle\psi_1|\phi_1\rangle\langle\psi_2|\phi_2\rangle
\end{equation}

\noindent holds, which essentially is a group orthonormality relation 
proved by Moyal. But, dropping 
$\langle\psi_1|$ and $|\psi_2\rangle$, this expression
exhibits the resolution of unity in the form 
$(2\pi)^{-1}\int U(p,q)|\phi_1\rangle\langle\phi_2|U^\dagger(p,q)dp\,dq
=\langle\phi_2|\phi_1\rangle1\!\!1$. \\

{\bf Spin coherent states}:\\

The spin coherent states are the group-defined coherent states of the SU(2)
group the generators of which satisfy the spin algebra $[S_l,S_m]=
i\epsilon_{lmn}S_k$, $l,m,n\in\{1,2,3\}$, and $\epsilon$ is the totally 
antisymmetric tensor (Levi-Civita tensor) and summation over repeated indices
is understood. One strongly continuous unitary irreducible representation
is given by $U(\theta, \phi, \psi):=e^{-i\phi S_3}e^{-i\theta S_2}
e^{-i\psi S_3}$ 
, $0\leq\theta\leq\pi$, $0\leq\phi<2\pi$, $0\leq\psi<2\pi$, 
which is called the Euler angle characterization.

The operator $U$ acts on 
a $(2s+1)$-dimensional Hilbert space, where $s$ is the spin. 
If the fiducial vector is chosen as an eigenstate $|m\rangle$ of $S_3$ then the 
operator $e^{-i\psi S_3}$ becomes a simple phase and can be omitted leading to

\begin{equation}
U(\theta, \phi)|m\rangle=e^{-i\phi S_3}e^{-i\theta S_2}|m\rangle
=|\theta\phi\rangle
\end{equation}

\noindent closely resembling the case of the canonical coherent states since
it involves only two variables.

The expression $\sin\theta d\theta d\phi$ is proportional to the 
invariant
group measure and the group volume is $4\pi$ which is finite since
the group is compact. Therefore the resolution of unity becomes

\begin{equation}
1\!\!1={\textstyle\frac{2s+1}{4\pi}}\int|\theta\phi\rangle\langle\theta\phi|
\sin\theta d\theta d\phi
\end{equation}

\noindent and the coherent state overlap reads

\begin{eqnarray}
\langle\theta\phi|\theta^\prime\phi^\prime\rangle
&=&\langle m|e^{i\theta S_2}e^{i(\phi-\phi^\prime)S_3}e^{-i\theta^\prime S_2}
|m\rangle\nonumber\\
&=&\sum_{n=-s}^s\langle m|e^{i\theta S_2}|n\rangle\langle n|e^{-i\theta^\prime
S_2}|m\rangle e^{i(\phi-\phi^\prime)n}
\end{eqnarray}

\noindent which is recognized to be expressed in terms of reduced Wigner 
coefficients
of the spin-s representation. The overlap reduces to a simple expression 
for extremal weight vectors ($m=\pm s$). For $m=s$ it is

\begin{eqnarray}
\langle\theta\phi|\theta^\prime\phi^\prime\rangle
&=&[\cos(\theta/2)\cos(\theta^\prime/2)e^{i/2(\phi-\phi^\prime)}]^{2s}\nonumber\\
&&\times[1+\tan(\theta/2)\tan(\theta^\prime/2)e^{-i(\phi-\phi^\prime)}]^{2s}
\end{eqnarray}

If $m=s$ then every operator ${\cal H}$ is uniquely determined by 
its upper
symbol $H(\theta,\phi)$ and thus every operator admits a diagonal representation,
i.e. the lower symbol exists \cite{Klauder}\cite{Sudarshan}. This is not 
necessarily true
for a choice of fiducial vector other than the extremal weight ones\footnote{
See \cite{Klauder}, page 34, for a counterexample.}.

Observe that the Euler angle characterization is given in non-canonical 
coordinates. 
For canonical coordinates, $p=s^{1/2}\sin\theta$ and $q=s^{1/2}\phi$, the 
overlap becomes

\begin{eqnarray*}
\langle pq|p^\prime q^\prime\rangle&=&{\textstyle \frac{2s+1}{4\pi s}}
[{\textstyle \frac{1}{2}}(1+s^{-1/2}p)^{1/2}(1+s^{-1/2}p^\prime)^{1/2}
\exp\{is^{-1/2}(q-q^\prime)/2\}\\
&&+{\textstyle \frac{1}{2}}(1-s^{-1/2}p)^{1/2}(1-s^{-1/2}p^\prime)^{1/2}
\exp\{-is^{-1/2}(q-q^\prime)/2\}]^{2s}
\end{eqnarray*}

\noindent and in this notation it can be seen that in the limit 
$s\rightarrow\infty$
the overlap reduces to the overlap for canonical coherent states, equation
(\ref{OverlapCCS}), for oscillator frequency $\omega=1$. \\

{\bf Affine coherent states}:\\

The ``ax+b"-group, or affine group,
is the set $M_+:=\mathbb{R}^+\times\mathbb{R}$ with the
group law $(a^\prime, b^\prime)(a,b)=(a^\prime a, b^\prime+a^\prime b)$. The
group has two nontrivial inequivalent irreducible unitary representations 
$U_\pm$ with 
realizations on $L^2(\mathbb{R}^+)$: $[U_\pm(a,b)\psi](x)=a^{1/2}e^{\pm ibx}
\psi(ax)$ \cite{DKP}.

Although still referred to as the ``ax+b"-group, new coordinates (labels)
$a\rightarrow q^{-1}$ and $b\rightarrow p$ will be used throughout this work.

The generators of the representations $U_\pm$ are the affine kinematical
variables $Q$ with $Q>0$ (position operator) and $D$ (dilation operator) 
satisfying the affine commutation relation

\begin{equation}\label{AffineCommRel}
[Q,D]=iQ
\end{equation}

\noindent Then the representations in group coordinates of the second kind are

\begin{equation}
U_\pm(p,q)=e^{\pm ipQ}e^{-i\ln q\,D}
\end{equation}

\noindent Only $U_+$ will later be important, so $U:=U_+$.

As in the canonical and spin case, the main interest will concentrate on 
the group action on extremal weight vectors. 
The minimum uncertainty states, satisfying the Heisenberg uncertainty equation
$\Delta Q\Delta P=\langle Q\rangle/2$, are a 2-parameter family given in
x-representation as \cite{KlauderWCS} 
$\eta_{\alpha,\beta}(x)=N_{\alpha,\beta}x^\alpha 
e^{-\beta x}$.

Setting the mean of $Q$ to 1 will lead to a 1-parameter family of minimum 
uncertainty states

\begin{equation}\label{AffineFiducialVector}
\eta_\beta(x)=N_\beta x^{\beta-1/2}e^{-\beta x}
\end{equation}

\noindent with the normalization 
$N_\beta=(2\beta)^\beta \Gamma^{-1/2}(2\beta)$.

The vectors $|\eta_\beta\rangle$ satisfy $(Q-1+i\beta^{-1}D)|\eta_\beta\rangle
=0$, which is the analog of $(\omega^{1/2} Q+i\omega^{-1/2}P)|0\rangle=0$ 
in the canonical case.

This again leads to a complex polarization condition (as in the canonical
case) and to analytic properties of the affine coherent states 
(most easily recognized in the overlap).

The affine coherent states are defined as 

\begin{equation}
|pq\rangle=U(p,q)|\eta_\beta\rangle
\end{equation}

\noindent and the $\beta$-dependence will often be left implicit, 
i.e. $|\eta_\beta\rangle=:|\eta\rangle$.

For the whole parameter range $0<\beta$ the overlap is

\begin{equation}\label{AffineOverlap}
\langle pq|rs\rangle=(qs)^{-\beta}2^{-2\beta}[(q^{-1}+s^{-1})+i\beta^{-1}
(p-r)]^{-2\beta}
\end{equation} 

The left-invariant group measure is proportional to $dp\,dq$ whereas 
the
right-invariant group measure is proportional to $dp\,dq/p$. This is the first
example where the two measures differ and it follows from the general discussion 
that square integrability of the representation, necessary for a 
resolution of unity, can impose restrictions on the choice of fiducial vectors.
This indeed happens, and there is a fiducial vector admissibility condition
\cite{Aslaksen}\cite{DKP}\cite{KlauderWCS}

\begin{equation}\label{FVAC}
\langle Q^{-1}\rangle=\int_0^\infty x^{-1}|\eta_\beta(x)|^2dx<\infty
\end{equation}

\noindent This condition is violated if $0<\beta\leq 1/2$ and fulfilled 
for $\beta>1/2$
in which case the resolution of unity reads

\begin{equation}
1\!\!1=(1-{\textstyle\frac{1}{2\beta}})(2\pi)^{-1}
\int|pq\rangle\langle pq|dp\,dq
\end{equation}

Observe that adding the phase factor $\exp\{-i\beta^{1/2}(p-r)\}$
to the 
overlap and translating $q$ to $q+\beta^{1/2}$ one obtains in the limit
$\beta\rightarrow\infty$ the overlap of the canonical coherent states (again
for unit angular frequency) \cite{KlauderQisG}. \\

\begin{table}[tb]
{\small
\begin{tabular}{|rl|}
\hline
&{\bf Canonical coherent states}\\
\hline
generators&$(P,Q)$\\
phase space variables&$(p,q)$\\
coherent states&$|pq\rangle=\exp\{-iqP\}\exp\{ipQ\}|0\rangle$\\
overlap&$\langle pq|p^\prime q^\prime\rangle= e^{i/2(p^\prime q-q^\prime p)}
e^{-1/4[(p-p^\prime)^2+(q-q^\prime)^2]}$\\
canonical 1-form&$i\langle pq|d|pq\rangle=p\,dq$\\
induced metric&$d\sigma^2=2[\parallel\!\! d|pq\rangle\!\!\parallel^2-|\langle pq|d|
pq\rangle|^2]=dp^2+dq^2$\\
\hline\hline
&{\bf Spin coherent states}\\
\hline
generators&$(S_2,S_3)$\\
phase space variables&$(\theta,\phi)$ or $(p,q)$\\
coherent states& $|\theta\phi\rangle=e^{-i\phi S_3}e^{-i\theta S_2}|s\rangle$\\
overlap {\scriptsize (non-can. coord.)}
&$\langle\theta\phi|\theta^\prime\phi^\prime\rangle
\sum_{n=-s}^s\langle m|e^{i\theta S_2}|n\rangle\langle n|e^{-i\theta^\prime
S_2}|m\rangle e^{i(\phi-\phi^\prime)n}$\\
overlap {\scriptsize (can. coord.)}&$\langle pq|p^\prime q^\prime\rangle=
{\textstyle \frac{2s+1}{4\pi s}}
[{\textstyle \frac{1}{2}}(1+s^{-1/2}p)^{1/2}(1+s^{-1/2}p^\prime)^{1/2}$\\
&$\phantom{\langle pq|p^\prime q^\prime\rangle={\textstyle \frac{2s+1}{4\pi s}}}
\times\exp\{\frac{i}{2}s^{-1/2}(q-q^\prime)\}$\\
&$\phantom{\langle pq|p^\prime q^\prime\rangle={\textstyle \frac{2s+1}{4\pi s}}}
+{\textstyle \frac{1}{2}}(1-s^{-1/2}p)^{1/2}(1-s^{-1/2}p^\prime)^{1/2}$\\
&$\phantom{\langle pq|p^\prime q^\prime\rangle={\textstyle \frac{2s+1}{4\pi s}}}
\times\exp\{-\frac{i}{2}s^{-1/2}(q-q^\prime)\}]^{2s}$\\
non-canonical 1-form&$i\langle \theta\phi|d|\theta^\prime\phi^\prime\rangle
=s\sin\theta d\theta d\phi$\\
canonical 1-form&$i\langle pq|d|pq\rangle=p\,dq$\\
metric {\scriptsize (non-canonical)}
&$d\sigma^2=s(d\theta^2+\sin^2\theta d\phi^2)$\\
metric {\scriptsize (canonical)}&$d\sigma^2=(1-p^2/s)^{-1}dp^2+(1-p^2/s)dq^2$\\
\hline\hline
&{\bf Affine coherent states}\\
\hline
generators&$(Q,D)$\\
phase space variables &$(p,q)$\\
coherent states&$|pq\rangle=\exp\{ipQ\}\exp\{-i\ln qD\}|\eta\rangle$\\
overlap&$\langle pq|rs\rangle=
(qs)^{-\beta}2^{-2\beta}[(q^{-1}+s^{-1})+i\beta^{-1}(p-r)]^{-2\beta}$\\
canonical 1-form&$i\langle pq|d|pq\rangle=p\,dq$\\
induced metric&$d\sigma^2=2[\parallel\!\! d|pq\rangle\!\!\parallel^2-|\langle pq|d|
pq\rangle|^2]=\beta^{-1}q^2dp^2+\beta q^{-2}dq^2$\\
\hline
\end{tabular}
}
\caption{Coherent states}\label{CoherentStates}
\end{table}

\subsection{The standard coherent state path integral}\label{CSsCSPI}

Let ${\cal H}$ be the Hamiltonian operator for a quantum mechanical system
which is supposed not to depend on time explicitly\footnote{
Generalization to time-dependent Hamiltonians is straightforward and will 
involve the time-ordering operator $\mathbf{T}$.}. Then the time
evolution operator is $\hat{U}(t^{\prime\prime}-t^\prime)=\exp\{-i{\cal H}
(t^{\prime\prime}-t^\prime)$ and 
w.l.o.g. $t^\prime=0$, $t^{\prime\prime}=T$. Then $|\psi(t)\rangle
=\hat{U}(t)|\psi(0)\rangle$ solves the Schr\"{o}dinger
equation.

The standard approach to a path integral is the factorization of the 
evolution operator $\exp\{-i{\cal H}T\}$ into $N$ equal terms
$\exp\{-i{\cal H}\epsilon\}...\exp\{-i{\cal H}\epsilon\}$ with $N\epsilon=T$.
Then resolutions of unity are inserted between these factors, in 
this case given by the expression $1\!\!1=\int|l\rangle\langle l|\delta l$ 
belonging the coherent states labeled by $l$. This leads
to 

\begin{equation*}
\langle l^{\prime\prime};T|l^\prime;0\rangle\\
\equiv \int\prod_{k=0}^N\langle l_{k+1}|e^{-i\epsilon{\cal H}}|l_k
\rangle\prod_{k=1}^N\delta l
\end{equation*}

\noindent and since this is an identity it holds in the limit 
$\epsilon\rightarrow 0$
taken as a final step. As usual $l_{N+1}:=l^{\prime\prime}$ 
and $l_0:=l^\prime$. 
For small $\epsilon$ the approximation for each factor

\begin{eqnarray}\label{FactorApproximation}
&&\langle l_{k+1}|e^{-i\epsilon {\cal H}}|l_k\rangle
\approx \langle l_{k+1}|(1-i\epsilon {\cal H})|l_k\rangle\nonumber\\
&=&\langle l_{k+1}|l_k\rangle[1-i\epsilon
\langle l_{k+1}| {\cal H}|l_k\rangle
\langle l_{k+1}|l_k\rangle^{-1}]\nonumber\\
&\approx& \langle l_{k+1}|l_k\rangle\exp\{-i\epsilon
\langle l_{k+1}| {\cal H}|l_k\rangle
\langle l_{k+1}|l_k\rangle^{-1}\}
\end{eqnarray}

\noindent is correct to lowest order in $\epsilon$ \cite{KlauderPI}. 
With 
$H(l_{k+1};l_k):=\langle l_{k+1}|l_k\rangle^{-1}
\langle l_{k+1}| {\cal H}|l_k\rangle$
it follows that

\begin{eqnarray}\label{sCSPILattice}
&&\langle l^{\prime\prime};T|l^\prime ;0\rangle
=\lim_{\epsilon\rightarrow 0}\int\prod_{k=0}^N[
\langle l_{k+1}|l_k\rangle
\exp\{-i\epsilon 
H(l_{k+1};l_k)\}]\prod_{k=1}^N\delta l
\end{eqnarray}

\noindent This is a well-defined expression if the integrals 
converge\footnote{If
the integrals do not converge, 
the last approximation in Eq. (\ref{FactorApproximation}) 
- although correct in the integrand - simply did not result in an integrable 
function
and hence can not be made.}. It is customary
at this point to interchange the integrations and the limit although this
is not justified and replace the variables $l_k$ by real functions
$l(t)$. With

\begin{equation*}
\prod_{k=0}^N\langle l_{k+1}|l_k\rangle
=\prod_{k=0}^N[1-\langle l_{k+1}(|l_{k+1}\rangle-|l_k\rangle)]
\approx \exp\{-\sum_{k=0}^N
\langle l_{k+1}(|l_{k+1}\rangle-|l_k\rangle)\}
\end{equation*}

\noindent valid to first order in the difference vectors the (formal) path 
integral is given by

\begin{equation}\label{StandardCSPI}
\langle l^{\prime\prime};T|l^\prime ;0\rangle
={\cal N}\pmb{\int}\exp\{i{\textstyle{\int}}[i\langle l(t)|\dot{\l(t)\rangle}
-H(l)]dt\}{\cal D}l
\end{equation}

\noindent where $H(l)=\langle l(t)|{\cal H}|l(t)\rangle$ is the upper symbol, 
${\cal D}l=\lim_{\epsilon\rightarrow0}\prod_{k=1}^N\delta l_k$ and ${\cal N}$ is
a (formal, possibly infinite, possibly zero) normalization constant.

There is a second derivation of a coherent state path integral \cite{Klauder},
this time
involving the lower symbol $h(l)$ which is assumed to exist for the 
Hamiltonian ${\cal H}$ under consideration. Given some restrictions on 
$h(l)$\footnote{namely $\int|h(l_1)h(l_2)\langle l|l_1\rangle\langle l_1
|l_2\rangle\langle l_2|l\rangle|\delta l_1\delta l_2<\infty$ } the 
operator $F(t):=\int e^{-ith(l)}|l\rangle\langle l|\delta l$ is bounded by
unity and the Cauchy-Schwarz inequality gives
$|\langle\phi|F(t)|\psi\rangle|\leq\parallel\!\!|\psi\rangle\!\!\parallel\cdot
\parallel\!\!|\phi\rangle\!\!\parallel$ for all $|\phi\rangle$ and 
$|\psi\rangle$. 
The strong limit $s-\lim_{t\rightarrow0}[1\!\!1-F(t)]/(it)$ exists and
is ${\cal H}$ on the coherent states. This is enough \cite{Chernoff}
to ensure $s-\lim_{N\rightarrow\infty}[F(T/N)]^N=\exp\{-iT{\cal H}\}$ on the whole
Hilbert space $\mathfrak{H}$. Application of this, and setting
$\epsilon:=T/N$, $l_{N+1}:=l^{\prime\prime}$, $l_0:=l^\prime$, leads to 

\begin{equation}
\langle l^{\prime\prime}|\exp\{-iT{\cal H}\}|l^\prime\rangle
=\lim_{N\rightarrow\infty}\int \prod_{n=0}^N\langle l_{n+1}|l_n\rangle
\prod_{n=1}^N \exp\{-i\epsilon h(l_n)\}\delta l_n
\end{equation}

\noindent which is a valid (discrete) ``path integral". 

Interchange of the integrations and the limit results in an expression

\begin{equation}
\langle l^{\prime\prime}|\exp\{-iT{\cal H}\}|l^\prime\rangle
=\int\exp\{i\int[i\langle l(t)|\dot{l}(t)\rangle-h(l)]dt\}{\cal D}l
\end{equation}

\noindent for the coherent state propagator which is different from equation 
(\ref{StandardCSPI}) since in general $H(l)\neq h(l)$ [compare the 
definitions of the upper and lower symbol, equations (\ref{UpperSymbol}),
(\ref{LowerSymbol})]. 

This difference clearly shows the formal character of the standard path 
integral since it
suggests the equality of two expressions which obviously are not equal. 
The discrete lattice formulations, on the other hand, are rigorously derived 
and valid.

One remark about the connection of classical and quantum mechanics is in place
here.
Consider the quantum action functional $I=\int\langle\psi|i\partial_t-
{\cal H}|\psi\rangle dt$ from which the Schr\"{o}dinger equation
$i\dot{|\psi\rangle}={\cal H}|\psi\rangle$ can be derived by 
requesting stationarity under unrestricted 
variations of the vector $|\psi(t)\rangle$. W.l.o.g the
variation can be carried out over unit vectors only, also excluding
mere phase variations $|\psi_t\rangle=\exp\{i\alpha(t)\}|\psi\rangle$
\cite{KlauderPI}.

The action in the exponent of the path integral, 
$I=\int[i\langle l(t)|\dot{\l(t)\rangle}-H(l)]dt$, is the classical action
if the upper symbol
is regarded as the classical Hamiltonian according to the weak correspondence
principle. Variation will lead to the classical equations
of motion. 
This classical action functional is identical in form with the quantum action
functional, the only difference being that the variations range only over the 
special set of coherent state unit vectors 
instead of over all unit vectors. Regarded in this light, and
referring to an analog with thermodynamic equilibrium ,
 ``quantum mechanics is just classical mechanics with all 
stops removed" \cite{KlauderPI}

\subsection{Mathematical aspects of path integration}\label{CSMathPI}

In the last section it was already stressed that the path integral, as it 
stands, is only a formal expression. The same (coherent state) propagator
was expressed in two different path integrals. And it gets worse. The same
path integral expression can be shown to stem from completely different 
propagators. 

To realize this, consider Eq. (\ref{StandardCSPI}) and let the arbitrary 
coherent states
$|l\rangle$ now be the coherent states of the Heisenberg-Weyl group 
$|pq\rangle$ which are supposed to be centered, i.e. 
$\langle pq|Q|pq\rangle=0$ and $\langle pq|P|pq\rangle=0$. These could
be the canonical coherent states. 
Because of the translation property Eq. (\ref{TranslationOperators}), one has
$id|pq\rangle=id(e^{-iqP}e^{ipQ}|\psi_0\rangle)
=[-Qdp+(P+p)dq]|pq\rangle$, and hence
$\langle pq|i\partial_t|pq\rangle=p\dot{q}$ holds.

The action functional reads $I=\int[p\dot{q}-H(p,q)]dt$ and the 
path integral is now

\begin{equation}
{\cal N}\pmb{\int}\exp\{i{\textstyle{\int}}[p\dot{q}-H(p,q)]dt\}{\cal D}p
{\cal D}q
\end{equation}

\noindent However, this is the very same expression one obtains for the 
conventional
phase space path integral even though the propagators are not the same.
The coherent state propagator 
$\langle p^{\prime\prime}q^{\prime\prime};T|p^\prime q^\prime;0\rangle$ 
is certainly not equal to
$\langle q^{\prime\prime};T|q^\prime;0\rangle$. Even the interpretation of
``${\cal D}p$" and ``${\cal D}q$" is different. In the conventional
path integral the ``$p$'s" are integrated out (``there is always one more
$dp$ than $dq$") whereas in the coherent state interpretation there are 
the same number of $p$- and $q$-integrations\footnote{This has 
deep consequences concerning the geometric nature of 
the formulation. See section (\ref{CSMathTools}).}. 
This does not mean that the expressions
are simply wrong since it was said before that the path integral at this
level is merely formal. The last rigorous expressions, at the ``lattice
level", are quite different. For the coherent state propagator this is 
Eq. (\ref{sCSPILattice}), now with $p$ and $q$ as labels:

\begin{eqnarray}
&&\langle p^{\prime\prime}q^{\prime\prime};T|p^\prime q^\prime;0\rangle
=\lim_{\epsilon\rightarrow 0}\int{\textstyle\prod_{k=0}^N}\Bigl[
\langle p_{k+1}q_{k+1}|p_kq_k\rangle\nonumber\\
&&\phantom{\langle p^{\prime\prime}q^{\prime\prime};T|p^\prime q^\prime;0\rangle}
\times\exp\{-i\epsilon 
H(p_{k+1},q_{k+1};p_k,q_k)\}\Bigr]{\textstyle\prod_{k=1}^N}
{\textstyle\frac{dp_kdq_k}{2\pi}}
\end{eqnarray}

\noindent The conventional propagator, on the other hand, is expressed as

\begin{eqnarray*}
\langle q^{\prime\prime};T| q^\prime;0\rangle & =
& \lim_{N\rightarrow\infty}(2\pi)^{-N+1}
{\pmb \int}\exp\Bigl{\{}i{\textstyle \sum}_{l=0}^{N}
\Bigl[p_{l+1/2}(q_{l+1}-q_l)\\
&&-\varepsilon H\bigl(p_{l+1/2},{\textstyle\frac{1}{2}}(q_{l+1}+q_l)\bigr)
\Bigr]\Bigr{\}} 
{\textstyle \prod}_{l=0}^N dp_{l+1/2}{\textstyle \prod}_{l=1}^N dq_l 
\end{eqnarray*}

\noindent where $q_{N+1}=q^{\prime\prime}$ and $q_0=q^\prime$ and the indices 
$l+1/2$
indicate that the resolutions of unity for the $p$-representation have been 
inserted at different (intermediate) times. Otherwise the Heisenberg
uncertainty principle would be violated. This is because $p$ and $q$ are 
``sharp" eigenstates in their own representations as opposed to the 
coherent state representation, where they are both mean values which can 
be specified at the same time.

Consider now the path integral 
${\cal N}\pmb{\int}\exp\{i{\textstyle{\int}}[p\dot{q}-H(p,q)]dt\}{\cal D}p
{\cal D}q$ in its own right, i.e. no matter what propagator it was derived 
from. What can be said about this expression? 

First, note that the 
functional ``measures" ${\cal D}p$ and ${\cal D}q$ are modeled in a way to
be an analog of the homogeneous, translation invariant Lebesgue measure. 
But it is 
known \cite{Cameron1}\cite{Cameron2} that such a measure does not exist. 
It lacks the
property of countable additivity (it is only finitely additive). Hence, the
so-called path integral is not an integral but a linear functional. 

Second, even if the measure existed, the integral would be improper since
it is of the form ${\cal N}{\pmb\int}\exp\{iI\}{\cal D}p{\cal D}q$ which is 
not absolutely
integrable. So some sort of regularization is necessary to remove ambiguities
that go along with this fact.

The Feynman path integral (the conventional real time path integral) has been
constructed to be a formulation of quantum mechanics 
equivalent
to the Schr\"{o}dinger formulation \cite{Feynman}. 
It was pointed out in the very beginning
of the present work that the Schr\"{o}dinger formulation resides in Cartesian
coordinates only. Not surprisingly, the Feynman path integral has the same
limitations. Superficially, the path integral 
${\cal N}\pmb{\int}\exp\{i{\textstyle{\int}}[p\dot{q}-H(p,q)]dt\}{\cal D}p
{\cal D}q$ looks covariant under canonical transformations, but it is not. 
On the ``lattice level", the last rigorous
formulation defining the formal path integral, there is no such covariance. 
Naturally, canonical transformations can be performed, but the path integral
has to be rederived in every case and will generally look quite different
then\footnote{See e.g. \cite{GroscheSteiner}, page 63-66 for the path integral
in spherical coordinates, or page 67-78 for the path integral in 
general coordinates.}. 

The coherent states on the other hand merely change labels under canonical
transformations (apart from phase factors). Does that mean that the latter can
be performed in the path integrand? Certainly not. The same things said above
about the Feynman path integral hold true here as well. The standard coherent 
state path integral is defined by the lattice
approach and there is no covariance under canonical transformations. 

A last remark about the formal path integral: It is well-known that operator
ordering ambiguities arise when, according to the ``royal route", i.e. 
Schr\"{o}dinger quantization, classical variables are promoted to operators.
Where is this ordering ambiguity in the path integral? It is concealed in 
the formal expression but clearly present on the lattice level. There,
an expression like $q_k^2p_{k-1/2}q_{k-1}$ would correspond to an 
operator $Q^2PQ$ whereas $q_k^3p_{k-1/2}$ would correspond to $Q^3P$ etc.
 
It might be unjust to say the path integral hides more truth than it
exhibits since it gives a certain amount of intuitive insight and raises many
interesting questions. However, it would be foolish to try to rigorously 
derive anything
from an expression that is so ill-defined. Questions like ``what is the
nature of the paths" can be answered heuristically but are, strictly speaking,
meaningless (\cite{Klauder}, pages 67-68).

\subsection{Mathematical tools}\label{CSMathTools}

How to overcome all the aforementioned ambiguities will
be the content of section (\ref{CSCSPI}). The formulation
of a well-defined coherent state path integral will require some 
mathematical background not previously encountered in standard path integration.
This section introduces these mathematical tools. 
For simplicity everything will be formulated for a single degree of freedom.\\

{\bf The ``shadow metric"}\\

Classical mechanics is described by Hamilton's equations of motion

\begin{equation}\label{HamiltonEquations}
\dot{q}=\partial_ph(p,q)\hspace{1cm}\dot{p}=-\partial_qh(p,q)
\end{equation}

\noindent where $h(p,q)$ is the classical Hamiltonian. 
These equations are derived from an
equivalence class of action functionals, $I=\int pdq+dF_0-hdt$, by 
requirement of stationarity under variations which hold both endpoints
of phase-space paths fixed. Thus the $C^1$ function $F$ makes no contribution
to the equations of motion.

Note that Hamilton's equations are often derived from an action $S=\int
[p\dot{q}-H(p,q)]dt$, holding only the position $q$ fixed at the end-points. The
reason for this is that in the standard path integral holding both endpoints
fixed would lead to an overspecification since the $p$'s are completely 
integrated
out (remember: ``there is always one more $dp$ than $dq$"). Certainly, this
can not be a geometric formulation when the phase space variables are treated on 
an unequal footing. In this approach, Hamilton's equations can not be 
derived from 
an action $S=\int[-q\dot{p}-H(p,q)]dt$ without first changing the problem by 
partial integration to the standard action given above. 
The problem can not be overcome with the traditional 
coherent state path integral
since the path ``integral" is ill-defined as it stands. But,
if there is a sort of well-defined coherent state path integral, then the fact
that $p$ and $q$ are treated in the same way promises the reward of a geometric 
nature of quantization. 

The equations of motion are covariant under canonical ($C^1$ coordinate)
transformations \cite{Goldstein}, which imply the existence of an 
underlying geometrical structure \cite{Abraham}: Phase space is a
differentiable manifold $M$ endowed with a nondegenerate, closed, symplectic 
two-form $\omega$. So locally $\omega=d(\theta+dF_0)$, and the action 
functional is $I=\int (\theta+dF_0-hdt)$. By Darboux's theorem
local coordinates can be introduced such that $\theta=pdq$, $F_0=F_0(p,q)$
and $\omega=dp\wedge dq$ leading to the equations of motion
(\ref{HamiltonEquations}). 

Covariance under canonical (coordinate) transformations implies on the other
hand that coordinatized expressions, e.g. for the Hamiltonian, have no 
definite physical meaning by themselves. 
Consider a Hamiltonian of the form $h(p,q)
=p^2+q^2$ which is supposed to correspond physically to the harmonic oscillator.
In a different coordinate system (here: polar coordinates in phase space)
the same Hamiltonian becomes $\bar{h}(\bar{p},\bar{q})=\bar{p}$. 
Without further specifications each of these expressions would not be
connected to any particular physical system. 

On the other hand, 
if the phase space $(M,\omega)$ is augmented by a Riemannian metric
$d\sigma^2$ then mathematical expressions like $h(p,q)$ gain a unique 
physical meaning. Consider the example above: $(M,\omega, d\sigma)$ with 
metric $d\sigma^2=dp^2+dq^2$ (Cartesian metric) would show that 
$h(p,q)=p^2+q^2$ is the harmonic oscillator Hamiltonian whereas $h(p,q)=p$
clearly is not. On the other hand choosing the polar metric
$d\sigma^2=(2p)^{-1}dp^2+(2p)dq^2$ would imply that $h(p,q)=p$ is the
harmonic oscillator Hamiltonian and $h(p,q)=p^2+q^2$ would correspond to 
something more exotic.

The metric $d\sigma^2$ is often called a ``shadow" or ``secret" metric 
\cite{KlauderUQ}.\\

{\bf Brownian motion}\\

A shadow metric appended to phase space has another effect, namely it supports
Brownian motion (sometimes called the Wiener process) on the Riemannian 
phase space manifold. The symplectic 
structure alone was not enough since only with a metric can one define the 
Laplace-Beltrami operator 

\begin{equation}\label{DeltaLB}
\Delta_{LB}:=\sqrt{g}^{-1}\partial_a \sqrt{g}g^{ab}\partial_b
\end{equation}

\noindent where $g^{ab}$, $a,b\in\{p,q\}$, is the metric tensor 
($d\sigma^2=g^{ab}dx_adx_b$) and $g$ is its
determinant. Let $K(p^{\prime\prime},q^{\prime\prime},p^\prime,q^\prime,t)
:=[\exp\{t\Delta_{LB}\}](p,q,p^\prime,q^\prime)$ 
be the heat kernel, i.e. the solution to the diffusion equation 

\begin{equation}\label{Diffusion}
\partial_tK={\textstyle\frac{1}{2}}\nu\Delta_{LB}K
\end{equation}

\noindent with initial condition 
$K(p,q,p^\prime,q^\prime;0)=\delta(p-p^\prime)\delta
(q-q^\prime)$. The parameter $\nu$ is called the diffusion constant. Then

\begin{equation}
\int d\mu^{\nu;T}_{W;p^{\prime\prime},q^{\prime\prime},p^\prime,q^\prime}
:=K(p^{\prime\prime},q^{\prime\prime},p^\prime,q^\prime;\nu T)
\end{equation}

\noindent unambiguously defines a measure pinned at $(p^\prime,q^\prime)$ for 
$t=0$ and at $(p^{\prime\prime},q^{\prime\prime})$ for $t=T$ \cite{DKP}.
It is called the
pinned  or conditional\cite{Roepstorff} Wiener measure 
with diffusion constant $\nu$ and is a true measure
on path space concentrated on continuous but nowhere differentiable paths. 
Often the parameters $p^{\prime\prime},q^{\prime\prime},p^\prime$ and 
$q^\prime$ will be suppressed or - in a slight abuse of notation - be 
written on the integral sign as upper and lower limits of the integration.

So the paths are continuum random walks or, better, stochastic processes (See
\cite{Roepstorff} for an elementary introduction or \cite{Simon} for a more
mathematical approach). To be more precise they are so-called Brownian bridges
and are intimately related to
Brownian motion which is the key to understand the nature of the 
paths\footnote{Brownian processes enter the picture naturally since
they are connected to random walks with diffusion, which in turn are connected
to the Laplace-Beltrami operator by the diffusion equation.}. 

Standard Brownian motion (or the Wiener process) is the family $(X_t)_{0\leq t}$
of centered, i.e. mean zero, Gaussian random variables with covariance 
$\langle X_tX_s\rangle= \mbox{min}(t,s)$. Here $\langle\cdot\rangle$ denotes
the expectation value. Since the process is Gaussian, the first 
two moments specify all higher moments. This fact can be expressed
in the form of the moment-generating functional $\langle e^{ibX}\rangle
=e^{-b^2\langle X^2\rangle/2}$.

It was already stated that the Brownian paths are continuous but nowhere 
differentiable. This can be seen as follows:

\begin{figure}
\includegraphics{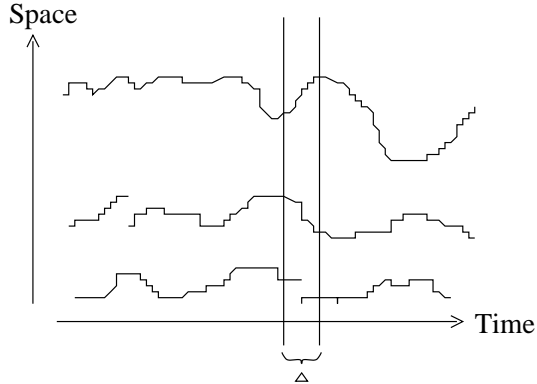} 
\caption{Brownian Paths}\label{BrownianPaths}
\end{figure}

From the generating functional the second and fourth moment are readily
computed as $\langle (X_t-X_s)^2\rangle=|t-s|$ and $\langle(X_t-X_s)^4\rangle
=3|t-s|^2$. Consider now $\langle(X_{t+\Delta}-X_t)^2\rangle=\Delta$ and 
write it as $\Delta=A\Delta+B\Delta$. The first contribution comes from
continuous paths and here $\Delta$ can be interpreted as arising from 
the slope of the path on the interval $[t,t+\Delta]$. 
The second contribution comes from discontinuous paths and in this case
$\Delta$ can be seen as being proportional to 
the capturing probability for such a path (see 
figure \ref{BrownianPaths}). The aim is to show that there are no contributions
from discontinuous paths, i.e. $B=0$. Continuing the argumentation with the 
$\Delta$-parts coming from continuous/discontinuous paths one would expect
an expression $\langle(X_{t+\Delta}-X_t)^4\rangle=A^\prime\Delta^2+B^\prime
\Delta$ since in the continuous part the slope would be squared now whereas
the capturing probability stays the same, i.e. linear in $\Delta$. But the
fourth moment really is
$\langle(X_{t+\Delta}-X_t)^4\rangle=3\Delta^2$ showing that there is no
contribution from discontinuous paths (and hence the same is true for
the second moment as well, i.e. $B=0$). The conclusion is that
Brownian paths are continuous with probability one.

The statement about continuity can be proved in a less pictorial way by
recognizing that the fourth moment fulfills the properties required for a 
theorem of Kolmogorov \cite{Skorokhod} which says: If there are $\alpha>0$,
$\beta>0$, $C>0$ such that $\langle|Z_{t_2}-Z_{t_1}|^\alpha\rangle
\leq C|t_2-t_1|^{1+\beta}$ for all $t_1,t_2$ 
then the process $Z$ is continuous with probability one. 

To show that Brownian paths are almost nowhere differentiable, 
the moment generating functional $\langle e^{itX}\rangle$  is integrated  
over $t$ in the following way: $\int e^{-t^2/2}
\langle e^{itX}\rangle dt=\int e^{-t^2/2}e^{-t^2\langle X\rangle/2}dt
\propto (1+\langle X^2\rangle)^{-1/2}=\langle e^{-X^2/2}\rangle$. These 
relations hold since $X$ is Gaussian. Consider the stochastic process
$Y:=\frac{X(t)-X(s)}{|t-s|^\beta}$ which becomes the derivative of $X$ 
for $\beta=1$ and $t\rightarrow s$. Then
$\langle \exp\{-\frac{1}{2}\frac{[X(t)-X(s)]^2}{|t-s|^{2\beta}}\}\rangle
=(1+|t-s|^{1-2\beta})^{-1/2}$. For $\beta=1$ and $t\rightarrow s$ this
assumes the form $\langle e^{-[\dot{X}^2(t)]/2}\rangle=0$ which says that
$\dot{X}$ is infinity almost everywhere. Thus Brownian paths are almost 
nowhere differentiable.

General Brownian motion is connected to standard Brownian motion by
$B_t=\sqrt{\nu}X_t+c$ where the offset $c$ is a constant and so is $\nu$ 
(often called the diffusion constant).

If $X_t$ is standard Brownian motion with $X_0=0$, then scaling shows that 
$X^*_t:=l^{-1}(X_{s+tl^2}-X_s)$ is Brownian motion again with $X_0^*=0$. The 
standard Brownian bridge is defined as the stochastic process

\begin{equation}\label{BrownianBridge}
\bar{X}_t:=X^*_t-tX_1^*
\end{equation}

\noindent and has mean zero and covariance\footnote{The Brownian 
bridge can also
be defined as the centered Gaussian process with this covariance. 
Then the other
properties follow, see e.g. \cite{Simon}.}

\begin{equation}
\langle \bar{X}_t\bar{X}_s\rangle=\mbox{min}(t,s)-ts
\end{equation}

\noindent It follows that $\bar{X}_0=\bar{X}_1=0$, and this is why the 
standard Brownian bridge is sometimes called
``circular" Brownian motion. 

The general Brownian bridge is connected to the standard one first by 
rescaling of the time (to get an arbitrary time-interval instead of 
$[0,1]$): $\tilde{X}_t:=X^*_t-\frac{t}{T}X_T^*$ and hence
$\langle \tilde{X}_t\bar{X}_s\rangle=\mbox{min}(t,s)-ts/T$. Then the 
deterministic classical path with initial point $x^{\prime}$ at time $0$ and
final point $x^{\prime\prime}$ at time $T$ is added leading to the expression
$\bar{B}_t=x^\prime(T-t)/T+x^{\prime\prime}t/T+\tilde{X}_t$.

In the path integral the paths will be restricted by initial and final
conditions. This is exactly why the Brownian bridge rather than 
Brownian motion will be relevant.\\

{\bf Stochastic integrals}\\

Let $X$ be Brownian motion. It was stated above that Brownian paths are
nowhere differentiable. But since $\partial_t\partial_s\mbox{min}(s,t)
=\delta(t-s)$ one can talk about a formal derivative $W_t=\dot{X}_t$ with
covariance $\langle W_sW_t\rangle=\delta(t-s)$ termed ``white noise". This 
distribution, called a generalized stochastic process,
can produce bona-fide stochastic variables when integrated 
against suitable test functions. Let $f$ be such a function
then 

\begin{equation}
W(f):=\int_0^\infty f(t)W_tdt=:\int_0^\infty f(t)dX_t
\end{equation}

\noindent is well-defined. The last expression can also be understood in 
another way. 
The problem is that $dX_t$ is of unbounded variation
on any time interval and does not represent a true measure. But by partial 
integration one obtains $\int\dot{f}X_tdt$ which can be interpreted as a
Lebesgue-Stieltjes integral (provided integrability conditions on $\dot{f}$
hold).

Let $Y$ be another Brownian motion, possibly independent of $X$.
The integral

\begin{equation}
\int XdY
\end{equation}

\noindent can not be given meaning by partial integration 
($dX$ would be just as unsuitable
a measure as $dY$). One has to
adopt a rule on how to define this expression. The most common rules are the 
It\^o and Stratonovich interpretations. Both rules define the integral as
the continuum limit of discrete random walks on a lattice. 
The It\^o rule says $\int XdY:=\lim\sum X_l(Y_{l+1}-Y_l)$ whereas Stratonovich
uses the ``midpoint rule"

\begin{equation}
\int XdY:=\lim\sum{\textstyle\frac{1}{2}}(X_{l+1}+X_l)(Y_{l+1}-Y_l)
\end{equation}

\noindent Because of the
unbounded variations of the Brownian paths these two expressions are generally
different (as opposed to a ``normal" deterministic integral, where they 
would coincide). 

The It\^o rule is favored by most mathematicians since it
has theoretical advantages which are based on the independent increments 
of a Brownian process \cite{Roepstorff}. 
This leads directly to the
famous It\^o calculus (see e.g. \cite{Simon}), a stochastic differential 
calculus. One of its rules says that the square of the stochastic differential
is deterministic, namely $dX_t^2=dt$. This is more than saying that the
mean of this squared differential is $dt$, which was already 
recognized by Feynman
in his original work on path integration \cite{Feynman}: ``Although the average
value of the displacement of a particle in the time $dt$ is $vdt$, where
$v$ is the mean velocity, the mean of the square of this displacement is not
of order $dt^2$, but only of order $dt$". An application of this rule on 
the metric $d\sigma^2=dx_ag^{ab}dx_b$ shows that, as a stochastic variable,
$d\sigma^2=2\nu dt$, where $\nu$ is the diffusion constant and the factor
$2$ indicates the dimensionality of the phase space manifold \cite{KlauderQiG}.

On the
other hand the Stratonovich rule has the advantage that common rules of 
calculus still apply (which is generally not true for the It\^o-integral,
see e.g. \cite{Simon}). Canonical transformations can be performed on 
stochastic processes in formally the same way as for deterministic
functions when the Stratonovich rule is applied.

For more information on Brownian motion, other stochastic processes and
stochastic calculus and integration see e.g. 
\cite{Roepstorff}\cite{Simon}\cite{Skorokhod}.\\

{\bf Functional integration with Wiener measure}\\

A standard way to get a well-defined path integral is to analytically 
continue the time-evolution operator to imaginary time (Wick rotation), thus
obtaining the operator $\exp\{-TH\}$ characteristic for the diffusion equation.
Here $H$ is the Hamilton operator and if it is of the form
$H=H_0+V$, where $H_0=-\frac{1}{2}\Delta$ is the Hamiltonian for a free
particle with unit mass, 
one is led to the Euclidean configuration space path integral 
\cite{Roepstorff}\cite{GroscheSteiner}

\begin{eqnarray*}
&&K_E(x^{\prime\prime},x^\prime;T)=\langle x^{\prime\prime}|
\exp\{-T[H_0+V(x)]\}|x^\prime\rangle\\
&=&\lim_{N\rightarrow\infty}(2\pi\epsilon)^{-N/2}\prod_{k=1}^{N-1}
\int dx_k\exp\{-\sum_{j=0}^{N-1}[{\textstyle \frac{1}{2\epsilon}}
(x_{j+1}-x_j)^2+\epsilon V(x_j)]\}\\
&=:&\pmb{\int}\exp\{-\int[{\textstyle\frac{1}{2}}\dot{x}^2+V(x)]dt\}{\cal D}x
=:\pmb{\int}\exp\{-\int Vdt\} \:d\mu_W
\end{eqnarray*}

\noindent where $d\mu_W=
\exp\{-\int {\textstyle\frac{1}{2}}\dot{x}^2\}
{\cal D}x$ is a
Wiener measure pinned at the endpoints.

This result is known as the Feynman-Kac formula and one may speak of functional
integration to emphasize its well-definedness. Observe, however, that 
the Wiener measure arises only because of the special form of the Hamiltonian
as a free Hamiltonian plus a potential. This approach has nothing
to do with the coherent state path integral of the next section 
except that it
uses the same mathematical tools.

For further studies, e.g. for a connection between such path integrals and 
stochastic expectations, see e.g. \cite{Roepstorff}. \\

{\bf Improper integrals}\\

The main problems with the path integral were the ill-definedness of the 
measure and the ambiguity arising from the fact that, even with a true 
measure, the path integral in its standard form would be an improper integral 
of the form ${\pmb\int}\exp\{iI\}{\cal D}l$. 

To give a well-defined meaning to an improper integral one has to adopt 
a particular regularization. Consider the toy example

\begin{equation}
\int_{-\infty}^\infty e^{iy^2/2}dy:=\lim_{\nu\rightarrow\infty}
\int_{-\infty}^\infty e^{iy^2/2-y^2/(2\nu)}dy=\sqrt{2\pi i}
\end{equation}

\noindent Here a regularization factor was introduced which rendered the 
integrand absolutely integrable. The improper integral was defined by this 
regularization. Although the result $\sqrt{2\pi i}$ seems
natural (in the sense that if one simply denied the problem and integrated
$e^{-y^2/(2i)}$ according to the usual rule for Gaussian integrals one would
be led to the very same answer), it must be stressed that different 
regularizations could very well lead to different results.

And different results for different regularizations are exactly
what will be encountered in the path integral in the next section.
The idea of a continuous regularization (or convergence)
factor should be kept in mind.

\subsection{The coherent state path integral with Wiener measure}\label{CSCSPI}

Consider again the expression for the coherent state path integral 
[Eq. (\ref{StandardCSPI})]:

\begin{equation*}
\langle l^{\prime\prime};T|l^\prime ;0\rangle\nonumber\\
={\cal N}\pmb{\int}\exp\{i{\textstyle{\int}}[i\langle l(t)|\dot{\l(t)\rangle}
-H(l)]dt\}{\cal D}l
\end{equation*}

\noindent where $H(l)$ is again the upper symbol. 
When trying to apply the stationary phase approximation to the action in the
integrand, it turns out that the resulting first-order extremal equations are 
generally incompatible with the boundary conditions $l^{\prime\prime}=l(T)$
and $l^\prime=l(0)$ \cite{KlauderPIspha} \cite{KlauderPI}. The idea is to
introduce an extra term $\frac{1}{2}i\epsilon\langle \dot{l}|(1-|l\rangle 
\langle l|)|\dot{l}\rangle$ into the action which changes the extremal equations
to second order, compatible with the boundary conditions, 
and taking the limit $\epsilon
\rightarrow0$ as a final step. The fact 
that this term is chosen and no other has a 
reason which will be clarified with the example of the canonical coherent 
states. The coherent state overlap for a short time interval (and with the
notations from the path integral) is 
$\langle p_{l+1}q_{l+1}|p_lq_l\rangle=\exp\{i(\frac{p_{l+1}+p_l}{2})
(q_{l+1}-q_l)-\frac{\epsilon}{4}[\frac{p_{l+1}-p_l}{\epsilon^2}
+\frac{q_{l+1}-q_l}{\epsilon^2}]\epsilon$. In the limit $N\rightarrow\infty$,
i.e. $\epsilon\rightarrow 0$ the second term in the exponent vanishes. But 
letting one $\epsilon$ ``by force" not become an infinitesimal $dt$ the term
turns out to be the regularization factor $e^{-\frac{\epsilon}{4}
\int(\frac{dp^2}{dt^2}+\frac{dq^2}{dt^2})dt}$ introduced above (here for the
canonical coherent states). At this level it is ``bad mathematics" to 
``save" one $\epsilon$ but it 
works! Later this procedure will be be put on a firm mathematical ground.

But one has to point out that the correctness of this procedure can only be
shown a posteriori which means after evaluating the stationary phase method
for the well-defined lattice expressions and comparison with the 
results obtained from this continuous selection method \cite{KlauderPI}.

Assuming correctness (which was shown for the canonical and spin coherent states
in \cite{KlauderPIspha}) the path integral assumes the form

\begin{eqnarray}\label{CSPIepsilon}
&&\langle l^{\prime\prime};T|l^\prime ;0\rangle\nonumber\\
&=&\lim_{\epsilon\rightarrow0}{\cal N}\pmb{\int}\exp\{i {\textstyle \int}
[i\langle l|\dot{l}\rangle+{\textstyle \frac{1}{2}}i\epsilon\langle
\dot{l}|(1-|l\rangle\langle l|)|\dot{l}\rangle
-H(l)]dt\}{\cal D}l
\end{eqnarray}

\noindent Note that in the exponent two terms of clear geometric character 
appear:
First, $i\langle l|d|l\rangle$ is recognized to be the canonical
one-form $\theta$ [see the discussion in \ref{CSMathTools}], which for 
the canonical coherent states becomes $\theta=pdq$. Second,
${\textstyle \frac{1}{2}}(\langle l|d)(1-|l\rangle\langle l|)(d|l\rangle)$
is a Riemannian metric $d\sigma^2$, which for the canonical coherent states
becomes $d\sigma^2=dp^2+dq^2$. 

Here appears for the first time a new structure, a metric, which in this 
case is induced by the coherent states themselves. For the canonical coherent
states this turns out to be the Cartesian metric, so there is reason to believe
that they are in some way associated with a flat phase-space. This point
will be discussed later in a more general context. 

The things to be learned from Eq. (\ref{CSPIepsilon}) and to be kept in 
mind are\\

\noindent 1) the association of a metric structure with the phase space\\
2) a continuous regularization in the path integral (as opposed to the 
usual time-slicing and discretization)\\

Now the continuous regularization of the path integral by a 
factor including a metric on phase space is taken as a starting point. 
To begin with, the metrics under consideration will be the flat, spherical
or hyperbolic metric. Later, generalizations to arbitrary
metrics on 2-dimensional phase space manifolds without symmetry will be 
discussed. Instead
of the abstract notation $i\langle l|\dot{l}\rangle$ and 
${\textstyle \frac{1}{2}} i\epsilon\langle
\dot{l}|(1-|l\rangle\langle l|)|\dot{l}\rangle$ for the canonical one-form and
the metric respectively, canonical coordinates will be used, i.e. those 
coordinates for which the canonical one-form is $pdq$.

In the following path integral the total time-derivative of an arbitrary 
function $G(p,q)$ is included which, for a simply connected phase space 
manifold,
will simply be an unimportant overall phase factor\footnote{For 
multiply connected manifolds the phase factor carries 
the Aharonov-Bohm phase.}. The
classical Hamiltonian is written as $h(p,q)$. Why $h$ is chosen 
rather than $H$ will become apparent later:

\begin{equation}\label{FormalPIwithMetric}
{\cal N}_\nu\pmb{\int}\exp\{i{\textstyle \int}
[p\dot{q}-\dot{G}(p,q)-h(p,q)]dt\}\exp\{-{\textstyle \frac{1}{2\nu}}
{\textstyle \int}(d\sigma^2/dt^2)dt\}{\cal D}p{\cal D}q
\end{equation}

\noindent Here $\nu\in\mathbb{R}^+$. This expression is still a formal path 
integral. 
A very important step on the 
way to a well-defined path integral is to realize that

\begin{equation}
{\cal N}_\nu\exp\{-{\textstyle \frac{1}{2\nu}}
{\textstyle \int}(d\sigma^2/dt^2)dt\}{\cal D}p{\cal D}q=:
d\mu^\nu_W
\end{equation}

\noindent is a well-defined pinned Wiener measure. Mathematically one
defines the Wiener measure $d\mu^\nu_W$ 
by setting ${\pmb\int} d\mu^\nu_W:=
[\exp\{\nu T\Delta_{LB}\}](p^{\prime\prime}, q^{\prime\prime}, q^\prime, 
p^\prime)$, where the last expression is the kernel of the exponentiated 
Laplace-Beltrami operator (or heat kernel). 

As was pointed out in section (\ref{CSMathTools}) the Wiener measure 
is concentrated 
on Brownian motion-like paths. So some integrals like $\int qdp$ have to 
be understood as stochastic integrals, and the Stratonovich rule is adopted
for them. Then 

\begin{equation}\label{PIwithWienerMeasure}
\pmb{\int}\exp\{-i{\textstyle \int}[qdp+d\tilde{G}(p,q)+h(p,q)dt]\}
\:d\mu^\nu_W
\end{equation}

\noindent is a well-defined path integral and 
expression (\ref{FormalPIwithMetric}) can
be given meaning by equating it to this path integral (with $\tilde{G}=
-(G+pq)$).

The question is: What happens in the limit of diverging diffusion constant
$\nu$ (where the regularization formally becomes unity)? 

If this limit leads to a well-defined path integral then further questions 
follow:
What kind of path integral is it, i.e. to which states is it connected? What
is the relation of the classical and quantum Hamiltonian\footnote{Remember: 
Due
to the formal nature of the standard path integrals there were ambiguities
concerning this point. But a well-defined path integral has to select 
a specific connection between the classical and quantum Hamiltonian.}?

For the case of the flat and spherical metric these questions were answered
by Daubechies/Klauder \cite{DK} and for the hyperbolic metric by 
Daubechies/Klauder/Paul (DKP) \cite{DKP}:

In the limit of diverging diffusion constant $\nu$ the path integral
(\ref{PIwithWienerMeasure}) reduces to the coherent state matrix element
$\langle p^{\prime\prime}q^{\prime\prime}|\exp\{-iT{\cal H}\}|p^\prime 
q^\prime\rangle$ of the unitary time-evolution operator and the specific metric 
determines the coherent states in question. 
The flat metric is inevitably connected 
with the coherent states of the Heisenberg-Weyl group, and in the canonical,
Cartesian form, it is connected to the canonical coherent states. The 
spherical metric is associated with the coherent states of the SU(2) group and
the hyperbolic metric leads to the coherent states of the affine group. And
with each group comes a set of quantum kinematical operators. Thus
one can say that the choice of geometry augmenting the classical phase space
manifold determines the quantum kinematical operators uniquely!
Furthermore, the classical Hamiltonian that goes with the quantum Hamiltonian
${\cal H}$ is given by the lower symbol (explaining why $h$ was used earlier
instead of $H$ to characterize the classical Hamiltonian). 

For a better understanding of how these results emerge it is instructive to 
point out the idea of the proof which is similar in the three cases and 
shall be illustrated for the affine group since this group will be of the
most interest in the later work.

The construction of the coherent state path integral with Wiener 
measure for the affine group proceeds in two steps. First, the path integral 
for a zero Hamiltonian
is derived; second, a non-zero Hamiltonian is introduced. The key to the
first task is a linear (complex) polarization condition, an idea borrowed from 
the program of
``geometric quantization". Why this polarization 
condition has to be linear will become apparent. Observe, that the
minimum uncertainty states $|\eta\rangle$ from section (\ref{CSGroupCS}) are
extremal weight vectors which fulfill
$(Q-1+i\beta^{-1}D)|\eta_\beta\rangle=0$. Using the definition of the
affine coherent states, this leads directly to a linear first order
differential operator

\begin{equation}
B=-iq^{-1}\partial_p+1+\beta^{-1}q\partial_q
\end{equation}

\noindent which annihilates the coherent
state overlap $\langle pq|rs\rangle=2^{-2\beta}(qs)^{-\beta}[(q^{-1}+s^{-1})
+i\beta^{-1}(p-r)]^{-2\beta}$. As a positive-definite function the overlap can
be taken as a reproducing kernel and thus $B$ annihilates the whole 
reproducing kernel Hilbert space ${\cal C}_\beta$ built from the overlap
\cite{KlauderWCS}. This condition $B{\cal C}_\beta=0$ is the linear complex
polarization condition referred to above.

At this point specialize to a parameter range $\beta>1/2$ for the fiducial
vector $|\eta\rangle$. Then the fiducial vector admissibility condition
(\ref{FVAC}) is fulfilled and the representation is square-integrable. This
means ${\cal C}_\beta$ is a subspace of $L^2(M_+)$. Define the operator

\begin{eqnarray}\label{OperatorA}
A&:=&{\textstyle\frac{1}{2}}\beta B^\dagger B\nonumber\\
&=&{\textstyle\frac{1}{2}}
\{-\beta^{-1}\partial_qq^2\partial_q-\beta q^{-2}\partial^2_p-1+\beta
-2i\beta q^{-1}\partial_p\}
\end{eqnarray}

\noindent Naturally one has $A{\cal C}_\beta=0$. 
DKP \cite{DKP} showed that $A$ is a non-negative, self-adjoint 
operator\footnote{To compare the work of DKP to the one here it must be 
noticed that they 
generally chose the $a$-$b$ notation rather than the $p$-$q$ notation. 
At the point where they talk
about the connection between their $a$-$b$ notation and a certain 
$p$-$q$ notation one
has to interchange $p$ and $q$ again to match the notation in the present work. 
Another difference is their choice of the one-parameter family 
$|\eta_\beta\rangle$. 
Whereas they chose to set the mean of $Q$ equal to $\beta$ the choice is here
$Q$=1 and so their $\beta$ is physically different from the one here
but mathematically things turn out to be the same apart from some factors
$\beta$ here and there. See appendix (\ref{AppendixA1}).}.
For $\beta>1/2$ the operator
$A$ has a discrete eigenvalue $0$, which is clear since $A$ annihilates the 
subspace ${\cal C}_\beta$. 
Furthermore, the spectrum of $A$ was derived by reducing the 2-dimensional
problem to a 1-dimensional one involving the Morse operator, the spectrum of 
which is well-known \cite{Morse}. Thus, DKP additionally showed that $0$ is
an isolated eigenvalue and concluded that the semigroup $\exp\{-\nu TA\}$
strongly converges to the projection operator $P_0$ onto the subspace
${\cal C}_\beta$ in the limit of diverging diffusion constant $\nu$.
This
ensures convergence of the kernels in a distributional sense: 
$[\exp\{-\nu TA\}](p^{\prime\prime},q^{\prime\prime}, p^\prime, q^\prime)
=\exp\{-\nu TA\}\delta(p-p^\prime)\delta(q-q^\prime)|_{p=p^{\prime\prime},
q=q^{\prime\prime}}\rightharpoonup[P_0]
(p^{\prime\prime},q^{\prime\prime}, p^\prime, q^\prime)
=(1-\frac{1}{2\beta})(2\pi)^{-1}
\langle p^{\prime\prime}q^{\prime\prime}|p^\prime q^\prime\rangle$. It is
the author's point of view that a discrete eigenvalue $0$ is already enough 
and that the question about a gap separating $0$ from the rest of
the spectrum is not relevant. Assume that there is no gap, i.e.
the continuous spectrum starts at $0$, too. Then, ignoring degeneracy,
$\exp\{-\nu TA\}=\sum\exp\{-\nu Tn\}|n\rangle\langle n|+\int\exp
\{-\nu T\lambda\}
|\lambda\rangle\langle\lambda|d\lambda$ and in the 
limit $\nu\rightarrow\infty$ this 
reduces to $|0\rangle\langle 0|=P_0$ since the single non-zero point in the 
integral,
coming from the continuous ``eigenvalue" $0$, does not affect the zero result of
the integral. As a
next step, pointwise convergence of the kernels was established. Finally
the Feynman-Kac-Stratonovich representation of the operator $\exp\{-\nu TA\}$ 
(or its kernel, to be more precise) was derived\footnote{The calculation is
lengthy, see subsection (\ref{WCSRegularizing}) 
where it is done for a modified operator 
$A_\varepsilon$. Set $\varepsilon=0$ to get back to the operator $A$}. It 
is 

\begin{equation}\label{51}
{\cal N}_\nu\pmb{\int}\exp\{i{\textstyle \int}
[p\dot{q}]dt\}\exp\{-{\textstyle \frac{1}{2\nu}}
{\textstyle \int}(\beta q^{-2}\dot{q}^2+\beta^{-1}q^2\dot{p}^2)dt\}
{\cal D}p{\cal D}q
\end{equation}

\noindent which is Eq. (\ref{FormalPIwithMetric}) for the special case
of the hyperbolic metric and zero Hamiltonian. Equation (\ref{51})
can be turned into
a well-defined
path integral by introducing the Wiener measure 

\begin{equation}\label{WienerMeasure}
d\mu_W^\nu:=
{\cal N}_\nu\exp\{-{\textstyle \frac{1}{2\nu}}
{\textstyle \int}(\beta q^{-2}\dot{q}^2+\beta^{-1}q^2\dot{p}^2)dt\}
{\cal D}p{\cal D}q
\end{equation}

\noindent Combining this with the foregoing arguments, it was
proved that the path integral for zero Hamiltonian

\begin{equation}
\pmb{\int}\exp\{-i{\textstyle \int}qdp\}\:d\mu^\nu_W
\end{equation}

\noindent is proportional to the coherent state overlap in the limit of 
diverging diffusion 
constant. At this point the requirement of a linear (complex) polarization
condition can be explained: Only a linear polarization condition for the 
fiducial vectors leads to a first-order differential operator $B$ and, ergo,
to a second-order differential operator $A$. The Wiener measure emerges only
because $A$ is second-order. A higher-order polarization condition for the 
fiducial
vector would not lead to a Wiener measure path integral.

Finally, moving along the same lines outlined for zero Hamiltonian, DKP showed
the well-definedness of the coherent state path integral with Wiener measure
for a wide (dense) class of Hamiltonians (at least polynomial). This proof is 
not so important for the work at hand and the reader is referred
to \cite{DK} \cite{DKP}. Suffice it to say, the classical Hamiltonian
associated with the quantum Hamiltonian turns out to be determined by the 
lower symbol.

The mathematical side of the new coherent state path integral with Wiener 
measure has been discussed for the three 2-dimensional homogeneous spaces of 
constant curvature and the well-definedness of the integral has been
established. But the path integral 

\begin{eqnarray}
&&\langle p^{\prime\prime}q^{\prime\prime}|\exp\{-iT{\cal H}\}|p^\prime 
q^\prime\rangle\nonumber\\
&=&2\pi(1-{\textstyle\frac{1}{2\beta}})^{-1}
\lim_{\nu\rightarrow\infty}\pmb{\int}\exp\{-i{\textstyle \int}[qdp+dG(p,q)
+h(p,q)dt]\}
\:d\mu^\nu_W\nonumber\\
&=:&2\pi(1-{\textstyle\frac{1}{2\beta}})^{-1}
\lim_{\nu\rightarrow\infty}{\cal N}_\nu\pmb{\int}\exp\{-i{\textstyle \int}
[q\dot{p}+\dot{G}(p,q)+h(p,q)]dt\}\nonumber\\
&&\phantom{2\pi(1-{\textstyle\frac{1}{2\beta}})^{-1}
\lim_{\nu\rightarrow\infty}{\cal N}_\nu}
\times\exp\{-{\textstyle \frac{1}{2\nu}}
{\textstyle \int}(d\sigma^2/dt^2)dt\}{\cal D}p{\cal D}q
\end{eqnarray}

\noindent has more to offer.

First, observe that
the metric in the regularization factor\footnote{or taking the more mathematical
approach: the metric that enters the definition of the
Wiener measure through the Laplace-Beltrami operator} gives physical meaning
to the coordinatized expression for the action
$-\int qdp+dG+h(p,q)dt$, which otherwise is simply not there. (Compare the 
discussion in section (\ref{CSMathTools}) where the concept of a shadow metric 
on the classical phase space was introduced.) In the standard path integrals
this question did not have to be raised since the lattice, on which the
formal integral was defined, automatically interpreted the Hamiltonian.

Second, by virtue of the Stratonovich interpretation of the stochastic
integrals, canonical (coordinate) transformations can be performed. The 
coherent states are (apart from relabeling and unimportant phase factors) 
unaffected by these transformations \cite{DK}. The quantization is 
truly geometric in nature. (Since the term ``geometric quantization" has
already been claimed, it is for obvious reasons sometimes called 
``metrical quantization"). The connection of a phase space manifold with
a specific shadow metric and the groups generating the coherent states are 
fundamental and independent of the coordinatization. The only thing that changes
under canonical (coordinate) transformations is the way the groups are 
parameterized. Generally, the group will not be in group coordinates of the 
first or second kind anymore). 

Finally note that a generalization of the above to 2-dimensional phase space 
manifolds with arbitrary choice of geometry is 
possible \cite{Maraner}\cite{KlauderWCS}. 
Since it will not come up again in later sections only the
result is given here. The path integral is

\begin{eqnarray*}
&&\langle\xi^{\prime\prime},T|\xi^\prime ,0\rangle
 =  \langle\xi^{\prime\prime}|
e^{-i{\cal H}T}|\xi^\prime\rangle \\
& = &
\lim_{\nu\rightarrow\infty}{\cal N}_\nu\int\exp\{i\int[a_j(\xi)\dot{\xi}^j-
h(\xi)]dt\}\exp\{-{\textstyle\frac{1}{2\nu}}
\int g_{jk}(\xi)\dot{\xi}^j\dot{\xi}^k dt\} \\
& &\phantom{\lim_{\nu\rightarrow\infty}{\cal N}_\nu}
 \times\exp\{{\textstyle\frac{\nu}{4}}
\int\sqrt{g(\xi)}\varepsilon^{jk}f_{jk}(\xi)dt\
\prod_t\sqrt{g(\xi)}d\xi^1d\xi^2 
\end{eqnarray*}

\noindent where $\xi^j$, $j=1,2$ are coordinates, $g_{jk}$ the metric tensor, 
$g$ its determinant,
$a_j$ is a vector, $f_{jk}=\partial_ja_k-\partial_ka_j$ the field tensor
and $h$ the Hamiltonian.

In general there will be no group to define the coherent states. The metric
tensor $g_{jk}$ added to phase space may well differ from the metric induced
by the coherent states (whereas in the case of a homogeneous space they
coincided). The symplectic form need no longer be proportional to the 
volume element. 
Still, the states $|\xi\rangle$ are coherent states (so 
$1\!\!1=\int|\xi\rangle\langle\xi|
\sqrt{g(\xi)}d\xi^1d\xi^2$, resolution of unity) and the quantum Hamiltonian
is associated to the classical Hamiltonian given by the lower symbol: 
${\cal H}=\int h(\xi)|\xi\rangle\langle\xi|\sqrt{g(\xi)}d\xi^1d\xi^2$.

\newpage
\section{Weak Coherent States and the Weak Coherent State Path Integral}
\label{WCS}

If the second defining property of coherent states, the resolution of unity, 
is relaxed in a certain way one obtains what will be called the Klauder 
states, an even further
generalization of the coherent states. Those Klauder states which are not
coherent states in the old sense will be named weak coherent states.  
The definition of Klauder states and, therefore, weak coherent states will be
given in section (\ref{WCSWCS}). The main part of this work is the study of the 
existence of path integrals for weak coherent states which follows in section
(\ref{WCSWCSPI}). 

Before this program is carried out some motivating remarks are in place about
why a generalization of coherent states and their path integrals are of 
interest.\\

{\bf Affine quantum gravity} \\

The spatial part of the space-time metric $g_{\mu\nu}(x)$ is strictly positive
definite. Upon quantization this $3\times3$ matrix-valued field variable 
becomes a field operator which preserves this property and, hence, can not be
an unbounded operator satisfying canonical commutation relations. 
The positivity property strongly suggests affine commutation relations, the
simple analog of which is Eq. (\ref{AffineCommRel}) for one degree of
freedom. Whereas in this
simple case affine commutation relations follow from the canonical 
commutation relations by 
multiplication with $Q$, this is not true in the case of fields (due to a 
certain infinite product and infinite rescaling). To 
choose affine commutation relations means then to discard canonical ones and
the quantization is noncanonical \cite{KlauderAQG1}\cite{KlauderAQG2}. \\

Taking the affine field operators as generators one can define coherent 
states, or to be more cautious, Klauder states, since effectively it turns
out that the overlap will in the simplest case of constant fields have
a functional form which is analogous to the one for the affine weak coherent
states. In this way one is naturally led to study the problem for the 
affine weak coherent states. Whereas in the latter case one can explicitly 
describe the family of fiducial vectors $(|\eta_\beta\rangle)$ this is not
possible for the general case of non-constant fields. There the overlap is
central to the analysis \cite{KlauderAQG1} 
and the fiducial vector is implicitly contained in
the functional form of the overlap. 

The natural question is whether there is
a path integral representation for this overlap. For the special case of 
constant fields, which is equivalent to the affine (weak) coherent state 
problem for one degree of freedom, 
it is known that there is a well-defined path integral 
for a certain parameter range 
($\beta<1/2$), namely the
one presented in section (\ref{CSCSPI}). 
On the other hand it turns out that for the parameter range
$0<\beta\leq1/2$ the states do not admit a resolution of unity and thus
they are weak coherent states. 

Hence, an important step on the way to establish a path integral representation 
for the case of the affine fields is to prove the existence of a well-defined
affine weak coherent state path integral for one degree of freedom. 
Certainly there can be no conventional way to define this path integral 
since the construction
of the standard path integral requires a resolution of unity. But the concepts
of the coherent state path integral with Wiener measure might be the key
for an approach to a weak coherent state path integral. This was 
proposed by Klauder \cite{KlauderWCS}\cite{KlauderAQG1} and it is the aim of
this work to shed some light on the existence of weak coherent state path
integrals.

\subsection{Weak coherent states}\label{WCSWCS}

If one relaxes the second property which defines coherent states, namely that
they possess a resolution of unity, and postulates only that the new states
span the Hilbert space one arrives at a generalization called Klauder states.
In mathematical terms Klauder states are defined by\\

{\bf 1) Continuity}: The states $|l\rangle$ are a strongly continuous 
vector-valued function of the label $l$.\\

{\bf 2) Completeness}: The family of vectors $(|l\rangle)$ is total, i.e.
the closed linear span of $(|l\rangle)$ is the whole Hilbert space 
$\mathfrak{H}$.\\

The Klauder states which are not coherent states, i.e. which do not have a 
resolution of unity, are named weak coherent states. 

The first question is: Is the class of Klauder states really bigger than
the class of coherent states? Intuitively this seems clear, but it is proved
once one can specify one set of weak coherent states explicitly. This is
done subsequently and the weak coherent states in question will be 
group-defined.

From the discussion in section (\ref{CSGroupCS}) it is clear that groups 
for which left- and right-invariant group measure coincide
can not create weak coherent states, since all
their representations are square integrable. Of the three group-defined
sets of coherent states - canonical, spin and affine - which were given 
a closer look at (and which are associated with a flat, spherical and
hyperbolic phase space respectively in the path integral) 
only the affine ones can possibly be weak coherent states. Equation
(\ref{FVAC}) states a fiducial vector admissibility condition which must
be fulfilled to ensure a resolution of unity for affine coherent states, 
namely the parameter $\beta$ must be bigger than 
$1/2$ \footnote{With the connection to the Lobachevsky half-plane as the 
underlying phase space manifold this amounts to the requirement that the 
(scalar) curvature
of the hyperbolic manifold, which is $R=-2/\beta$, must be less than $-4$}.

Choosing a fiducial vector from the set of non-admissible vectors
[Eq. (\ref{AffineFiducialVector})], i.e. 

\begin{equation}
\eta_\beta(x)=N_\beta x^{\beta-1/2}e^{-\beta x}
\end{equation}

\noindent for $\beta\leq1/2$, one indeed obtains weak coherent states. 
Joint continuity in the overlap [Eq. (\ref{AffineOverlap})]

\begin{equation}
\langle pq|rs\rangle=(qs)^{-\beta}2^{-2\beta}[(q^{-1}+s^{-1})+i\beta^{-1}
(p-r)]^{-2\beta}
\end{equation} 

\noindent is clear. Also, the states span the whole space which follows
from the irreducibility of the group representation.

No studies of other weak coherent states have been made so far, but it seems
highly probable that every group with similar properties can define weak
coherent states when the fiducial vector is chosen such that the representation
is non-square integrable. But whether they or possibly non-group-defined weak 
coherent states are of physical interest has yet to be determined.

\subsection{The affine weak coherent state path integral}\label{WCSWCSPI}

In section (\ref{CSCSPI}) the coherent state path integral with Wiener measure
was derived. For the affine coherent states the parameter range for the
fiducial vectors $|\eta\rangle$ had to be 
limited to $\beta>1/2$. This ensured the resolution of unity and the discrete
(isolated) eigenvalue $0$ for the operator $A$ needed in the course of 
the development of the path integral. 

Things change very much when the parameter range is $0<\beta\leq1/2$. As 
was shown in section (\ref{WCSWCS}) the states generated by the affine group 
acting
on fiducial vectors with $\beta\in(0,1/2]$ are no longer coherent states, but
weak coherent states.

It is clear that a weak coherent state path integral - should it exist - can
not be constructed in the usual way, since the standard procedure of time
slicing requires a resolution of unity. One way is to start 
from the well-defined path integral with Wiener 
measure, and to try to adapt it to the new situation of weak coherent states.

For $0<\beta\leq1/2$ 
the coherent state overlap is no longer in $L^2(M_+)$. Consequently, the
reproducing kernel Hilbert space ${\cal C}_\beta$ is not in the Hilbert space
$L^2(M_+)$, and so 
$0$ can not be a discrete eigenvalue any more. In fact it will be shown that 
only for 
$\beta=1/2$ the vectors in ${\cal C}_\beta$ are generalized eigenvectors to
the eigenvalue $0$. 
The question why $0$ is not in the (continuous) spectrum in spite of the
fact that still $A{\cal C}_\beta=0$ for all $\beta\in(0,1/2)$ 
will be investigated.

If 0 is in the continuous spectrum
of a non-negative operator $X$ one has to face the problem that 
$\lim_{\nu\rightarrow\infty}e^{-\nu TX}=0$
[$e^{-\nu TX}=\int_0^\infty e^{-\nu T\lambda}|\lambda\rangle\langle\lambda|
\,d\lambda\stackrel{\nu\rightarrow\infty}{\longrightarrow}0$ since only one
point in the integrand ``survives" in the limit (namely $\lambda=0$)
which does not change the value of the integral]. Thus, it is not possible
to arrive at a valid path integral representation with Wiener measure in 
the same way as before since the reduction to a projection
operator onto the ground state in the limit of diverging diffusion constant
was crucial in its construction [section (\ref{CSCSPI})].

Klauder \cite{KlauderWCS} proposed a procedure to prevent this collapse onto 0 
by 
rescaling with a $\nu$-dependent factor before taking the limit 
(which amounts to putting a $\delta$-function
weight on the ``eigenvalue" 0). He assumed that this procedure to isolate
the space ${\cal C}_\beta$ could be carried out for
the whole parameter range $0<\beta\leq1/2$. Since $0\in spec(A)$ only
for $\beta=1/2$ this idea, which will henceforth be called the spectral 
approach, can and will be developed in this case exclusively.

In the following subsection (\ref{WCSIsolating}) the isolating procedure is
first explained on a toy model. The procedure is then developed for an 
arbitrary self-adjoint operator. After the questions about the spectrum of the 
operator $A$ have been clarified, the general formulae are applied to the 
affine case for $\beta=1/2$. The corresponding weak coherent state path 
integral for zero Hamiltonian is formulated. Dynamics and symbols are 
introduced in subsection (\ref{WCSIsolatingDynamics}). 
The case $0<\beta<1/2$ is treated afterwards
with a different method, namely a regularization approach. Again a 
zero Hamiltonian is assumed first in subsection (\ref{WCSRegularizing}) 
before dynamics are 
introduced in subsection (\ref{WCSRegularizingDynamics}).

\subsubsection{Spectral approach}\label{WCSIsolating}

To explain the procedure to isolate the space ${\cal C}_\beta$ in the 
spectral approach a simple example, given by Klauder \cite{KlauderWCS}, is
good for illustration:\\

{\bf Toy example}\\

Let 
$\tilde{A}:=\frac{1}{2}\tilde{B}^2$,\ $\tilde{B}:=-i\partial_x$ on $L^2(
\mathbb{R})$. Then $\tilde{B}$ is self-adjoint, and so is $\tilde{A}$ with
$\tilde{A}\geq0$. The eigenvalue 
$\lambda=0$ lies in the continuous spectrum of $\tilde{A}$.
The aim is to isolate the space of generalized eigenfunctions
(non-square integrable functions) belonging to the eigenvalue $\lambda=0$ of
the operator $e^{-\nu T\tilde{A}}$
in the limit $\nu\rightarrow\infty$. 

Klauder says at this point: ``The desired space of functions is composed of
those for which $\psi(x)=
\mbox{const}$, namely a one-dimensional space. We can choose the reproducing
kernel for this space to be identically one."
 
A few comments on this statement are helpful, since one can easily find
a two dimensional space of solutions of $\tilde{A}\psi=0$, the space
spanned by $\{1,x\}$, and ask the question:
Why does the second solution $x$  not come into the game?

The answer to this is twofold. First, one has to understand what is meant by 
``the desired space of functions". Why can one simply talk about such a space 
and ignore the other solution? Since the goal is
to find a valid path integral representation for only this kernel
one should almost certainly be justified in ignoring the other solution. 
The kernel ($1$ in this case) is the starting point! To
make this clear consider the original problem. Whatever the (weak coherent 
state) path integral may be, 
the coherent state overlap must be the kernel for zero
Hamiltonian ($\langle pq|e^{-iHt}|p^\prime q^\prime\rangle= 
\langle pq|p^\prime q^\prime\rangle$ for $H=0$ independent of what the 
path integral looks like). So what one is trying to do is to ``inflate" the 
simple coherent state overlap to a meaningful path integral. The same is valid
for the toy example. The question is how to build a well-defined path integral 
representation of $1$, but then it is unlikely that an ``$x$-part" 
should appear in the process.

If such an ``$x$-part" really appeared it would have to cancel exactly 
or vanish when
taking limits. So one can still insist on the question about a second solution.
There is even more reason to do so, since 
later things will be approached from the other 
direction, i.e. some path integral will be taken as a starting point and its
reduction  
(in the limit of diverging diffusion constant) to the kernel will be shown. 

But indeed, the ``second solution" can not appear and there is quite a
subtle point to be learned on the way. 

Before tackling the main problem remember that often some
solutions of certain equations (differential or other) are ruled out by 
additional requirements like positivity, square integrability etc. So one 
could ask the new question: What could be such an additional condition here 
to prevent the ``second solution" from entering the picture?

This issue is addressed first on Hilbert space. Assume
$B^\dagger B\psi=0$. Take 
$0=\langle\psi|B^\dagger B\psi\rangle=
\langle B\psi|B\psi\rangle$, so $B\psi=0$ and thus 
$B^\dagger B\psi=0\Leftrightarrow B\psi=0$ and in this case no more solutions
can exist.

For the toy example, change the kernel $1$ to an
(unnormalized) Hilbert space vector by
multiplying it with $e^{-\frac{1}{2}\varepsilon x^2}$ 
(the limit $\varepsilon
\rightarrow 0$ will be taken as the last step). Asking the question
which operator annihilates this new kernel would lead to an
$\varepsilon$-modified version of $\tilde{B}$ and, therefore, of $\tilde{A}$.
(And, indeed, this is the right way to proceed. The kernel is the starting
point and determines $\tilde{B}_\varepsilon$ and $\tilde{A}_\varepsilon$!)
One gets $\tilde{B}_\varepsilon=-i\partial_x-i\varepsilon x$ and
$\tilde{A}_\varepsilon=\frac{1}{2}(-\partial_x^2+\varepsilon^2x^2-\varepsilon)$. 
By the little Hilbert space theorem above there can not be more solutions
than $e^{-\frac{1}{2}\varepsilon x^2}$ and the limit will certainly take things back
to the original kernel $1$. 

The regularizing approach in subsection (\ref{WCSRegularizing}), 
which will solve the problem of the affine weak coherent states for 
the parameter range $0<\beta<1/2$, is just such an
$\varepsilon$-modification. In whatever way the modification might complicate
things, one certainly does not have to worry about a ``second solution". 
But there is still more to say:

What would this procedure look like for the ``second solution" $x$? Consider
the unnormalized Hilbert space vector $xe^{-1/2\varepsilon x^2}$. The 
$\varepsilon$-modified version of $\tilde{B}$ would be $\tilde{B}_\varepsilon=
-i\partial_x-i\varepsilon x+i/x$. But, this would not reduce to $\tilde{B}$ 
in the limit $\varepsilon\rightarrow0$, or else 
the second and first solution would be 
the same. For $\tilde{B}_\varepsilon$ to make sense 
one has to impose $\psi(0)=0$ for the solutions! This will be called 
a Dirichlet boundary condition (D.B.C.), although $0$ is, strictly speaking,
not a boundary here.
In the limit $\tilde{A}_\varepsilon=\frac{1}{2}\tilde{B}^\dagger\tilde{B}
=\frac{1}{2}(-\partial_x^2|_{D.B.C.}+\varepsilon^2x^2-3\varepsilon)$ 
does not reduce 
to $\tilde{A}$ since the $\varepsilon$-modified version carries the D.B.C. and
is positive definite only because the D.B.C. excludes the
ground state of this oscillator (see e.g. \cite{Simon}).

So the final answer is that by asking the naive question about a ``second
solution" the additional constraint of boundary conditions was overlooked. 
Since a path integral is a solution to a differential equation, it
implicitly carries a boundary condition. That is why neither in the toy 
example nor in the case of the affine weak coherent states 
one will have to worry about more 
solutions than the ones already given by $\tilde{B}\psi=0$ (or later $B\psi=0$).
A solution of $B^\dagger B\psi=0$ for which $B\psi\neq0$ would carry a 
different boundary condition.

This interesting point is connected to the problem of operator 
extensions (Friedrichs' extension, Krein extension etc. \cite{Strook}).
 
Returning to the toy example one finds
by direct calculation\footnote{Fourier representation of the $\delta$-function
and action of the operator will lead to simple Gaussian integrals.}
(or with the aid of the Mehler formula \cite{Roepstorff} for the oscillator 
path integral
followed by taking the oscillator frequency to $0$):

\begin{equation*}
\langle x^{\prime\prime}|e^{-\nu T\tilde{A}}|x^\prime\rangle
=e^{-\nu T\tilde{A}(x)}\delta(x-x^\prime)\mid_{x=x^{\prime\prime}}
=\frac{1}{\sqrt{2\pi\nu T}}e^{-\frac{(x^{\prime\prime}-x^\prime)^2}{2\nu T}}
={\pmb\int} d\mu_W^\nu(x)
\end{equation*}

It is the factor $(2\pi\nu T)^{-1/2}$ which is responsible for the 
collapse onto $0$. By the rescaling with $\sqrt{2\pi\nu T}$, one gets

\begin{equation}
\lim_{\nu\rightarrow\infty}\sqrt{2\pi\nu T}{\pmb\int} 
d\mu_W^\nu(x)
=\lim_{\nu\rightarrow\infty}e^{-\frac{(x^{\prime\prime}-x^\prime)^2}{2\nu t}}
=1
\end{equation}

\noindent which gives the correct reproducing kernel for the desired subspace.\\

Also, notice that 

\begin{equation}
e^{-\frac{(x^{\prime\prime}-x^\prime)^2}{2\nu T}}
=\frac{{\pmb\int} d\mu_W^\nu(x)}
{{\pmb\int}_{x^\prime=0}^{x^{\prime\prime}=0}
d\mu_W^\nu(x)}
\end{equation}

\noindent is a self-consistent way to determine the proper rescaling 
factor.\\

{\bf The general case}\\

Let X be a non-negative, self-adjoint operator on a certain Hilbert space and
assume $0$ is in its continuous, but not in its discrete spectrum. 
The operator $X$ generates
a semigroup $e^{-\nu TX}$ which has a spectral representation
$e^{-\nu XT}=\int_0^\infty e^{-\nu\lambda T}d\mathbb{ E}(\lambda)$ or
$\langle x^{\prime\prime}|e^{-\nu TX}|x^\prime\rangle=
\int_0^\infty e^{-\nu\lambda T}d\langle x^{\prime\prime}|\mathbb{ E}(\lambda)|
x^\prime\rangle$. 

Since only rather well-behaved potentials will eventually be of interest, the
reasonable assumption is made that the measure
$d\langle x^{\prime\prime}|\mathbb{ E}(\lambda)|
x^\prime\rangle$ has an absolutely continuous, but no singularly 
continuous part.
Then the spectral family can be written as a (weighted) integral over 
one-dimensional projection operators 
$\mathbb{ E}(\lambda)=\int_{-\infty}^\lambda|E\rangle\langle E| 
\rho(E)dE$ \footnote{For a singularly continuous measure this would not be
possible: $\mu_{sc}(x)=\int_{-\infty}^x d\mu_{sc}(y)\neq
\int_{-\infty}^x(d\mu_{sc}/dy)dy=0$ since $d\mu_{sc}/dy=0$ almost everywhere.}.
If the generalized eigenstates $|E\rangle$ are $\delta$-orthonormalized, then
$\rho(E)=1$. This follows from the fact that every operator 
$\mathbb{E}(\lambda)$ 
is a projection operator, and 
hence
$[\mathbb{E}(\lambda)]^2=\int_{-\infty}^\lambda\int_{-\infty}^\lambda
|E\rangle\langle E|E^\prime\rangle\langle E^\prime|\rho(E)\rho(E^\prime)dE\,
dE^\prime=\int_{-\infty}^\lambda\int_{-\infty}^\infty
|E\rangle\delta(E-E^\prime)\langle E^\prime|\rho(E)$ $\rho(E^\prime)dE\,
dE^\prime=\int_{-\infty}^\lambda|E\rangle \langle E|\rho^2(E)dE
=\mathbb{E}(\lambda)=\int_{-\infty}^\lambda|E\rangle\langle E|\rho(E)dE$. 
Since $\rho$ is a non-negative function (and not identically zero 
almost everywhere) $\rho(E)\equiv 1$ follows.

The matrix element of $e^{-\nu TX}$ can then be written as

\begin{equation}\label{OperatorX}
\langle x^{\prime\prime}|e^{-\nu TX}|x^\prime\rangle=
\int_0^\infty e^{-\nu\lambda T}\psi_\lambda(x^{\prime\prime})
\psi^*_\lambda(x^\prime)\rho(\lambda)d\lambda
\end{equation}

\noindent and the $\psi_\lambda$ are
continuous in $\lambda$. Moreover, $\rho$ - being part of the measure - is
at least right-continuous and for $\delta$-orthonormalized wavefunctions
$\rho(\lambda)\equiv 1$. 

The goal is to find the rescaling factor which saves equation 
(\ref{OperatorX}) from becoming trivial in the limit of diverging diffusion
constant $\nu$. Since for very large $\nu$ the factor $e^{-\nu T\lambda}$
suppresses everything but very small $\lambda$, the behavior of 
$f_{x^\prime, x^{\prime\prime}}(\lambda):=\psi_\lambda(x^{\prime\prime})
\psi^*_\lambda(x^\prime)\rho(\lambda)$ is all that matters. 
To give an example assume that $f_{x^\prime,x^{\prime\prime}}(\lambda)
\propto \lambda^a$ for small $\lambda$. 
Then the proper rescaling factor can be determined
such that $\int_0^\infty e^{-\nu\lambda T}\psi_\lambda(x^{\prime\prime})
\psi^*_\lambda(x^\prime)\rho(\lambda)d\lambda$ will not collapse to $0$ but 
instead go to the desired $\psi_0(x^{\prime\prime})\psi_0(x^\prime)$.
In the example this factor is
 
\begin{equation}\label{Example}
\int_0^\infty d\lambda \lambda^a e^{-\nu\lambda T}=
\frac{\Gamma(a+1)}{(\nu T)^{a+1}}
\end{equation}

\noindent which goes to $0$ for $\nu\rightarrow\infty$. 
After rescaling with the inverse one gets 
$\frac{(\nu T)^{a+1}}{\Gamma(a+1)}\lambda^a
e^{-\nu\lambda T}\stackrel{\nu\rightarrow\infty}{\longrightarrow}
\delta(\lambda)$, which is the desired $\delta$-function weight on $0$. 

As in the toy example, the rescaling factor can be computed self-con\-sistently
and the general formula becomes

\begin{equation}\label{GeneralSpectralApproach}
\lim_{\nu\rightarrow\infty}\frac{\int_0^\infty e^{-\nu\lambda T}
\psi_\lambda(x^{\prime\prime})\psi^*_\lambda(x^\prime)
\rho(\lambda)d\lambda}{\int_0^\infty e^{-\nu\lambda T}
\psi_\lambda(0)\psi^*_\lambda(0)
\rho(\lambda)d\lambda}\stackrel{\nu\rightarrow\infty}{-\!\!\!\rightharpoonup}
\psi_0(x^{\prime\prime})
\psi^*_0(x^\prime)
\end{equation}

\noindent The last expression, $\psi_0(x^{\prime\prime})
\psi^*_0(x^\prime)$, is the kernel of the desired projection 
operator onto the ground state. The convergence is in a distributional sense
(denoted by the symbol $\rightharpoonup$). If the functional form of
$\psi_0(x^{\prime\prime})\psi^*_0(x^\prime)$ is known to be continuous then
the convergence is pointwise. 

Observe in the example with $f_{x^\prime,x^{\prime\prime}}(\lambda)=\lambda^a$
one must have $a>-1$ or else the rescaling factor would be identically $0$ 
(since the integral would be infinity). But since the rescaling factor can
be determined self-consistently, i.e. by the denominator of equation
(\ref{GeneralSpectralApproach}), which always exists, there is no hidden 
``trap" to look out for. Moreover the evaluation of the denominator
need not necessarily be at the point $x^{\prime\prime}=x^\prime=0$. It
could be at any point $x^{\prime\prime}=x^\prime=b$, $b\in\mathbb{R}$ or even
$b=\pm\infty$ as long as the function $\psi_\lambda(x)$ is not $0$ at $b$.  
Whatever gives the easiest result is the preferred choice. Sometimes this choice
even makes no difference as in the toy example above, since the function there
depends on $x^{\prime\prime}-x^\prime$ only. But in general there is an 
arbitrariness, but this arbitrariness exists for a good reason. Assume that
some expression $K$ serves as a reproducing kernel. Let $a$ be a constant, then
$aK$ is just as good a reproducing kernel, but the inner product in the 
reproducing kernel Hilbert space must be redefined. Sometimes, on the other 
hand, there clearly is a preferred choice. Take the case of the affine coherent
states overlap ($\beta>1/2$). The overlap is a kernel for a reproducing kernel
Hilbert space ${\cal C}_\beta$, but since there is a resolution of unity which
gives rise to an integral definition of the inner product, the factor in front 
of the reproducing kernel is fixed (or at least it would not make much sense to
define another inner product on ${\cal C}_\beta$). This factor used to be
$(1-\frac{1}{2\beta})(2\pi)^{-1}$ and appeared frequently in the affine
coherent state path integral. For the case of the weak coherent state path 
integral ($\beta\leq1/2$) on the other hand, the factor is no longer fixed, 
since the resolution of unity does not exist anymore.

The function
$f_{x^\prime,x^{\prime\prime}}(\lambda)=
\langle x^{\prime\prime}|\delta(\lambda -X)|x^\prime\rangle$ is connected 
to the Green's function and its determination
proceeds as follows.

Define
$\delta_+(y):=\lim_{\varepsilon\rightarrow 0}\int_0^\infty\frac{1}{2\pi}
e^{iyt-\varepsilon t}dt=\lim_{\varepsilon\rightarrow 0}\frac{1}{2\pi}\frac{1}
{\varepsilon -iy}$. Write
$\frac{1}{\varepsilon-iy}=\frac{1}{\varepsilon-iy}
\frac{\varepsilon+iy}{\varepsilon+iy}=\frac{iy}{\varepsilon^2+y^2}+
\frac{\varepsilon}{\varepsilon^2+y^2}$ and since $\lim_{\varepsilon\rightarrow 0}
\int_{-\infty}^\infty\frac{\varepsilon dy}{\varepsilon^2+y^2}=
\int_{-\infty}^\infty\frac{du}{1+u^2}=\pi$ one has 
$\lim_{\varepsilon\rightarrow 0}\frac{1}{\varepsilon-iy}=
i{\cal P}(\frac{1}{y})+\pi\delta(y)$
where the principal value ${\cal P}$ is defined by 
$\int{\cal P}[\frac{f(y)}{y}]dy:=\lim_{\varepsilon\rightarrow 0}
(\int_\varepsilon^\infty+\int_{-\infty}^{-\varepsilon})\frac{f(y)}{y}dy$.
Hence, the $\delta_+$-function can be expressed as 
$\delta_+(y)=\frac{1}{2}\delta(y)+\frac{i}{2\pi}{\cal P}
(\frac{1}{y})$.

The Green's function $G=\lim_{\varepsilon\rightarrow0}
\int_0^\infty e^{i(E+i\varepsilon)T}\langle x^{\prime\prime}|e^{-iHT}|
x^\prime\rangle dT$ is, after integration, seen to be 
$G=\lim_{\varepsilon\rightarrow0}
\langle x^{\prime\prime}|\frac{i}{(E-H)+i\varepsilon}|x^\prime\rangle$.
By the above $\lim_{\varepsilon\rightarrow 0}\frac{i}{(E-H)+i\varepsilon}=
\pi\delta(E-H)+i{\cal P}(\frac{1}{E-H})$ and so by taking the real part one
has

\begin{eqnarray}\label{GreenAndEigenfunctions}
&&\Re\int_0^\infty dT e^{i(E+i\varepsilon)T}\langle x^{\prime\prime}|e^{-iHT}|
x^\prime\rangle
 =  \Re\langle x^{\prime\prime}|\int_0^\infty dT e^{iT(E-H)-
\varepsilon T}|x^\prime\rangle\nonumber\\ 
& = & \Re\langle x^{\prime\prime}|\frac{i}{(E-H)+i\varepsilon}|x^\prime\rangle 
 \stackrel{\varepsilon\rightarrow 0}{\longrightarrow}  
\pi\langle x^{\prime\prime}|
\delta(E-H)|x^\prime\rangle=\pi f_{x^\prime,x^{\prime\prime}}(E)
\end{eqnarray}

If the function $f_{x^\prime,x^{\prime\prime}}(E)$ is known then the 
rescaling factor can be determined - at least in principle - with the
aid of the denominator of Eq. (\ref{GeneralSpectralApproach}). 

The spectral approach works for all operators which possess the same
properties as the operator $X$. \\

{\bf The affine case}\\

The foregoing is now applied to the case of the affine weak coherent
states [see especially sections (\ref{CSGroupCS}), (\ref{CSCSPI}) and 
(\ref{WCSWCS})].

First, it is necessary, to see in which situation the spectral approach
can possibly work. DKP \cite{DKP} gave the spectrum of the operator $A$ 
[which was defined in Eq. (\ref{OperatorA})] as 

\begin{eqnarray}\label{SpecA}
spec(A)&=&\{(\beta-{\textstyle\frac{1}{2}})^2-
(\beta-{\textstyle\frac{1}{2}}-n)^2; n\in\mathbb{N}, 
n<\beta-{\textstyle\frac{1}{2}}\}
\nonumber\\
&&\cup \bigl[(\beta-{\textstyle\frac{1}{2}})^2,\infty\bigr)
\end{eqnarray}

\noindent Thus, there is no discrete spectrum for $0<\beta\leq1/2$. 
And only for
$\beta=1/2$ does the continuous spectrum include $0$. 

The question is: Why is $0$ not in the spectrum even if all functions
$\psi\in{\cal C}_\beta$ fulfill $A\psi=0$ for the whole parameter range
$\beta>0$?

An equation $A\psi=\alpha\psi$ need not imply $\alpha\in spec(A)$.
If $\psi$ is square integrable then $\alpha$ is in the discrete spectrum. But
if $\psi$ is not square integrable, then $\alpha$ may or may not be in the
continuous spectrum. The criterium is the following:
There must be an (unnormalized) square integrable
sequence $(\tilde{\psi}_n)\in
L^2(M_+)^\mathbb{ N}$ for which $\tilde{\psi}_n(x)
\rightarrow\psi(x)$ for $n\rightarrow\infty$ 
(at least in a distributional sense)
such that $\psi_n:=N_n\tilde{\psi}_n$ is a normalized
sequence (and $N_n$ is a normalization constant) and  
$\parallel\!\! (A-\alpha)\psi_n\!\!\parallel\longrightarrow 0$.


A toy example is again helpful to see this before the case on the 
Lobachevsky plane is discussed.

Let $\psi_\alpha=e^{i\alpha x}$ where $\alpha=p+ip^\prime$. 
Take $P=-i\partial_x$,
then $P\psi_\alpha=\alpha\psi_\alpha$. The claim is nonetheless that  
$\alpha\not\in spec(A)$, except when $p^\prime=0$.

Consider a special sequence first: $\psi_n=N_n e^{i\alpha x}e^{-x^2/n^2}$. 

Calculation shows: 
$1\stackrel{!}{=}\parallel\!\!\psi_n\!\!\parallel^2
=N_n^*N_n e^{n^2(p^\prime)^2/2}(\frac{\pi}{2})^{1/2}n$ 
and with that one computes 
(partial integration and Gaussian integrals): 
$\parallel\!\!(P-\alpha)\psi_n\!\!\parallel^2=1/n^2+(p^\prime)^2
=\parallel\!\!(P-p)\psi_n
\!\!\parallel^2+\parallel\!\! p^\prime\psi_n\!\!\parallel^2$ 
which does not go to 0 for 
$n\longrightarrow\infty$ except for $p^\prime=0$. This proves that for $\alpha
=p$, i.e. $p^\prime=0$, $\alpha\in spec(A)$.

To complete the proof that $\alpha\not\in spec(A)$ for $p^\prime\neq 0$ the
above 
must  be shown for every sequence $(\psi_n)\in L^2(M_+)^\mathbb{ N}$ having the
required properties. This will not be further studied here, but be discussed
for the real problem with the operator $A$ on the Lobachevsky half-plane.

It is clear now that, even if 
$A\psi(p,q)=0$ for all $\psi\in{\cal C}_\beta$, one still has to
find out whether there is a normalized 
sequence $(\psi_n:)\in L^2(M_+)^\mathbb{ N}$ such that
$\parallel\!\! A\psi_n\!\!\parallel\rightarrow 0$ for $n\rightarrow\infty$. 
Since 
$\psi(p,q)=\langle pq|(\sum_{j=1}^J\alpha_j|p_jq_j\rangle)$ (or limits of 
corresponding Cauchy sequences), it is enough to consider $\psi(p,q)=\langle
pq|rs\rangle=(1/2)^{-2\beta}(qs)^{-\beta}[(q^{-1}+s^{-1})+i\beta^{-1}(p-r)]
^{-2\beta}$.

For large $q$ this expression is proportional to $q^{-\beta}$, and since 
$0<\beta\leq1/2$, a regularization factor, which is effective at 
infinity, is required to 
produce Hilbert space vectors again.
Now, for $0<4\beta\leq 1$: 
$\int_{-\infty}^{\infty}(c^2+p^2)^{-2\beta}dp=\infty$ (where $c$ is a constant), 
and one 
must in this case regularize in $p$ as well, 
whereas for $1<4\beta<2$ this extra regularization is not
needed. A regularization in $p$ makes a regularization 
in $q$ for small $q$ necessary. This leads, in turn, to more complicated terms.

The case $1<4\beta<2$ will be examined first. Most of the details are 
transferred to the appendix, since they consist of lengthy
calculations which offer no great insight. Likewise,
the case $0<4\beta<1$ is studied entirely in appendix (\ref{AppendixA3}). 
The case $\beta=1/2$ will be treated
separately, and since $0$ will then turn out to be in the spectrum,  
the isolating procedure referred to earlier will be carried out.

Note: Since the interest lies in the $n$-dependence of every expression, 
constants
are generally absorbed in a factor ``$const$" which can change from line
to line.\\

{\bf I) ${\mathbf 1{\boldsymbol{<}}4\boldsymbol{\beta}{\boldsymbol{<}}2}$}\\

The goal is to show $0\not\in spec(A)$. 
The proof starts with a special sequence: 
$\psi_n(p,q):=N_n(1/2)^{-2\beta}(qs)^{-\beta}[(q^{-1}+s^{-1})+i\beta^{-1}(p-r)]
^{-2\beta} e^{-q/n}$, the norm of which is

\begin{eqnarray*}
\parallel\!\!\psi_n\!\!\parallel^2 & = & 
N_n^*N_ns^{-2\beta}4^{2\beta}\int_0^\infty 
dq\, e^{-2q/n}q^{-2\beta}\\
&&\phantom{N_n^*N_ns^{-2\beta}4^{2\beta}}\times
\int_{-\infty}^\infty dp\, [(q^{-1}+s^{-1})^2+\beta^{-2}
(p-r)^2]^{-2\beta} \\
& = & N_n^*N_ns^{-2\beta}4^{2\beta}\int_0^\infty dq\, e^{-2q/n}q^{-2\beta}
(q^{-1}+s^{-1})^{1-4\beta}\\
&&\phantom{N_n^*N_ns^{-2\beta}4^{2\beta}}
\times\int_{-\infty}^\infty dp\, [1+\beta^{-2}p^2]^{-2\beta} \\
& = & const\cdot N_n^*N_n\int_0^\infty dq\, e^{-2q/n}q^{-2\beta}
(q^{-1}+s^{-1})^{1-4\beta} \\
& = & const\cdot N_n^*N_n n^{1-2\beta}\int_0^\infty dq\, e^{-2q}q^{-2\beta}
[(qn)^{-1}+s^{-1}]^{1-4\beta}
\end{eqnarray*}

\noindent This expression will always be smaller than 
$const\cdot N_n^*N_n n^{1-2\beta}$, 
which is the correct asymptotic behavior for large $n$. 
So $N_n^*N_n>n^{2\beta-1}\times const^{-1}$.

Note that in the step from line 1 to 2 the substitution 
$p\rightarrow p/(q^{-1}+s^{-1})$ was made. 
The resulting $p$-integral can be found in \cite{Gradsheteyn},page 297, 
integral \#3.252, 11.

With $[+]:=[\frac{1}{2}(q^{-1}+s^{-1})+\frac{1}{2}i\beta^{-1}(p-r)]$ 
one finds in this case:
$A\psi_n=\{-q^2n^{-2}+2\beta[+]^{-1}n^{-1}+(2-2\beta)qn^{-1}\}\psi_n$ 

Use 
$[-]:=[+]^*=[\frac{1}{2}(q^{-1}+s^{-1})-\frac{1}{2}i\beta^{-1}(p-r)]$
and $[\ \ ]:=[-][+]=\frac{1}{4}[(q^{-1}+s^{-1})^2+\beta^{-2}(p-r)^2]$ and 
$[+]+[-]= (q^{-1}+s^{-1})$ then calculation of 
$\parallel\!\! A\psi_n\!\!\parallel^2$ yields:

\begin{eqnarray*}
& & \parallel\!\! A\psi_n\!\!\parallel^2 \\
& = & N_n^*N_n\int_{-\infty}^\infty dp\int_0^\infty dq\,
e^{-2q/n}q^{-2\beta}\{{\textstyle\frac{1}{4}}
[(q^{-1}+s^{-1})^2+\beta^{-2}(p-r)^2]\}^{-2\beta} \\
& &\phantom{ N_n^*N_n\int_{-\infty}^\infty dp\int_0^\infty}
\times\{-q^2n^{-2}+2\beta n^{-1}[-]^{-1}+2(1-\beta) qn^{-1}\}\\
&&\phantom{ N_n^*N_n\int_{-\infty}^\infty dp\int_0^\infty}
\times\{-q^2n^{-2}+2\beta n^{-1}[+]^{-1}+2(1-\beta) qn^{-1}\} \\
& = & \int_0^\infty dq\,[-q^2n^{-2}+2(1-\beta)qn^{-1}]^2
\int_{-\infty}^\infty dp\, \psi_n^*\psi_n \\
& & +\int_0^\infty dq\int_{-\infty}^\infty dp\, 
\{-2\beta q^2([+]^{-1}+[-]^{-1})n^{-3}+4\beta^2[+][-]n^{-2} \\
& &\phantom{+\int_0^\infty dq\int_{-\infty}^\infty dp \{}
 +4\beta(1-\beta)q([+]^{-1}+[-]^{-1})n^{-2}\}\psi_n^*\psi_n\\
& = & \int_0^\infty dq\,[-q^2n^{-2}+2(1-\beta)qn^{-1}]^2
\int_{-\infty}^\infty dp\, \psi_n^*\psi_n \\
& & +\int_0^\infty dq\int_{-\infty}^\infty dp\, 
\{-2\beta q^2([+]+[-])[\ \ ]^{-1}n^{-3}+4\beta^2[\ \ ]n^{-2} \\
& &\phantom{+\int_0^\infty dq\int_{-\infty}^\infty dp \{}
+4\beta(1-\beta)q([+]+[-])[\ \ ]^{-1}n^{-2}\}\psi_n^*\psi_n
\end{eqnarray*}

One has to make sure that cancelations do not occur.
The first summand (the ``$q$-part") goes to a constant, whereas the second
(the ``[\ \ ]-part") goes to zero:

\begin{eqnarray*}
& & N_n^*N_n\int_0^\infty dq\,[-q^2n^{-2}+2(1-\beta)qn^{-1}]^2 
\int_{-\infty}^\infty dp \,(N_n^*N_n)^{-1} \psi_n^*\psi_n \\
& > & n^{2\beta-1}\int_0^\infty dq\,[-q^2n^{-2}+2(1-\beta)qn^{-1}]^2
e^{-2q/n}q^{-2\beta}(q^{-1}+s^{-1})^{1-4\beta} \\
& = & \int_0^\infty dq\,[-q^2+2(1-\beta)q]^2e^{-2q}q^{-2\beta}
[(qn)^{-1}+s^{-1}]^{1-4\beta}
\end{eqnarray*}

\noindent This will not go to zero as $n\rightarrow\infty$ but to 
a constant $>0$! The second term is

\begin{eqnarray*}
& & N_n^*N_n\int_0^\infty dq\, e^{-2q/n}q^{-2\beta} \int_{-\infty}^\infty dp\,
\{-2\beta q^2([+]+[-])[\ \ ]^{-1}n^{-3} \\
&&\phantom{N_n^*N_n\int_0^\infty}
+4\beta(1-\beta)q([+]+[-])[\ \ ]^{-1}n^{-2}
+4\beta^2[\ \ ]n^{-2}\}[\ \ ]^{-2\beta} \\
& = & const\cdot N_n^*N_n\int_0^\infty dq\, e^{-2q/n}q^{-2\beta} \\
& & \phantom{const\cdot N_n^*N_n\int_0^\infty}
\times\{[-2\beta q^2n^{-3}+4\beta(1-\beta)qn^{-2}]
(q^{-1}+s^{-1})^{-4\beta}\\
&&\phantom{const\cdot N_n^*N_n\int_0^\infty\times}
+4\beta^2(q^{-1}+s^{-1})^{-1-4\beta}n^{-2}\} \\
& = & const\cdot N_n^*N_n n^{1-2\beta}\int_0^\infty dq\, e^{-2q}q^{-2\beta} \\
& &\phantom{const\cdot N_n^*N_n\int_0^\infty}
\times\{[-2\beta q^2n^{-1}+4\beta(1-\beta)qn^{-1}]
[(nq)^{-1}+s^{-1}]^{-4\beta}\\
&&\phantom{const\cdot N_n^*N_n\int_0^\infty\times}
+4\beta^2n^{-2}[(nq)^{-1}+s^{-1}]^{-1-4\beta}\}
\end{eqnarray*}

\noindent and the bottom line will go to zero due to the $1/n$, $1/n^2$ terms.

It has been shown that convergence fails for this special sequence.
As in the toy example above, this is not yet a complete proof. One has to show
such behavior for all possible $\psi_n$ which fulfill the 
necessary requirements. This can be found in appendix (\ref{AppendixA3}) 
where the case
$0<4\beta<1$ and the limiting case $4\beta=1$ are treated as well.\\

{\bf The case ${\mathbf{\boldsymbol{\beta}}{\boldsymbol{=}}1{\boldsymbol{/}}2}$}\\

In this case 
the claim is $0\in spec(A)$ (so it is enough to find one special sequence
$\psi_n$ that will do the job):

\begin{eqnarray*}
\parallel\!\!\psi_n\!\!\parallel^2
& = &const\cdot N_n^*N_n \int_0^\infty dq\,e^{-2q}q^{-1}
[(nq)^{-1}+s^{-1}]^{-1} \\
& = & const\cdot N_n^*N_n\{\int_0^c dq\,e^{-2q}q^{-1}[(nq)^{-1}+s^{-1}]^{-1}\\
&&\phantom{const\cdot N_n^*N_n\{}
+\int_c^\infty dq\,e^{-2q}q^{-1}[(nq)^{-1}+s^{-1}]^{-1}\} \\
&\leq& const\cdot N_n^*N_n\{\int_0^c dq\,q^{-1}(nq)+const(c)\}\\
&=&const\cdot N_n^*N_n[nc+const(c)]
\end{eqnarray*}

\noindent So $N_n^*N_n\geq const\cdot[nc+const(c)]^{-1}$ and this is the 
correct
asymptotic behavior. Observe how this is not the
limit of the usual $n^{2\beta-1}$ as $\beta\!\nearrow1/2$, which would be 
constant. As $q^{-1}$ is no longer integrable at $q=0$, one picks up an 
extra term of $1/(nc+const(c))\sim 1/n$ for large $n$.

It is alright not to bother about the ``$[\ \ ]$-part" in 
$\parallel\!\! A\psi_n\!\!\parallel^2$
which was going to $0$ even in the case $0<\beta<1/2$ (compare above). Now, 
the ``$q$-part" will 
also go to $0$ by virtue of the new $1/n$-dependence:

\begin{eqnarray*}
& & N_n^*N_n\int_0^\infty dq\,[-q^2n^{-2}+qn^{-1}]^2e^{-2q/n}q^{-1}
(q^{-1}+s^{-1})^{-1} \\
&=& N_n^*N_n\int_0^\infty dq\,[-q+1]^2qe^{-2q}
[(nq)^{-1}+s^{-1}]^{-1}
\end{eqnarray*}

\noindent The integral goes to a constant again 
(trivially,
since the ``$q$-factors" have taken care of the $q^{-1}$). 
But the whole expression
goes to zero because of the $1/n$-dependence of the normalization
factors $N_n^*N_n$!
This completes the proof that $0\in spec(A)$ for $\beta=1/2$. 

Indeed, what was established is exactly what was expected from the knowledge 
of the 
spectrum of A given by DKP [Eq. ({\ref{SpecA})].

Whereas there is no way to isolate the space ${\cal C}_\beta$ 
with the spectral approach in
the case $0<\beta<1/2$ (simply because $0$ is not in the spectrum), this 
is possible for $\beta=1/2$ since the operator $A$ fulfills
all the requirements for the application of the general theory developed 
earlier. This does not seem like a major advance 
compared to the DKP-article, which treated $\beta>1/2$, but in the case 
$\beta=1/2$ one is dealing with weak coherent states for the first time!\\

{\bf The spectral approach and the path integral for 
${\mathbf {\boldsymbol{\beta}}{\boldsymbol{=}}1{\boldsymbol{/}}2}$}\\

As the operator $A$ posesses all required properties in the case
$\beta=1/2$ (non-negativity, self-adjointness, $0$ is in the continuous, but
not in the discrete spectrum), it follows
from the general theory that the procedure to isolate the
subspace ${\cal C}_{1/2}$ can be carried out. This is enough to ensure
that a weak coherent state path integral can be constructed which will 
subsequently be done, at first for a zero Hamiltonian. Dynamics will be 
introduced in the following subsection.
   
For the the present problem a certain connection (established in \cite{DKP}) 
between the operator $A$ and the Morse operator $H_{Morse}$ exists and makes
the explicit functional form of the generalized eigenfunctions available. With
the aid of the latter one can compute the rescaling factor explicitly. 
With these
calculations at hand it turns out to be just a short step to study the
limit of diverging diffusion constant explicitly as well, confirming the
general theory.

The problem to find the eigenfunctions of the operator $A$ 
is reduced to a problem on 
$L^2(\mathbb{ R}^+)$ and then to a problem on $L^2(\mathbb{ R})$ leading 
to the Morse operator:

\begin{eqnarray}\label{ConnectionAH}
& & A\langle U(p,q)\phi|\psi\rangle=\langle A^*U(p,q)\phi|\psi\rangle\nonumber\\
& = &{\textstyle\frac{1}{2}}\langle 
[-\beta^{-1}\partial_q q^2\partial_q-\beta 
q^{-2}\partial^2_p
-2i\beta q^{-1}\partial_p+\beta-1]
e^{ipQ}e^{-i\ln qD}\phi|\psi\rangle\nonumber\\
&=&{\textstyle\frac{1}{2}}
\beta^{-1}\langle e^{ipQ}e^{-i\ln qD}\{D^2+iD+\beta^2Q^2\
-2\beta^2Q+\beta^2-\beta\}(Q^{1/2}\phi^\prime)
|\psi\rangle\nonumber\\
&=&{\textstyle\frac{1}{2}}\beta^{-1}\langle e^{ipQ}e^{-i\ln qD}Q^{1/2}\{D^2
+\beta^2Q^2\
-2\beta^2Q+(\beta-{\textstyle\frac{1}{2}})^2\} \phi^\prime|\psi\rangle
\end{eqnarray}

\noindent Here $e^{i\ln qD}Qe^{-i\ln qD}=qQ$ and 
$[D,Q^\alpha]=-i\alpha Q^\alpha$ were used, 
so $DQ^{1/2}=Q^{1/2}(D-i/2)$ and $D^2Q^{1/2}=Q^{1/2}(D-i/2)^2$. Observe that in 
the first line the operator $A$ on the Lobachevsky plane is not an abstract
Hilbert space operator ${\cal A}$, but already a representation thereof. Thus, 
it acts merely as a linear
functional on the inner product, and, because of anti-linearity in the first 
position, it comes inside complex conjugated.

Under the unitary transformation 

\begin{equation}\label{U}
(\tilde{U}\psi)(x)=e^{x/2}\psi(e^x)
\end{equation}

\noindent the operator in
braces in the last line of Eq. (\ref{ConnectionAH}}) [called $H$ in DKP] 
is transformed to the Morse operator:

\begin{equation}
H_{Morse}=-{\textstyle\frac{d^2}{dx^2}}+\beta^2(e^{2x}-2e^x)
+(\beta-{\textstyle\frac{1}{2}})^2
\end{equation}

The Green's function associated with the Morse operator (and derived from a
conventional configuration space path integral) is given by 
Grosche and Steiner (\cite{GroscheSteiner}, 6.4.2, page
228). 

\begin{eqnarray*}
& & \int_0^\infty \!dT\, e^{i(E+i\varepsilon)T/\hbar}
{\pmb\int}_{x(t^\prime)=x^\prime}^{x(t^{\prime\prime})=x^{\prime\prime}}
\!\exp\{{\textstyle\frac{i}{\hbar}}
{\textstyle\int}_{t^\prime}^{t^{\prime\prime}}
[{\textstyle\frac{m}{2}} \dot{x}^2-{\textstyle\frac{\hbar^2V_0^2}{2m}}
(e^{2x}-2\alpha e^x)]dt\} {\cal D}x\\
&=& -{\textstyle\frac{im}{2\hbar V_0}}
\Gamma(1/2+\sqrt{-2mE}/\hbar-\alpha V_0)\Gamma^{-1}(1+
2\sqrt{-2mE})e^{(x^\prime+x^{\prime\prime})/2}\\
& &\times W_{\alpha V_0,\sqrt{-2mE}/\hbar}(2V_0e^{x_>})
M_{\alpha V_0,\sqrt{-2mE}/\hbar}
(2V_0e^{x_<})
\end{eqnarray*}

\noindent Here $M$ and $W$ are Whittaker functions, the definition of which is:

\begin{eqnarray}\label{Whittaker}
M_{\mu,\nu}(z)&=&e^{-z/2}z^{1/2+\nu} \,_1F_1(1/2-\mu+\nu,1+2\nu,z)\nonumber\\
W_{\mu,\nu}(z)&=&e^{-z/2}z^{1/2+\nu} U(1/2-\mu+\nu,1+2\nu,z)
\end{eqnarray}

\noindent where $_1F_1$ and $U$ are confluent hypergeometric functions 
(of the first and second kind):

\begin{eqnarray*}
_1F_1(a,b,z)&=&\Gamma(b)\Gamma^{-1}(a)\Gamma^{-1}(b-a)\int_0^1
e^{zt}t^{a-1}(1-t)^{b-a-1}dt\\
U(a,b,z)&=&\Gamma^{-1}(a)\int_0^\infty e^{-zt}t^{a-1}(1+t)^{b-a-1}dt
\end{eqnarray*}

In this work $\hbar=1$, and furthermore choose $m=1/2$, $\alpha=1$,
$V_0=\beta$ and finally $\beta=1/2$ \footnote{This is a critical 
case; for further details consult Grosche\cite{Grosche},
page 118.}

Grosche/Steiner \cite{GroscheSteiner} and Grosche \cite{Grosche} 
show that 
the Green's function can also be written as\footnote{The factor $-i$ is not
present in \cite{Grosche} since Grosche defines the Fourier transformation with
an imaginary factor; this rule is not being follwed here.}:

\begin{eqnarray}\label{GreensFunction0}
G(x^{\prime\prime},x^\prime,E)
& = & {\textstyle\frac{-i}{\pi^2}}\int_0^\infty d\lambda
{\textstyle\frac{\lambda\sinh 2\pi \lambda}{\lambda^2-E}}
|\Gamma(i\lambda)|^2 e^{-(x^\prime+x^{\prime\prime})/2}\nonumber\\
& &\phantom{{\textstyle\frac{-i}{\pi^2}}}
\times W_{1/2,i\lambda}(
e^{x^\prime})W_{1/2,i\lambda}(e^{x^{\prime\prime}})
\end{eqnarray}

\noindent First replace $\int_0^\infty$ by $\frac{1}{2}\int_{-\infty}^\infty$.
This is correct since the Whittaker functions have the property 
$W_{\mu,\nu}=W_{\mu,-\nu}$ and the rest of the integrand is even as well. 
If the integral is now regularized by the convention 
$E\rightarrow E+i\epsilon$ one can go 
to a contour integral from -R to R on the real line and on a semi-circle
over this (can be closed in the upper or lower half plane). By Cauchy's 
integral formula the contour integral is $2\pi i$ times the residue of the 
integrand (which has a simple pole). Unfortunately it is not easy to prove
that the integral over the semi-circle vanishes in the limit
$R\rightarrow\infty$, which is necessary to go back to the real-line integral. 
But since the final result will be shown to coincide with a result found by 
Grosche this point is not stressed here. The Green's function becomes

\begin{eqnarray}\label{GreensFunction}
G(x^{\prime\prime},x^\prime,E)
& = & {\textstyle\frac{1}{\pi}}\sinh (2\pi \sqrt{E})
|\Gamma(i\sqrt{E})|^2 e^{-(x^\prime+x^{\prime\prime})/2}\nonumber\\
& &\times W_{1/2,i\sqrt{E}}(
e^{x^\prime})W_{1/2,i\sqrt{E}}(e^{x^{\prime\prime}})
\end{eqnarray}

\noindent The connection of the Green's function
to the energy eigenfunctions is given by 
Eq. (\ref{GreenAndEigenfunctions}). By the property 
$W_{\mu,\nu}=W_{\mu,-\nu}$ of the Whittaker function 
Eq. (\ref{GreensFunction}) is already real, so the real part need not
be taken and the expression (\ref{GreensFunction}) is equal to

\begin{equation*}
\pi\psi_E(x^{\prime\prime})
\psi_E^*(x^\prime)\rho(E)
\end{equation*}

From the general discussion earlier it is known that $\rho(E)$ must be 
identically $1$ for $\delta$-orthonormalized wave functions. This can be
verified directly. Assume $\rho(E)=1$ then the energy eigenfunctions are

\begin{equation}\label{MorseWavefunctions}
\psi_E(x)=({\textstyle\frac{\sinh(2\pi\sqrt{E})}
{2\pi^2}})^{1/2}\Gamma(i\sqrt{E})e^{-x/2}W_{1/2,i\sqrt{E}}
(e^x)
\end{equation}
 
\noindent and they are $\delta$-orthonormalized. To see this 
use equations III.22 and III.23 of \cite{Grosche}, which 
for the case at hand show that $\int_0^\infty u^{-2}W_{1/2,i\lambda}(u)
W_{1/2,i\lambda^\prime}(u)du=2\pi|\frac{\Gamma(2i\lambda)}
{\Gamma(i\lambda)}|^2\delta(\lambda-\lambda^\prime)$. Further use
$\Gamma(iy)\Gamma(-iy)=\pi y^{-1}[\sinh(\pi y)]^{-1}$ for
real $y$ (\cite{Abramowitz}, 
6.1.29) and thus

\begin{eqnarray*}
& &\int_{-\infty}^\infty\psi_E(x)\psi_{E^\prime}^*(x)dx\\
&=& {\textstyle\frac{1}{2\pi^2}}
[\sinh(2\pi \sqrt{E})
\sinh(2\pi \sqrt{E^\prime})]^{1/2}|\Gamma(i\sqrt{E})|
|\Gamma(-i\sqrt{E^\prime})|\\
& &\times \int_{-\infty}^\infty dx e^{-x}W_{1/2,i\lambda}(e^x)
W_{1/2,i\lambda^\prime}(e^x)\\
&=& {\textstyle\frac{1}{2\pi^2}}\sinh(2\pi\sqrt{E})2\pi
|\Gamma(2i\sqrt{E})|^2\delta(\sqrt{E}-
\sqrt{E^\prime}) =\delta(E-E^\prime)
\end{eqnarray*}

Grosche \cite{Grosche} Fourier transforms Eq. (\ref{GreensFunction0})
and thus gets the Feynman Kernel from which he reads off the eigenfunctions
in momentum representation ($E=\lambda^2$ since $m=1/2$ was chosen): 

\begin{equation}\label{MorseWavefunctionsLambda}
\psi_\lambda(x)
=({\textstyle\frac{\lambda\sinh(2\pi\lambda)}
{\pi^2}})^{1/2}\Gamma(i\lambda)e^{-x/2}W_{1/2,i\lambda}
(e^x)
\end{equation}

\noindent He also proves that they are $\delta$-orthonormalized wave-functions. 
Since $E=\lambda^2$, the momentum wave-functions $\psi_\lambda$ are
connected to the energy wave-functions $\psi_E$ by 
$\int\psi_{\lambda^\prime}^*(x)\psi_\lambda(x)dx=\delta(\lambda-\lambda^\prime)
=\delta(\sqrt{E}-\sqrt{E^\prime})=2\sqrt{E}\delta(E-E^\prime)=
\int\psi_{E^\prime}^*(x)\psi_E(x)\cdot 2\sqrt{E}dx$ (change effected by 
$\delta$-function rule) and 
$\int\psi_{\lambda}(x^{\prime\prime})\psi_\lambda^*(x^\prime)
d\lambda=\delta(x^{\prime\prime}-x^\prime)=
\int\psi_E(x^{\prime\prime})\psi_E^*(x^\prime)
dE$ (change effected by the measure).

\begin{figure}
\includegraphics{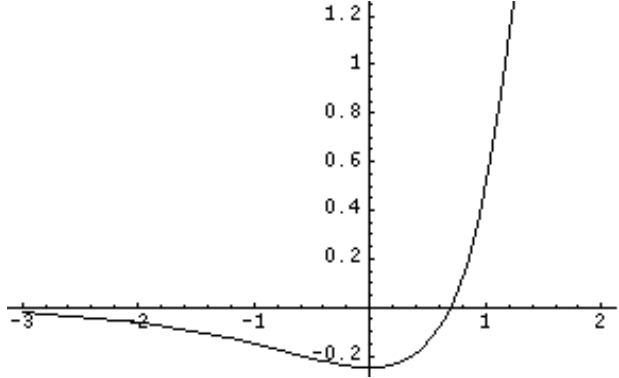} 
\caption{Morse Potential $\frac{1}{4}(\exp\{2x\}-2\exp\{x\})$}
\label{MorsePotential2}
\end{figure}

It was pointed out that, from a mathematical point of view, a density $\rho=1$ 
followed from the $\delta$-orthonormalization and the fact that the spectral 
family $[\mathbb{E}(\lambda)]$ is a family of projection operators. From a
physical point of view, the density $\rho(E)$ can be interpreted as a measure
for the degeneracy of the energy-wavefunctions. For the Morse potential
$\frac{1}{4}(\exp\{2x\}-2\exp\{x\})$ [see figure 
\ref{MorsePotential2}] one would expect no degeneracy or in other words
$\rho(E)=1$.

Since the Whittaker function $W_{1/2,0}(z)=e^{-z/2}z^{1/2}$ [as can easily
be seen from the definition (\ref{Whittaker})], the $x$-dependence of
$\psi_{E=0}(x)$ is $e^{-e^x/2}$. The rescaling factor is thus best be 
determined with the choice $x^{\prime\prime}=x^\prime=b=-\infty$ where this
function is $1$. For small $E$ the function $f_{-\infty,-\infty}(E)
=\psi_E(-\infty)\psi_E^*(-\infty)\rho(E)\approx \pi^{-1}E^{-1/2}$ because
$\sinh(2\pi\sqrt{E})\approx 2\pi\sqrt{E}$, and $|\Gamma(i\sqrt{E})|^2\approx
1/E$, and $\rho(E)=1$. Inserting this $E$-dependence into the general formula
(\ref{Example}) [set $a=-1/2$ there] one finds the inverse rescaling factor

\begin{equation}\label{InvRescalingFactor}
\int_0^\infty e^{-\nu TE}f_{-\infty,-\infty}(E)dE=(\pi\nu T)^{-1/2} 
\end{equation}

Because of the connection between the ``Morse"-level and the original 
problem [Eqs. (\ref{ConnectionAH}) and (\ref{U})] the rescaling
factor, computed on the ``Morse"-level, is valid for the original
problem as well. Nonetheless it is useful to write down the problem and
its solution on the ``A"-level.

The eigenfunctions for $A$ are\footnote{Observe
that strictly $\langle U(p,q)Q^{1/2}\psi_E^\prime|$ is a linear functional
on the Hilbert space $L^2(\mathbb{ R}^+)$, but the
corresponding ``ket" is not in the Hilbert space. But this ``ket" is 
much better behaved than e.g. $|x\rangle$ which usually
is used without the slightest hesitation. The $\psi_{E,\phi}(p,q)$ 
are not in $L^2(M_+)$, but are well-behaved functions (among them is e.g.
the coherent state overlap), and they can be taken as generalized eigenvectors.}:

\begin{eqnarray*}
A\psi_{E,\phi}(p,q)&:=& A\langle U(p,q)Q^{1/2}\psi_E^\prime|\phi\rangle
= \langle U(p,q)Q^{1/2}H\psi_E^\prime|\phi\rangle\\
&=& \langle U(p,q)Q^{1/2}U^{-1}UHU^{-1}U\psi_E^\prime|\phi\rangle\\
&=&\langle U(p,q)Q^{1/2}U^{-1}H_{Morse}\psi_E|\phi\rangle\\
&=& \langle U(p,q)Q^{1/2}U^{-1}E\psi_E|\phi\rangle
=E\langle U(p,q)Q^{1/2}U^{-1}\psi_E|\phi\rangle\\
&=& E\langle U(p,q)Q^{1/2}\psi^\prime_E|\phi\rangle 
=E\psi_{E,\phi}(p,q)
\end{eqnarray*}

As was stated above the $x$-dependence of $\psi_{E=0}$ is $e^{-e^x/2}$. 
The inverse of the unitary transformation $\tilde{U}$ 
[Eq. (\ref{U})]
is 
$\tilde{U}^{-1}: L^2(\mathbb{ R})\rightarrow 
L^2(\mathbb{ R}^+), \psi^\prime(x)=
(\tilde{U}^{-1}\psi)(x)=x^{-1/2}\psi(\ln x)$ and applying this to $\psi_{E=0}$
one has

\begin{equation*}
\psi_0^\prime(x)\propto x^{-1/2}e^{-\beta x}
\end{equation*}

\noindent Finally, by multiplying with $x^{1/2}$, one obtains:

\begin{equation*}
e^{-\beta x}
\end{equation*}

\noindent This is exactly the minimum uncertainty state $\eta_\beta$ for 
$\beta=1/2$ and the result is no surprise, since this means that the 
eigenfunctions of $A$ for eigenvalue $0$ are exactly $\psi_{0,\phi}(p,q)
=\langle pq|\phi\rangle$. 

Observe that overall factors (normalizations) are completely 
unimportant  
since a quotient is formed anyway [compare e.g.
Eq. (\ref{KernelConvergence})]. Even if they did ``survive" it would not
matter since a reproducing kernel multiplied with a factor is just as good
a kernel (only the inner product in the reproducing kernel Hilbert space 
has to be redefined). 

With an orthonormal basis $(|\phi_m\rangle)$ in $L^2(\mathbb{ R}^+)$ one can 
write the corresponding
eigenfunctions of $A$ as $\psi_{E,\phi_m}$, and again by the above: 
$\psi_{0,\phi_m}
=\langle pq|\phi_m\rangle$. 
Then use $\sum_m\int\psi_{E,\phi_m}
(p,q)\psi^*_{E,\phi_m}(p^\prime,q^\prime)\rho(E)dE=\delta(p-p^\prime)
\delta(q-q^\prime)$ [and still $\rho(E)=1$] to find:

\begin{eqnarray}\label{KernelConvergence}
&&\lim_{\nu\rightarrow\infty}K_\nu e^{-\nu TA}\delta(p-p^\prime)
\delta(q-q^\prime)\nonumber\\
&=& \lim_{\nu\rightarrow\infty}\frac{\sum_m\int_0^\infty e^{-\nu TE}
\psi_{E,\phi_m}(p,q)\psi^*_{E,\phi_m}(p^\prime,q^\prime)\rho(E)dE}
{\sum_{m^\prime}\int_0^\infty e^{-\nu TE}\psi_{E,\phi_{m^\prime}}(0,1)
\psi^*_{E,\phi_{m^\prime}}(0,1)
\rho(E)dE}\nonumber\\
&=&\sum_m\psi_{0,\phi_m}(p,q)\psi^*_{0,\phi_m}(p^\prime, q^\prime)\nonumber\\
&=& \sum_m\langle pq|\phi_m\rangle\langle\phi_m|p^\prime q^\prime\rangle
=\langle pq|p^\prime q^\prime\rangle
\end{eqnarray}

\noindent where $K_\nu^{-1}=e^{-\nu TA}\delta(p)\delta(q-1)|_{p=0,q=1}
=(\pi\nu T)^{-1/2}$ by Eq. (\ref{InvRescalingFactor}).

This is true at least in a distributional sense. Pointwise convergence on
the other hand is not a problem, since the functional form of the overlap is
known and is obviously continuous.

The last step is to formulate the weak coherent state path integral. Since 
the Feynman-Kac-Stratonovich representation of the operator $e^{-\nu TA}$\
is already known from section (\ref{CSCSPI}), this is quickly done:

\begin{eqnarray}\label{WCSPI}
&&\lim_{\nu\rightarrow\infty}K_\nu e^{-\nu TA}\delta(p-p^\prime)
\delta(q-q^\prime)|_{p=p^{\prime\prime},q=q^{\prime\prime}}\nonumber\\
&=&\lim_{\nu\rightarrow\infty}K_\nu {\cal N}_\nu {\pmb \int}\exp\{-i
{\textstyle \int}q\dot{p}dt\}\exp\{-\frac{1}{2\nu}{\textstyle \int}
[\beta^{-1}q^2\dot{p}^2+\beta q^{-2}\dot{q}^2]dt\}{\cal D}p{\cal D}q\nonumber\\
&=&\lim_{\nu\rightarrow\infty}K_\nu {\pmb \int}\exp\{-i
{\textstyle \int}q\,dp\}\:d\mu_W^\nu\nonumber\\
&=&\langle p^{\prime\prime}q^{\prime\prime}|p^\prime q^\prime\rangle
\end{eqnarray}

\noindent The rescaling factor is $K_\nu=(\pi\nu T)^{1/2}$ and the 
Wiener measure $d\mu_W^\nu$ is the same as in Eq.
(\ref{WienerMeasure}). This is the sought-for weak coherent state path integral
for $\beta=1/2$ when the Hamiltonian vanishes.\\

{\bf Direct derivation}\\

In the foregoing the weak coherent state path integral for $\beta=1/2$ and
a vanishing Hamiltonian was established. Its derivation relied on the 
general spectral approach. Since the (generalized) eigenfunctions of the Morse
operator are known, and because they are connected to the eigenfunctions of
$A$, the isolating procedure can be developed step by step, i.e. 
$\lim_{\nu\rightarrow\infty}K_\nu e^{-\nu TA}\delta(p-p^\prime)
\delta(q-q^\prime)$ can be computed directly. This will not lead to new 
knowledge, but since
Eq. (\ref{WCSPI}) is a central result it is good to confirm it:

\begin{eqnarray}\label{limitkernel}
&&\lim_{\nu\rightarrow\infty}K_\nu e^{-\nu TA}\delta(p-p^\prime)
\delta(q-q^\prime)\nonumber\\
&=& \lim_{\nu\rightarrow\infty}\frac{\sum_m\int_0^\infty e^{-\nu TE}
\psi_{E,\phi_m}(p,q)\psi^*_{E,\phi_m}(p^\prime,q^\prime)\rho(E)dE}
{\sum_{m^\prime}\int_0^\infty e^{-\nu TE}\psi_{E,\phi_{m^\prime}}
(0,1)\psi^*_{E,\phi_{m^\prime}}(0,1)\rho(E)dE}\nonumber\\
&=&\lim_{\nu\rightarrow\infty}\frac{\sum_m\int_0^\infty e^{-\nu TE}
\langle U(p,q)Q^{1/2}\psi_E^\prime|\phi_m\rangle\langle \phi_m|
U(p^\prime,q^\prime)Q^{1/2}\psi_E^\prime\rangle\rho(E)dE}
{\sum_{m^\prime}\int_0^\infty e^{-\nu TE}\langle Q^{1/2}\psi_E^\prime|
\phi_{m^\prime}\rangle
\langle \phi_{m^\prime}|Q^{1/2}\psi_E^\prime\rangle\rho(E)dE}\nonumber\\
&=&\lim_{\nu\rightarrow\infty}\{{\textstyle\int}_0^\infty e^{-\nu TE}
{\textstyle\int}_0^\infty dx (qq^\prime)^{-1/2}e^{-ix(p-p^\prime)}[(x/q)^{1/2}
\psi_E^\prime(x/q)]^*\nonumber\\
&&\phantom{\lim_{\nu\rightarrow\infty}
\{{\textstyle\int}_0^\infty e^{-\nu TE}{\textstyle\int}_0^\infty}
\times[(x/q^\prime)^{1/2}\psi_E^\prime(x/q^\prime)]\rho(E)dE\}\nonumber\\
&&\phantom{\lim_{\nu\rightarrow\infty}}
\times\{{\textstyle\int}_0^\infty e^{-\nu TE}{\textstyle\int}_0^\infty dx 
[x^{1/2}\psi_E^\prime(x)]^*
[x^{1/2}\psi_E^\prime(x)]\rho(E)dE\}^{-1}
\end{eqnarray}

Again $\rho(E)=1$ since the eigenfunctions are $\delta$-orthonormalized. 
The denominator of Eq. (\ref{limitkernel}) will be examined 
first and
it will be more convenient to work with the eigenfunctions $\psi_\lambda$
The connection between $\psi_\lambda$ and $\psi_E$ was demonstrated in Eqs.
(\ref{MorseWavefunctionsLambda}), (\ref{MorseWavefunctions}). Recall that,
by definition, 
$\psi_\lambda^\prime(x)=U^{-1}\psi_\lambda(x)$ with
$U^{-1}: L^2(\mathbb{ R})\rightarrow L^2(\mathbb{ R}^+), \psi^\prime(x)=
(U^{-1}\psi)(x)=x^{-1/2}\psi(\ln x)$. With $\beta=1/2$ one has:

\begin{equation*}
x^{1/2}\psi_\lambda^\prime(x)=x^{1/2}x^{-1/2}\psi_\lambda(\ln (x))
= ({\textstyle\frac{\lambda\sinh(2\pi\lambda)}
{\pi^2}})^{1/2}\Gamma(i\lambda)x^{-1/2}W_{1/2,i\lambda}(x)
\end{equation*}

\noindent For the $x$-integration, III.22 of Grosche \cite{Grosche} 
[or \cite{Gradsheteyn}
page 858] is used:

\begin{eqnarray*}
&&\int_0^\infty x^{\rho-1}W_{\kappa,\mu}(x)W_{\lambda,\nu}(x)dx\\
&=& {\textstyle\frac{\Gamma(1+\mu+\nu+\rho)\Gamma(1-\mu+\nu+\rho)\Gamma(-2\nu)}
{\Gamma(1/2-\lambda-\nu)\Gamma(3/2-\kappa+\nu+\rho)}}\\
&&\times \,_3F_2(1+\mu+\nu+\rho, 1-\mu+\nu+\rho, 1/2-\lambda+\nu;\\
&&\phantom{\times \,_3F_2(}
1+2\nu, 3/2-\kappa+\nu+\rho; 1)\\
&&+{\textstyle\frac{\Gamma(1+\mu-\nu+\rho)\Gamma(1-\mu-nu+\rho)\Gamma(2\nu)}
{\Gamma(1/2-\lambda+\nu)\Gamma(3/2-\kappa-\nu+\rho)}}\\
&&\times \,_3F_2(1+\mu-\nu+\rho, 1-\mu-\nu+\rho, 1/2-\lambda-\nu;\\
&&\phantom{\times \,_3F_2(}
1-2\nu, 3/2-\kappa-\nu+\rho; 1)
\end{eqnarray*}

\noindent Using $\rho=0$, $\kappa=\lambda=1/2$, $\mu=-\nu=i\lambda$, the 
equation becomes:

\begin{eqnarray*}
&&\int_0^\infty x^{-1}W_{1/2,i\lambda}(x)W_{1/2,i\lambda}(x)dx\\
&=&{\textstyle \frac{\Gamma(1)\Gamma(1-2i\lambda)\Gamma(2i\lambda)}
{\Gamma(i\lambda)\Gamma(1-i\lambda)}}\\
&&\times \,_3F_2(1, 1-2i\lambda, -i\lambda;
1-2i\lambda, 1-i\lambda; 1)\\
&&+{\textstyle\frac{\Gamma(1+2i\lambda)\Gamma(1)\Gamma(-2i\lambda)}
{\Gamma(-i\lambda)\Gamma(1+i\lambda)}}\\
&&\times \,_3F_2(1+2i\lambda,1, i\lambda;
1+2i\lambda, 1+i\lambda; 1)\\
\end{eqnarray*}

\noindent With $\Gamma(1+x)=x\Gamma(x)$ this relation simplifies:

\begin{eqnarray*}
&&2\frac{\Gamma(2i\lambda)\Gamma(-2i\lambda)}{\Gamma(i\lambda)
\Gamma(-i\lambda)}\\
&& [\,_3F_2(1, 1-2i\lambda, -i\lambda;
1-2i\lambda, 1-i\lambda; 1)\\
&&+\,_3F_2(1+2i\lambda,1, i\lambda;
1+2i\lambda, 1+i\lambda; 1)]
\end{eqnarray*}

Observe that by $\Gamma(iy)\Gamma(-iy)=\frac{\pi}{y\sinh(\pi y)}$ 
[e.g.
\cite{Abramowitz}, 6.1.29, p. 256], the prefactor cancels
the normalization of the wave functions, except for a factor $\pi$. 

Now, using the definition of the generalized hypergeometric function 
(or Barnes's extended hypergeometric function) 

\begin{eqnarray*}
_pF_q(\alpha_1,...,\alpha_p;\beta_1,...,\beta_q;z)
&:=& \sum_{n=0}^\infty \frac{(\alpha_1)_n...(\alpha_p)_n}{(\beta_1)_n...
(\beta_q)_n}\frac{z^n}{n!}
\end{eqnarray*}

\noindent with $(\alpha)_n=\Gamma(\alpha+n)/\Gamma(\alpha)$ 
(Gegenbauer coefficients), one can proceed formally

\begin{eqnarray*}
&&[\,_3F_2(1, 1-2i\lambda, -i\lambda;
1-2i\lambda, 1-i\lambda; 1)\\
&&+\,_3F_2(1+2i\lambda,1, i\lambda;
1+2i\lambda, 1+i\lambda; 1)]\\
&=&[\,_2F_1(1, -i\lambda; 1-i\lambda; 1)\\
&&+\,_2F_1(1, i\lambda; 1+i\lambda; 1)]\\
&=& \sum_{n=0}^\infty \frac{-i\lambda}{-i\lambda+n}+\sum_{n=0}^\infty 
\frac{i\lambda}{i\lambda+n}\\
&=:& \sum_{n=0}^\infty \frac{2\lambda^2}{n^2+\lambda^2}
= \pi \lambda \coth (\pi \lambda)+1
\end{eqnarray*}

It is important to notice that the first and second equality are merely
symbolic since the expressions are undefined (complex infinite terms). 
Because 
$|z|=1$ is the radius of convergence for the hypergeometric series defining
$\,_2F_1(a,b;c;z)$ (\cite{Abramowitz}, 15.1.1), only
conditional convergence exists 
for $-1<\Re(c-a-b)\leq 0$ (the point $z=1$ is even excluded).
Meaning was given to the undefined sums of infinite terms by setting up
an ordering prescription, thus arriving at the last line. The same ordering
prescription arises if one replaces the argument 1 by $\delta<1$, sums the
now absolutely convergent series and finally 
takes the limit $\delta\rightarrow 1$.

The complete denominator has become

\begin{eqnarray*}
\pi^{-1}\int_0^\infty e^{-\nu T\lambda^2}[\pi \lambda \coth (\pi \lambda)+1]
d\lambda&=:&h(\nu T)
\end{eqnarray*}

\noindent This integral exists and $h(\nu T)$ is well-defined. 
Using the power-series
expansion for the hyperbolic tangent the result is:

\begin{equation}\label{hNuT}
h(\nu T)=\pi^{-1/2}(\nu T)^{-1/2}[1+\pi^2/(12\nu T)+O((\nu T)^{-2})]
\end{equation}

\noindent The first term in the expansion is equal to the inverse rescaling
factor computed in Eq. (\ref{InvRescalingFactor}). 

There is yet another way to treat 
the denominator. The idea is to extract the $\lambda$-dependence
by series expansions, but the calculations are more complicated and so 
this part is presented in appendix (\ref{AppendixA4}).

The remaining task is to compute (again the $\lambda$-notation
is chosen rather than the $E$-notation):

\begin{eqnarray*}
&&\lim_{\nu\rightarrow\infty}\pi^{1/2}\sqrt{\nu T}
(1-\frac{\pi^2}{12\nu}+O(\nu^{-2}))\\
&&\times\int_0^\infty d\lambda\, e^{-\nu T\lambda^2}\int_0^\infty dx\, 
(qq^\prime)^{-1/2}
e^{-ix(p-p^\prime)}[(x/q)^{1/2}\psi_{\lambda}^\prime(x/q)]^*\\
&&\phantom{\times\int_0^\infty d\lambda e^{-\nu T\lambda^2}\int_0^\infty}
\times[(x/q^\prime)^{1/2}\psi_{\lambda}^\prime(x/q^\prime)]
\end{eqnarray*}

\noindent First, notice that
for $p=p^\prime$ and $q=q^\prime$, 
the $x$-integral will be the same as the one encountered in 
the denominator (make the transformation $x\rightarrow xq$) and thus
the whole expression is exactly 1 [which can already
be seen by symbolic manipulation in the third line of Eq. 
(\ref{limitkernel}).]. This result was expected, since 
$\langle pq|pq\rangle=1$.

Unfortunately it is very hard (if not impossible) to find an exact solution 
for the $x$-integral. 

Even for $q=q^\prime$ and after expanding $e^{-ix(p-p^\prime)}$ into a power
series, it does not seem possible to use the integration formula for the 
Whittaker functions and powers of $x$ and to sum things up using the 
expansions of the generalized hypergeometric functions, since this does not
converge. However, one still expects the integrals to exist.
This is a pity, since varying $p$ and $p^\prime$ alone would be enough, 
since the functions $\langle pq|p^\prime q\rangle$ already span the space.

But one can simply look at $[(x/q)^{1/2}\psi_{\lambda}^\prime(x/q)]^*
[(x/q^\prime)^{1/2}\psi_{\lambda}^\prime(x/q^\prime)]$ and write it as
$f_{x,q,q^\prime}(\lambda)$. This is continuous in $\lambda$, 
and one can write
$f_{x,q,q^\prime}(\lambda)=g_{x,q,q^\prime}+O_{x,q,q^\prime}
(\lambda)$. The 
$\lambda$-independent function g is exactly \\
$[(x/q)^{1/2}\psi_{0}^\prime(x/q)]^*
[(x/q^\prime)^{1/2}\psi_{0}^\prime(x/q^\prime)]=2\pi^{-1}
\eta(x/q)^*\eta(x/q^\prime)$, 
and the $\lambda$- and $x$-integrals will factorize. The first integral yields
$\frac{1}{2}\pi^{1/2}(\nu T)^{-1/2}$, while the second integral gives
$\pi^{-1}\langle pq|p^\prime q^\prime\rangle$. 
So,
the limit of diverging diffusion
constant yields the desired reproducing kernel.

One has to make sure that, for the higher orders $O_{x,q,q^\prime}(\lambda)$, 
the $x$-integrals exist. Otherwise the expansion followed by interchange
of summation and integration will not be justified.  
But this proof is simple since by the Cauchy-Schwarz inequality

\begin{eqnarray*}
&&|\int_0^\infty dx [q^{-1/2}e^{ixp}(x/q)^{1/2}\psi_{\lambda^2}^\prime(x/q)]^*
[q^{\prime-1/2}e^{ixp^\prime}(x/q^\prime)^{1/2}
\psi_{\lambda^2}^\prime(x/q^\prime)]|^2\\
&\leq& \int_0^\infty dx
|[q^{-1/2}e^{ixp}(x/q)^{1/2}\psi_{\lambda^2}^\prime(x/q)]|^2\\
&&\times \int_0^\infty dx
|[q^{\prime-1/2}e^{ixp^\prime}(x/q^\prime)^{1/2}
\psi_{\lambda^2}^\prime(x/q^\prime)]|^2\\
&=&q^{1/2}\int_0^\infty dx
|[x^{1/2}\psi_{\lambda^2}^\prime(x)]|^2
\cdot q^{\prime 1/2}\int_0^\infty dx
|[x^{1/2}\psi_{\lambda^2}^\prime(x)]|^2
\end{eqnarray*}

\noindent and the existence of the right-hand side has already been proved. 
These integrals are like the ones in the denominator (compare above). 
Once this is established, the exact
result of any of these integrations (which might be very hard to determine)
is of no consequence since integration over $\lambda$ gives in general 
($n\geq1$):

\begin{equation*}
\int_0^\infty e^{-\nu T\lambda^2}\lambda^n d\lambda=1/\sqrt{\nu T}\cdot 
\Gamma((n+1)/2)
(\nu T)^{-n/2}
\end{equation*}

\noindent Whatever the corresponding $x$-integral is, 
these terms will all vanish in the limit $\nu\rightarrow\infty$.

In this way, the isolating procedure has been directly computed and the results
(\ref{KernelConvergence}) and (\ref{WCSPI}) are confirmed.

\subsubsection{Introducing dynamics}\label{WCSIsolatingDynamics}

Since the only case in which the spectral approach worked was $\beta=1/2$, this
value is assumed throughout this subsection.
Dynamics are introduced by the quantum Hamiltonian ${\cal H}$, 
which is a function
of the basic kinematical operators $Q$ and $D$. The goal is to represent the
propagator $\langle p^{\prime\prime}q^{\prime\prime}|\exp\{-iT{\cal H}\}|
p^\prime q^\prime\rangle$ as a weak coherent state path integral. 

Klauder \cite{KlauderWCS} proposed 

\begin{eqnarray}
&&\langle p^{\prime\prime}q^{\prime\prime}|\exp\{-iT{\cal H}\}|
p^\prime q^\prime\rangle \nonumber\\
&=&\lim_{\nu\rightarrow\infty}K_\nu {\cal N}_\nu {\pmb \int}\exp\{-i
{\textstyle \int}[q\dot{p}+h_w(p,q)]dt\}\nonumber\\
&&\phantom{\lim_{\nu\rightarrow\infty}K_\nu {\cal N}_\nu {\pmb \int}}
\times\exp\{-{\textstyle\frac{1}{2\nu}}{\textstyle \int}
[\beta^{-1}q^2\dot{p}^2+\beta q^{-2}\dot{q}^2]dt\}{\cal D}p{\cal D}q\nonumber\\
&=&\lim_{\nu\rightarrow\infty}K_\nu {\pmb \int}\exp\{-i
{\textstyle \int}[q\,dp+h_w(p,q)dt]\}\:d\mu_W^\nu
\end{eqnarray}

\noindent as the path integral for a class of Hamiltonians which contains 
at least all Hamiltonians polynomial in $Q$ and $D$. The new symbol $h_w(p,q)$, 
interpreted as the classical Hamiltonian 
associated with the quantum Hamiltonian, is implicitly given by

\begin{equation}
\langle p^{\prime\prime}q^{\prime\prime}|{\cal H}|
p^\prime q^\prime\rangle
=\lim_{\nu\rightarrow\infty}K_\nu {\pmb \int}\exp\{-i
{\textstyle \int}q\,dp\}T^{-1}{\textstyle\int}h_w(p,q)dt
\:d\mu_W^\nu
\end{equation}

\noindent and will be called the weak symbol. 

The whole conjecture is based on the observation that, for a linear
Hamiltonian $RQ+SD$, the propagator can be reduced to a mere overlap. This can
be seen as follows: Write $\exp\{-it[RQ+SD]\}=\exp\{-iF(t)Q\}\exp\{-iG(t)D\}$.
Differentiation leads to the differential equation
$RQ+SD=(\dot{F}+F\dot{G})Q+\dot{G}D$, where $\exp\{-iFQ\}D\exp\{iFQ\}
=D+FQ$ was used at an intermediate step. The solution of this equation
with the initial values $F(0)=G(0)=0$ is $F(t)=R/S(1-\exp\{-St\})$ and
$G(t)=St$. With this,
one has

\begin{eqnarray*}
&&e^{i\ln q^{\prime\prime}D}e^{-ip^{\prime\prime}Q}e^{-i(RQ+SD)T}
=e^{i\ln q^{\prime\prime}D}e^{-ip^{\prime\prime}Q}e^{-iR/S
(1-e^{ST})Q}e^{-iSTD}\\
&=&e^{i\ln q^{\prime\prime}D}e^{-iSTD}e^{iSTD}
e^{-i[p^{\prime\prime}+R/S(1-e^{ST})]Q}e^{-iSTD}\\
&=&e^{i\ln(q^{\prime\prime}e^{-ST})D}e^{-ie^{ST}[p^{\prime\prime}
+R/S(1-e^{ST})]Q}\\
&=&e^{i\ln(q^{\prime\prime}e^{-ST})D}e^{-i[p^{\prime\prime}e^{ST}
+R/S(e^{ST}-1)]Q}
\end{eqnarray*}

\noindent and thus

\begin{equation}
\langle p^{\prime\prime}q^{\prime\prime}|e^{-i(RQ+SD)T}|p^\prime q^\prime
\rangle
=\langle p^{\prime\prime}e^{ST}+R/S\cdot(e^{ST}-1), q^{\prime\prime}
e^{-ST}|p^\prime q^\prime\rangle
\end{equation}

Consequently, the problem is already solved for a linear Hamiltonian
and it remains to determine the weak symbol associated with ${\cal H}=RQ+SD$. 
According to Eq. (\ref{WCSPI}) the path integral for this is

\begin{equation}
\lim_{\nu\rightarrow\infty}K_\nu {\pmb \int}^{p^{\prime\prime}e^{ST}
+R/S\cdot(e^{ST}-1), q^{\prime\prime}e^{-ST}}_{p^\prime, q^\prime}
\exp\{-i{\textstyle \int}q\,dp\}\:d\mu_W^\nu
\end{equation}

Since this is a well-defined functional integral, one can change
variables

\begin{eqnarray*}
&&p(t)\rightarrow p(t)e^{St}+R/S(e^{St}-1)\\
&&q(t)\rightarrow q(t)e^{-St}
\end{eqnarray*}

\noindent and the integrand becomes now $\exp\{-i\int(qe^{-St}
d[pe^{St}+R/S(e^{St}-1)]\}=\exp\{-i\int[q\,dp+(Rq+Spq)dt]\}$. The new measure
is\footnote{$(...)^\bullet$ means the time derivative of the expression
in parentheses}

\begin{eqnarray*}
d\tilde{\mu}_W^\nu
&=&{\cal N}_\nu\exp\{-{\textstyle \frac{1}{2\nu}}
{\textstyle \int}
[\beta^{-1}(qe^{-St})^2(pe^{St}+{\textstyle \frac{R}{S}}(e^{St}-1))^{\bullet 2}
\nonumber\\
&&+\beta(qe^{-St})^{-2}(qe^{-St})^{\bullet 2}]dt\}
{\cal D}[pe^{St}+{\textstyle \frac{R}{S}}(e^{St}-1)]{\cal D}(qe^{-St})
\end{eqnarray*}

\noindent But, since the symbolic measure is actually

\begin{eqnarray*}
{\cal D}[pe^{St}+{\textstyle \frac{R}{S}}(e^{St}-1)]&=&
\lim_{\epsilon\rightarrow0}{\textstyle \prod}_{k=1}^N d[p(t)e^{St}
|_{t=k\epsilon}+{\textstyle \frac{R}{S}}(e^{St}-1)|_{t=k\epsilon}]\\
&=&\lim_{\epsilon\rightarrow0}{\textstyle \prod}_{k=1}^N[dp(t)
e^{St}|_{t=k\epsilon}+(pSe^{St}+Re^{St})dt|_{t=k\epsilon}]\\
&=&\lim_{\epsilon\rightarrow0}{\textstyle \prod}_{k=1}^N[dp_k
e^{Sk\epsilon}+(p_kSe^{Sk\epsilon}+Re^{Sk\epsilon})\epsilon]\\
&=&\lim_{\epsilon\rightarrow0}{\textstyle \prod}_{k=1}^N dp_k
e^{Sk\epsilon}={\cal D}p{\textstyle\prod}_te^{St}
\end{eqnarray*}

\noindent and analogous ${\cal D}(qe^{-St})={\cal D}q\prod_t e^{-St}$, 
the new measure can be expressed in terms of the old one as

\begin{eqnarray*}
d\tilde{\mu}_W^\nu
&=&\exp\{-{\textstyle \frac{1}{2\nu}}{\textstyle \int}
[\beta^{-1}q^2((Sp+R)^2+2(Sp+R)\dot{p})\\
&&\phantom{\exp\{}
+\beta q^{-2}(S^2q^2-2Sq\dot{q})]dt\}
\:d\mu_W^\nu\\
&=&\exp\{-{\textstyle \frac{1}{2\nu}}{\textstyle \int}
[\beta^{-1}q^2((Sp+R)^2dt+2(Sp+R)dp)\\
&&\phantom{\exp\{}
+\beta q^{-2}(S^2q^2dt-2Sqdq)]\}
\:d\mu_W^\nu
\end{eqnarray*}

\noindent The first equality is again formal and gains meaning by the last line,
where the stochastic integrals have to be understood in the Stratonovich sense 
as usual. The change of variables has introduced additional 
terms in the exponent of the formal expression
which are at most linear in $\dot{p}$ or $\dot{q}$ respectively. These terms
are not critical, since, in the limit of diverging diffusion constant $\nu$, 
they 
will vanish. This means that the total change of the measure disappears
in the limit. Thus, one can write the path integral with the old measure
$d\mu_W^\nu$ (instead of with the new 
$d\tilde{\mu}_W^\nu$), and one finds

\begin{equation}
\langle p^{\prime\prime}q^{\prime\prime}|e^{-i(RQ+SD)T}|p^\prime q^\prime
\rangle=
\lim_{\nu\rightarrow\infty}K_\nu {\pmb \int}^{p^{\prime\prime}, 
q^{\prime\prime}}_{p^\prime, q^\prime}
\exp\{-i{\textstyle \int}[q\,dp+(Rq+Spq)dt]\}\:d\mu_W^\nu
\end{equation}

\noindent Now, the weak symbol can be read off:

\begin{equation}
h_w(p,q)=Rq+Spq
\end{equation}

The generalization to other Hamiltonians is based on the linearity, 
completeness and irreducibility
of the basic operators $Q$ and $D$ by virtue of which $\lim_{J\rightarrow\infty}
\sum_{j=1}^J\alpha_je^{-i(R_jQ+S_jD)}$ weakly converges to any 
(bounded) operator such as $e^{-i{\cal H}T}$. Thus, 

\begin{eqnarray}
&&\langle p^{\prime\prime}q^{\prime\prime}|e^{-i{\cal H}T}|p^\prime q^\prime
\rangle=\lim_{J\rightarrow\infty}\langle p^{\prime\prime}q^{\prime\prime}|
\sum_{j=1}^J\alpha_je^{-i(R_jQ+S_jD)}|p^\prime q^\prime\rangle\nonumber\\
&=&\lim_{J\rightarrow\infty}\lim_{\nu\rightarrow\infty}K_\nu {\pmb \int}
\exp\{-i{\textstyle \int}q\,dp\}\nonumber\\
&&\phantom{\lim_{J\rightarrow\infty}\lim_{\nu\rightarrow\infty}K_\nu {\pmb \int}}
\times[{\textstyle \sum}_{j=1}^J
\alpha_j \exp\{-i{\textstyle \int}(R_jq+S_jpq)dt\}]
\:d\mu_W^\nu
\end{eqnarray}

\noindent and the question, on which the next steps depend, is: can the
two limits be interchanged? In spite of some effort this question is not 
yet answered. 
Assuming that they can, however, one obtains

\begin{eqnarray}
&&\langle p^{\prime\prime}q^{\prime\prime}|e^{-i{\cal H}T}|p^\prime q^\prime
\rangle\nonumber\\
&=&\lim_{\nu\rightarrow\infty}K_\nu {\pmb \int}
e^{-i{\textstyle \int}q\,dp}[\lim_{J\rightarrow\infty}
{\textstyle \sum}_{j=1}^J
\alpha_j e^{-i{\textstyle \int}(R_jq+S_jpq)dt}]\:d\mu_W^\nu
\end{eqnarray}

\noindent The expression $[\lim_{J\rightarrow\infty}
{\textstyle \sum}_{j=1}^J
\alpha_j e^{-i{\textstyle \int}(R_jq+S_jpq)dt}]=:F[\int qdt, \int pqdt]$ 
is, unfortunately, not of
the form $e^{-i\int h_w(p,q)dt}$ for a general, local Hamiltonian $h_w$, e.g. 
$e^{-i\int q^2 dt}$ with Hamiltonian $q^2$. To produce local
Hamiltonians, one would need distributions $R(t)$ and $S(t)$ instead of the
constants $R$ and $S$. Then, taking e.g. $R(t)=\delta(t-\tau)$, 
one gets a local expression $q(\tau)$ and, by forming functions thereof, local
Hamiltonians. This was proposed in \cite{KlauderWCS}. However, the construction
of distributions from piecewise constant functions 
would require yet another limiting process, and, again, the
the interchangeability of the limits is questionable.



%
%

In the case of a linear Hamiltonian, the weak symbol was shown to be
$h_w(p,q)=Rq+Spq$. This is exactly what one would expect since the 
connection of the basic operators $Q$ and $D$ to classical variables is,
according to the weak correspondence principle,
$q$ and $pq$, respectively. But, the correspondence for a more general
Hamiltonian is not immediately clear and remains to be determined.

\subsubsection{Regularizing approach}\label{WCSRegularizing}

In the second to last subsection it could be seen that a rescaling procedure 
to isolate the 
reproducing kernel Hilbert space can - in the spectral approach - 
only work in the case $\beta=1/2$. 
To proceed to the final goal - a well-defined
weak coherent state path integral - in the case 
$0<\beta<1/2$ \footnote{or in other words,
since the (scalar) curvature of the Lobachevsky half-plane is $R=-2/\beta$: for
a higher negative curvature than $-4$} one has to discuss new methods.

DKP \cite{DKP} were able to use the full power of functional 
analysis and its operator techniques to prove the result in the
case $\beta>1/2$ since the problem was confined to the Hilbert space $L^2(M_+)$.
Instead of trying to work outside of Hilbert space, as was done in the 
previous two
subsections, one can try to bring the problem back into Hilbert space by 
introducing 
a new regularization parameter $\varepsilon$ which will go to zero 
in the very end. Two distinct proposals will be considered. 

The first idea is to reestablish the fiducial vector admissibility condition
$\langle Q^{-1}\rangle=\int_0^\infty x^{-1}|\eta_\beta(x)|^2 dx<\infty$, which 
was important to ensure a resolution of unity and which 
failed in the case $\beta\leq1/2$ since then $\int_0^\infty x^{-1}x^{2\beta-1}
e^{-2\beta x}=\infty$ due to lack of integrability at $0$. The regularization
factor $e^{-\varepsilon/x}$ takes care of that leading to a modified fiducial
vector $\eta_{\beta,\varepsilon}=N_{\beta,\varepsilon}x^{\beta-1/2}e^{-\beta x}
e^{-\varepsilon/x}$ \footnote{Observe that in the case $\beta=1/2$
a weaker regularization, namely $x^\varepsilon$, would be enough, which could be
interpreted as a deformation of the operator $Q$, thus leading to a deformed
version of the affine group.}. The affine group acts on this new fiducial vector
producing the group-defined coherent states $|pq\rangle_\varepsilon$. Their 
overlap is now

\begin{eqnarray*}
&&\langle pq|rs\rangle_\varepsilon\\
&=&\int_0^\infty dx\, (e^{ipx}e^{-\ln(q)(x\partial_x +1/2)}N_{\beta,\varepsilon}
x^{\beta-1/2}e^{-\beta x}e^{-\varepsilon/x})^*\\
&&\times(e^{irx}e^{-\ln(s)(x\partial_x +1/2)}N_{\beta,\varepsilon}
x^{\beta-1/2}e^{-\beta x}e^{-\varepsilon/x})\\
&=&|N_{\beta,\varepsilon}|^2(qs)^{-1/2}\\
&&\times\int_0^\infty dx\, 
e^{-i(p-r)x}({\textstyle\frac{x}{q}})^{\beta-1/2}
({\textstyle\frac{x}{s}})^{\beta-1/2}e^{-\beta(q^{-1}+s^{-1})x}e^{-\varepsilon (q+s)/x}
\end{eqnarray*}

\noindent with 
$|N_{\beta,\varepsilon}|^2=(\int_0^\infty dx\,x^{2\beta-1}e^{-2\beta}
e^{-2\varepsilon/x})^{-1}$

From the second to the third line 
$e^{-\alpha\partial_x}\phi(x)=\phi(x-\alpha)$ was used and thus 
$e^{-\ln(q)(x\partial_x)}\phi(x)=e^{-\ln(q)\partial_u}\phi(e^u)
=\phi(e^{u-\ln(q)})=\phi(e^{\ln(x)-\ln(q)})=\phi(x/q)$.

A new operator $B_\varepsilon$ which annihilates this overlap could be 
computed which would lead to a new operator $A_\varepsilon$.

But, there are reasons not to pursue this course. Retaining the group
definition of the coherent states on the positive side, one is confronted with
the problem that the group will no longer 
operate on an extremal weight vector (and minimum uncertainty state). 
Thus one can not expect any linear analog of the former 
$(Q-1+i\beta^{-1}D)|\eta_\beta\rangle=0$ to hold\footnote{And one 
has to give up analyticity, which previously held apart 
from a factor $(qs)^{-\beta}$.}.
This has serious implications for the new path integral, i.e., the path integral 
with Wiener measure. It was the linear complex polarization that led to a first
order differential operator $B$ and thus to a second order differential operator
$A=\frac{1}{2}\beta B^\dagger B$. It was due to this second order property 
that the Wiener measure emerged and the path integral became a well-defined 
functional integral [see \ref{CSCSPI}]. Whereas it would still be possible
to derive the old, standard path integral, the Wiener measure path integral
can not exist for fiducial vectors which are not extremal weight vectors.

The second possibility is to regularize the overlap directly by a multiplicative
$\varepsilon$-dependent factor which ensures square integrability on the 
Lobachevsky manifold.
From the second to last subsection and appendix (\ref{AppendixA3}) 
(where  
$0\not\in spec(A)$ for $\beta<1/2$) was established) it is known
that for $0<\beta<1/4$ one is forced to
regularize both in $p$ and $q$. So the case 
$1/4<\beta<1/2$, where no $p$-regularization is needed, will be examined 
first.\\

{\bf Case ${\mathbf 1{\boldsymbol{/}}4{\boldsymbol{<}}
{\boldsymbol{\beta}}{\boldsymbol{\leq}}1{\boldsymbol{/}}2}$}\\

Let 

\begin{equation} 
\langle pq|rs\rangle_\varepsilon=N_\varepsilon
\langle pq|rs\rangle e^{-(q+s)\varepsilon}\\
\end{equation}

\noindent where $N_\varepsilon$ is a normalization constant, which is
w.l.o.g. chosen to be real.
The extra factor $e^{-(q+s)\varepsilon}$ goes to $1$ in the limit 
$\varepsilon\rightarrow0$.

The modified overlap lies in the Hilbert space $L^2(M_+)$ and its
norm is computed to be: 

\begin{eqnarray*}
1&\stackrel{!}{=}&N_\varepsilon^*N_\varepsilon
\int_0^\infty dq\,(qs)^{-2\beta}e^{-2q\varepsilon}\int_{-\infty}^\infty dp\,
[(q^{-1}+s^{-1})^2+\beta^{-2}(p-r)^2]^{-2\beta}\\
&=& N_\varepsilon^*N_\varepsilon\int_0^\infty dq\,(qs)^{-2\beta}e^{-2q\varepsilon}
(q^{-1}+s^{-1})^{1-4\beta}\int_{-\infty}^\infty dp\, (1+\beta^{-2}p^2)^{-2\beta}\\
&=& N_\varepsilon^*N_\varepsilon\int_0^\infty dq\,\varepsilon^{-1}(qs)^{-2\beta}
\varepsilon^{2\beta}e^{-2q}(\varepsilon q^{-1}+s^{-1})^{1-4\beta}\cdot I_p\\
&\leq& N_\varepsilon^*N_\varepsilon I_p s^{2\beta-1}\varepsilon^{2\beta-1}
\int_0^\infty dq\, q^{-2\beta}e^{-2q}\\
&=& N_\varepsilon^*N_\varepsilon I_p s^{2\beta-1}\varepsilon^{2\beta-1}
2^{2\beta}\Gamma(1-2\beta)
\end{eqnarray*}

\noindent So $N_\epsilon^*N_\epsilon\geq const\cdot \varepsilon^{1-2\beta}$ 
and this is the correct asymptotic behavior. Instead of the explicit 
expression for the normalization constant one can use $\langle xy|
xy\rangle_\varepsilon=N_\varepsilon e^{-2y\varepsilon}$ (for arbitrary $x\in
\mathbb{R}$, $y\in\mathbb{R}^+$) 
and write $\langle pq|rs\rangle=\lim_{\varepsilon\rightarrow0}
\langle pq|rs\rangle_\varepsilon/\langle xy|xy\rangle_\varepsilon$ in a 
self-consistent way. For a shorter notation set

\begin{equation} 
\langle xy|xy\rangle_\varepsilon=:c_{\beta,\varepsilon}
\end{equation}

One could also use a different regularization like 
$e^{-(q^2+s^2)\varepsilon}$ or something similar. 
But, the expression $e^{-(q+s)\varepsilon}$ is simpler and thus the 
preferred choice.

Certainly, the new states $|pq\rangle_\varepsilon$ 
are not affine group-defined coherent states. The
group law for the affine group in $p$-$q$ notation is
$(p,q)*(p^\prime,q^\prime)=(q^{-1}p^\prime+p,qq^\prime)$. For the new 
$U_\varepsilon(p,q)=U(p,q)e^{-q\varepsilon}$, one would have to add
$q$ and $q^\prime$ in the 
extra factor instead of multiply them. Indeed, they are not group 
defined at all\footnote{For $\beta=1/2$ a 
weaker regularization is enough, namely 
$q^{-\varepsilon}=e^{-\ln(q)\varepsilon}$, leading to group-defined coherent
states with a modified dilation 
operator $D-i\varepsilon$ and thus to a modified affine group.}, 
but it will shortly be shown that they still share certain
properties of coherent states:

1) Continuity in the labels is no problem since joint continuity in 
the overlap is obvious.

2) The old $|pq\rangle$ spanned the space, i.e., any $|\psi\rangle$ 
could be expressed as $\sum_{n=0}^N\alpha_n|p_nq_n\rangle$ or Cauchy limits
thereof. The new $|pq\rangle_\varepsilon$ still
span the space since they are related to the old ones by 
$\sum\alpha_n|p_nq_n\rangle_\varepsilon=\sum\alpha_n N_\varepsilon^{1/2}
e^{-q_n\varepsilon}
|p_nq_n\rangle$ $=:\sum\beta_n|p_nq_n\rangle$.

Therefore, these new $|pq\rangle_\varepsilon$ certainly are Klauder states. 
However, there is ambiguity as to whether they are 
coherent states or, more likely, non-group-defined 
weak coherent states. Since they are only a mathematical aid used in 
intermediate steps this point will not be clarified here.

The new operator $B_\varepsilon$ which annihilates the 
modified kernel is derived by exploiting analyticity (up to a factor):
$[(q^{-1}+s^{-1})+i\beta^{-1}(p-r)]^{-2\beta}=:Y$ is analytic, so 
$\partial_{(q^{-1}-i\beta^{-1}p)}Y=(-\frac{1}{2}q^2\partial_q
+\frac{1}{2}i\beta\partial_p)Y=0$. 
Writing $Y$ as $e^{q\varepsilon}(qs)^{\beta}\langle pq|
rs\rangle_\varepsilon$, one moves $e^{q\varepsilon}(qs)^{\beta}$ to the left of
this operator. Then$e^{q\varepsilon}(qs)^{\beta}$ can be cancelled 
since this expression is everywhere 
nonzero. The result is the new operator

\begin{equation*} 
B_\varepsilon=(\beta^{-1}q\partial_q
+\beta^{-1}q\varepsilon+1-iq^{-1}\partial_p)
\end{equation*}

\noindent for which $B_\varepsilon\langle pq|rs\rangle_\varepsilon=0$.

Define $A_\varepsilon:=\frac{1}{2}\beta B_\varepsilon^\dagger B_\varepsilon$, 
then

\begin{eqnarray}
&&A_\varepsilon\nonumber\\
&=& {\textstyle\frac{1}{2}}\beta(-iq^{-1}\partial_p+1-\beta^{-1}\partial_qq+
\beta^{-1}q\varepsilon)(-iq^{-1}\partial_p+1+
\beta^{-1}q\partial_q+\beta^{-1}q\varepsilon)\nonumber\\
&=& {\textstyle\frac{1}{2}}\{\beta[-iq^{-1}\partial_p+1+\beta^{-1}q
\varepsilon]^2-\beta^{-1}\partial_qq^2\partial_q
-1-2\beta^{-1}q\varepsilon\}
\end{eqnarray}

\noindent Here $1/4<\beta\leq 1/2$ and $p$,$q$ are
the variables of the Lobachevsky half-plane $M_+$ ($q>0$, metric: $g^{pp}=
\beta q^{-2}$, $g^{qq}=\beta^{-1}q^2$, $g^{pq}=g^{qp}=0$, 
Laplace-Beltrami operator: $\beta^{-1}\partial_qq^2\partial_q
+\beta q^{-2}\partial^2_p$).

Setting $-i\partial_p=P_p$, $p=Q_p$ (does not appear), $-i\partial_q=P_q$,
and $q=Q_q$  
and writing the metric with a capital G to reflect its 
operator character ($a,b\in\{p,q\}$), one finds:

\begin{eqnarray*}
A_\varepsilon &=& {\textstyle\frac{1}{2}}
\{(P_a-{\cal A}_a)G^{ab}(P_b-{\cal A}_b)+V(Q_q,Q_p)\}\\
&=& {\textstyle\frac{1}{2}}
\{\beta Q_q^{-2}(P_p-{\cal A}_p)^2+\beta^{-1}(P_q-{\cal A}_q)
Q_q^2(P_q-A_q)+V(Q_q)\}
\end{eqnarray*}

where

\begin{eqnarray*}
{\cal A}_p&=&-Q_q(1+\beta^{-1}Q_q\varepsilon)\\
{\cal A}_q&=&0\\
V(Q_q)&=&-2\beta^{-1}Q_q\varepsilon-1
\end{eqnarray*}

\noindent Effectively the magnetic field has been changed (augmented by an
$\varepsilon$-de\-pen\-dent part), which alters the vector potential 
$\mathbf{\cal A}=({\cal A}_p,{\cal A}_q)$. 
Additionally, the explicit potential $V$ underwent a modification as well.

Next, define

\begin{equation}\label{Kzero}
K^0_{\nu,\varepsilon}(p^{\prime\prime},p^\prime, q^{\prime\prime},q^\prime):=
e^{\nu T/2}\pmb{\int} \exp\{-i{\textstyle\int}
(q+\beta^{-1}q^2\varepsilon)dp
+{\textstyle\frac{\nu}{2}}{\textstyle\int}\beta^{-1}qdt\,\varepsilon\}
\:d\mu_W^\nu
\end{equation}

\noindent with Wiener measure, pinned at $(p^\prime,q^\prime)$ for $t=0$ and
$(p^{\prime\prime},q^{\prime\prime})$ for $t=T$, [see Eq. 
(\ref{WienerMeasure})]: 

\begin{equation}\label{WienerMeasure2}
{\cal N}\exp\{-{\textstyle\frac{1}{2\nu}}
{\textstyle\int} [\beta^{-1}q^2\dot{p}^2+\beta q^{-2}\dot{q}^2]dt\}
{\cal D}p{\cal D}q=d\mu_W^\nu
\end{equation}

\noindent This is the Feynman-Kac-Stratonovich representation of the kernel of 
the operator $\exp\{-\nu TA_\varepsilon\}$ and it is derived in the following
way:

\begin{eqnarray*}
&&\exp\{-\nu TA_\varepsilon\}\delta(p-p^\prime)
\delta(q-q^\prime)|_{p=p^{\prime\prime},q=q^{\prime\prime}}\\
&=&\exp\{-{\textstyle\frac{1}{2}}\nu T
[\beta(-iq^{-1}\partial_p+1+\beta^{-1}q\varepsilon)^2-\beta^{-1}
\partial_qq^2\partial_q-1-2\beta^{-1}q\varepsilon]\}\\
&&\times\int e^{ix(p-p^\prime)-ik(q-q^\prime)}
{\textstyle\frac{dxdk}{(2\pi)^2}}|_{p=p^{\prime\prime},q=q^{\prime\prime}}\\
&=&e^{\nu T/2}\lim_{N\rightarrow\infty}[\exp\{-
{\textstyle\frac{1}{2}}\nu\delta[\beta(-i q^{-1}
\partial_p+1+\beta^{-1}q\varepsilon)^2-2\beta^{-1}q\varepsilon]\}\\
&&\times \exp\{-
{\textstyle\frac{1}{2}}\nu\delta(-\beta^{-1}\partial_qq^2\partial_q)\}]^N
\int e^{ix(p-p^\prime)-ik(q-q^\prime)}
{\textstyle\frac{dxdk}{(2\pi)^2}}|_{p=p^{\prime\prime},q=q^{\prime\prime}}\\
&=&\lim_{N\rightarrow\infty}e^{\nu T/2}
\int \exp\{i\sum x_{l+1/2}(p_{l+1}-p_l)-ik_{l+1/2}(q_{l+1}-q_l)\}\\
&&\times\exp\{-{\textstyle\frac{1}{2}}\nu\delta\sum
[\beta(q_l^{-1}x_{l+1/2}+1+\beta^{-1}q_l\varepsilon)^2-2\beta^{-1}q_l
\varepsilon]\}\\
&&\times \exp\{-{\textstyle\frac{1}{2}}
\nu\delta\sum\beta^{-1}k_{l+1/2}^2q_l^2\}\prod_{l=0}^N
\frac{dk_{l+1/2}dx_{l+1/2}}{(2\pi)^2}\prod_{l=1}^Ndp_ldq_l\\
&=:&e^{\nu T/2}{\cal N}\int \exp\{i\int(x\dot{p}-k\dot{q})dt\}\\
&&\times\exp\{-{\textstyle\frac{1}{2}}\nu\int
\{\beta(q^{-1}x+1+\beta^{-1}q\varepsilon)^2-2\beta^{-1}q\varepsilon+
\beta^{-1}k^2q^2\}dt\}\\
&&\times{\cal D}x{\cal D}k{\cal D}p{\cal D}q\\
&=&e^{\nu T/2}{\cal N}\int\exp\{i\int[(x-q-\beta^{-1}q^2\varepsilon)\dot{p}
-k\dot{q}]\}\\
&&\times\exp\{-{\textstyle\frac{1}{2}}
\nu\int(\beta q^{-2}x^2-2\beta^{-1}q\varepsilon
+\beta^{-1}k^2q^2)dt\}{\cal D}x{\cal D}k{\cal D}p{\cal D}q\\
&=&e^{\nu T/2}{\cal N}\int\exp\{-i\int(q+\beta^{-1}q^2
\varepsilon)\dot{p}dt\}\\
&&\times\exp\{{\textstyle\frac{1}{2}}\nu\int2\beta^{-1}q\varepsilon dt\}
\exp\{-{\textstyle\frac{1}{2\nu}}
\int[\beta^{-1}q^2\dot{p}^2+\beta q^{-2}\dot{q}^2]dt\}
{\cal D}p{\cal D}q
\end{eqnarray*}

\noindent with $N=T/\delta$. The Lie-Trotter product formula was used to go from
the second to the third equality. The indices $l+1/2$ and $l$ indicate that
the lattice points do not coincide for $x$,$p$ or $q$,$k$ respectively.
(This would violate the Heisenberg uncertainty principle.) For the
endpoints, the definitions
$p_0:=p^\prime$, $p_{N+1}:=p^{\prime\prime}$, $q_0:=q^\prime$ and
$q_{N+1}:=q^{\prime\prime}$ were made. Note that
$\exp\{\frac{1}{2}\nu\delta(-\beta^{-1}\partial_qq^2\partial_q)\}
\exp\{-ik(q-q^\prime)\}\approx \exp\{-\frac{1}{2}\nu\delta\beta^{-1}k^2q^2\}
\exp\{-ik(q-q^\prime)\}$ only to first order in $\delta$, but that was good 
enough for the path integral. From the second to 
last to the bottom line, $x$ was substituted by 
$x+q-\beta^{-1}q^2\varepsilon$ and the $x$- and $p$-integrations were 
carried out.
 
This formal expression becomes well-defined when understood in the sense of
Eq. (\ref{Kzero}) where the above Wiener measure. Also, the integrals over
the Brownian bridges have to be taken in the Stratonovich interpretation of
stochastic integrals.

The aim is to show that in the limit of diverging diffusion constant followed
by the limit $\varepsilon\rightarrow0$ the kernel 
$c_{\beta,\varepsilon}^{-1}K^0_{\nu,\varepsilon}$
reduces to the (weak coherent state) 
overlap $\langle p^{\prime\prime}q^{\prime\prime}|
p^\prime q^\prime\rangle$ for $1/4<\beta\leq1/2$.

This is similar to what was done in DKP \cite{DKP}. If one were to follow 
the proof of DKP [see \ref{CSCSPI}]
one would have to show these three 
properties:\\

\noindent 1)$A_\varepsilon$ is a non-negative, self-adjoint operator\\
2)$A_\varepsilon$ has an isolated eigenvalue 0\\
3)Pointwise convergence of the kernels 
$c_{\beta,\varepsilon}^{-1}K^0_{\nu,\varepsilon}$\\

To copy the proof is not possible. In the ``$A$-world" point 1) could be shown
by DKP using Kato-Rellich's theorem about stability of 
self-adjointness\footnote{If $T$ is self-adjoint, $V$ symmetric and $T$-bounded 
with
bound smaller than 1, then $T+V$ is self-adjoint.}. But in the 
``$A_\varepsilon$-world" the operator $A_\varepsilon$ is not 
$-\Delta_{LB}$-bounded. For point
2) and 3) DKP could exploit the connection to the Morse-operator. It is the 
author's view that the reason for the 
existence of this connection between the original 2-dimensional problem on
the Lobachevsky half-plane and the 1-dimensional Morse operator lies in the 
high symmetry of the homogeneous Lobachevsky manifold (constant negative 
curvature). This connection no longer exists with the 
$\varepsilon$-modified operator $A_\varepsilon$. 

Since other ways for the proof have to be found anyway, it makes sense to 
reexamine the points 1)-3): 

Point 1) certainly is a necessary condition
since it ensures non-negativity of $A_\varepsilon$\footnote{See the 
discussion earlier in this paragraph about operators of the 
form $B^\dagger B$ which were
non-negative only by requirement of certain boundary conditions.} 
and a real spectrum. It is
not obvious why DKP insisted on an isolated eigenvalue $0$ unless it was to
match the requirements for the application of some theorem which was not
mentioned in the article. 
Of course there must be a discrete ground state with
eigenvalue $0$ (which exists by construction), but in principle the continuous
spectrum could start at $0$ as well.
In the limit of diverging diffusion constant, the continuous part would 
disappear.
(This was the whole trouble in the preceding attempt to isolate the 
reproducing kernel for $\beta=1/2$ by a rescaling procedure.)

Although it is not clear why a gap should be necessary there is every 
reason to believe that a gap exists. From a physical point of view, the 
coexistence of a discrete ground state and a continuous state of the same
energy would require some conservation law to prevent the discrete
state from decaying into the continuum. However, this is invariably associated 
with
a symmetry property of the underlying phase space manifold. Even in the
case of the old operator $A$ associated with the Lobachevsky half-plane, a
space of high symmetry, there was no sign that anything like that 
could happen (e.g.
in the case $\beta=1/2$ the spectrum of $A$ started at $0$ but was purely 
continuous; the discrete ground state had disappeared). 

So why should one
believe that a symmetry which could result in a conservation law could
exist in the $A_\varepsilon$-world? One would expect that there is less 
symmetry than before! From a mathematical point of view one can look at
the function $f(\varepsilon)$ the values of which give the starting
point of the continuous spectrum. Certainly, $f(0)=(\beta-1/2)^2$ which is taken
from DKP \cite{DKP}. ``No gap" would mean, in this language, that this 
function $f$ must be discontinuous at $0$ and identically $0$ in an open
neighborhood of $0$.
It is very hard to imagine how such a functional behavior should arise. 
If, by any chance, this functional behavior turned up, then it
was connected with the special choice of the regularization.
But, the 
regularization is not fixed (the form chosen just seems to be the
most simple one) and could e.g. be $\exp\{-q^3\varepsilon^2\}$. This would
lead to a different $A_\varepsilon$. Also, the function $f$ would be different.
That the new function $f$ should display the same
functional behavior as the former one is even more unlikely if not impossible. 

To prove the third point DKP used ``heavy artillery" (operator estimates based
on the connection to the Morse operator) to ``shoot a rather small bird", namely
continuity of the limiting function - the reproducing kernel or coherent state
overlap\footnote{There is, of course, a reason why this intricate approach 
was used. The functional form of $\langle
pq|e^{-iT{\cal H}}|rs\rangle$ is not known for general Hamiltonian. But, with
the operator technique, the pointwise convergence could be shown in general.}.

Since the connection to the $1$-dimensional Morse operator is not 
available anyway, the natural (and much easier) way is to prove this continuity 
directly. But, apart from the continuous ``factor" $\exp\{-(q+s)\varepsilon\}
(qs)^{-\beta}$, the function is analytic in $q^{-1}+i\beta^{-1}p$ and, hence,
continuity of the whole function follows.

Summarizing the above, only the self-adjointness of $A_\varepsilon$ has to be
shown to establish the main result.

The aim is to show that the deficiency index equation 
$(A_\varepsilon^\dagger\psi)(p,q)
=\pm i\psi(p,q)$ has no solution for all 
$\psi\in D(A_\varepsilon^\dagger)$. This is 
equivalent to the statement ``$A_\varepsilon$ is 
essentially self-adjoint" \cite{Blank}. 
To see this a change of representation is made:

\begin{equation}
\psi(p,q)=(2\pi)^{-1/2}\int_{-\infty}^\infty e^{ipu}\tilde{\psi}(u,q)du
\end{equation}

In the new representation the operator $A_\varepsilon$ is given by
$(A_\varepsilon\phi)(u,q)=\frac{1}{2}\{\beta(uq^{-1}+1+\beta^{-1}q\varepsilon)^2
-1-2\beta^{-1}q\varepsilon-\beta^{-1}\partial_qq^2\partial_q\}\phi(u,q)$,
and the deficiency index equation becomes

\begin{equation}
\{\beta(uq^{-1}+1+\beta^{-1}q\varepsilon)^2
-1-2\beta^{-1}q\varepsilon-\beta^{-1}\partial_qq^2\partial_q\}\phi(u,q)
=\pm2i\phi(u,q)
\end{equation}

\noindent where $\phi\in D(A_\varepsilon)\Leftrightarrow[\phi\in L^2(
\mathbb{R}\times\mathbb{R}^+)\mbox{and}(A_\varepsilon\phi)\in L^2(
\mathbb{R}\times\mathbb{R}^+)]$.

In the following, $u$ is written as a parameter [$\phi_u(q)$] to indicate 
that only
$q$ is being dealt with. Multiplying the deficiency index equation 
from the left by 
$\phi_u^*(q)$, and integrating over $q$ with integration by parts,
one is led to

\begin{eqnarray}\label{IntegratedDIE}
&&\int \phi_u^*(q)A_\varepsilon\phi_u(q)dq
=\pm2i\int\phi_u^*(q)\phi_u(q)dq\nonumber\\
&\Leftrightarrow&\int \phi_u^*(q)\{\beta(uq^{-1}+1+\beta^{-1}q\varepsilon)^2
-1-2\beta^{-1}q\varepsilon-\beta^{-1}\partial_qq^2\partial_q\}\phi_u(q)dq
\nonumber\\
&&=\pm2i\int\phi_u^*(q)\phi_u(q)dq\nonumber\\
&\Leftrightarrow&\int[\beta(uq^{-1}+1+\beta^{-1}\varepsilon q)^2
-1-2\beta^{-1}\varepsilon q]|\phi_u|^2dq\nonumber\\
&&+\int\beta^{-1}|\partial_q(q\phi_u)|^2dq
-\beta^{-1}[q|\phi_u|^2]^\infty_0+
\beta^{-1}[q^2\phi_u^*\partial_1\phi_u]^\infty_0\nonumber\\
&&=\pm2i\int|\phi_u|^2dq
\end{eqnarray}

In the last equivalence the term on the right hand side is purely 
imaginary; all the terms on the left hand side of the equality sign
are purely real, except maybe for the term 
$\beta^{-1}[q^2\phi_u^*\partial_1\phi_u]^\infty_0$. The goal is to show that
this term is either purely real or vanishing. 
To exhibit this, one has to learn about the
asymptotic behavior of $\psi_u(q)$ for extremely large or small values of $q$.
 
The Wentzel-Kramers-Brioullin approximation (WKB) can tell about the asymptotic
behavior of one-dimensional functions. 
But, WKB approximately
solves $-\partial_x^2\psi+V(x)-E=0$ yielding the answer 
$\psi(x)\approx[V(x)-E]^{-1/4}\exp\{\pm\int^x\sqrt{V(x^\prime)-E}dx^\prime\}$
valid in the region to the right of the classically allowed region. 
The wave function with the positive sign
in the exponent is being discarded as unphysical as it is non-square integrable.
The
wave solution in the classically allowed region will not be needed. The solution
to the left of the classically allowed region is 
$\psi(x)\approx[V(x)-E]^{-1/4}\exp\{\pm\int_x\sqrt{V(x^\prime)-E}dx^\prime\}$
with the positive solution again ignored.

But the problem at hand is not yet in a form to be treated with the standard
WKB due to the kinetic energy term $-\beta^{-1}\partial_qq^2\partial_q$ which 
stems from the curved Lobachevsky half-plane. One could rather easily develop 
a new WKB suitable for this problem, but a change of variables,
$q\rightarrow a^{-1}$, will bring the problem into standard form. 
The kinetic energy
becomes $-\beta^{-1}a^2\partial_a^2$ and since $a>0$ one can divide by
$\beta^{-1}a^2$ to obtain the new deficiency equation

\begin{eqnarray}
&&[\varepsilon^2a^{-4}+\varepsilon(\beta-1)a^{-3}
+\beta(2\varepsilon u+\beta-1)a^{-2}+2\beta^2ua^{-1}+\beta^2u^2]
\phi_u(a)\nonumber\\
&&-\partial_a^2\phi_u(a)=\pm2ia^{-2}\phi_u(a)
\end{eqnarray}

\noindent For large $a$ ($\hat{=}$ small $q$), this becomes

\begin{equation}
-\partial_a^2\phi_u(a)+\beta^2u^2\phi_u(a)=0
\end{equation}

\noindent with the WKB solution

\begin{equation}
\phi_u(a)\propto\exp\{-\beta|u|a\}
\end{equation}

\noindent For small $a$ ($\hat{=}$ large $q$) the deficiency index equation
reads

\begin{equation}
-\partial_a^2\phi_u(a)+\varepsilon^2a^{-4}\phi_u(a)=0
\end{equation}

\noindent with WKB solution

\begin{equation}
\phi_u(a)\propto a\exp\{-\varepsilon a^{-1}\}
\end{equation}

Note how terms introduced by $\varepsilon$ dominate in the small $a$,
i.e., large $q$ region, just as one should expect since the regularization of
the overlap was aimed at large $q$. At this point it is already clear that the
functions $\phi_u$ will asymptotically be real, or, to be more precise, they
vanish asymptotically. Writing everything in the
$q$-notation again one finds

\begin{eqnarray}
&&\phi_u(q)\propto\exp\{\beta|u|q^{-1}\}\nonumber\\
&&\phi_u(q)\propto q^{-1}\exp\{-\varepsilon q\}
\end{eqnarray}

\noindent for the small $q$ and large $q$ asymptotic behavior, respectively.
One immediately sees that

\begin{eqnarray}
&&q^2\phi_u^*(q)\partial_q\phi_u(q)\propto \exp\{-2\beta|u|q^{-1}\}
\stackrel{q\rightarrow0}{\longrightarrow}0\nonumber\\
&&q^2\phi_u^*(q)\partial_q\phi_u(q)\propto \exp\{-2\varepsilon q\}
\stackrel{q\rightarrow\infty}{\longrightarrow}0
\end{eqnarray}

\noindent showing that the critical term in Eq. (\ref{IntegratedDIE}) 
indeed vanishes. Thus, the deficiency index equation has no solution. Hence,
$A_\varepsilon$ is essentially self-adjoint. 
From now on, and in a not perfectly pure but simpler notation, understand
$A_\varepsilon$ as the closure of the operator previously 
called $A_\varepsilon$.
The ``new" $A_\varepsilon$ is self-adjoint. This, together with the fact that 
$A_\varepsilon$ is of the form $B_\varepsilon^\dagger B_\varepsilon$, ensures
at the same time that $A_\varepsilon$ is non-negative.

Having established these results, one can conclude in the same way DKP did
that, for the parameter range 
$1/4<\beta\leq1/2$ and for arbitrary $r\in\mathbb{R}$, $s\in\mathbb{R}^+$,

\begin{eqnarray}
&&\lim_{\varepsilon\rightarrow 0}\lim_{\nu\rightarrow\infty}
c_{\beta,\varepsilon}^{-1}
K^0_{\nu\varepsilon}(p^{\prime\prime},q^{\prime\prime},p^\prime,q^\prime,T)
\nonumber\\
&=&\lim_{\varepsilon\rightarrow 0}\lim_{\nu\rightarrow\infty}
c_{\beta,\varepsilon}^{-1}
\exp\{-\nu A_\varepsilon T\}\delta(p-p^\prime)
\delta(q-q^\prime)|_{p=p^{\prime\prime}, q=q^{\prime\prime}}\nonumber\\
&=&\lim_{\varepsilon\rightarrow 0}\lim_{\nu\rightarrow\infty}
c_{\beta,\varepsilon}^{-1}
\exp\{\nu T/2\}\pmb{\int} \exp\{-i{\textstyle\int}(q+\beta^{-1}q^2
\varepsilon)dp\nonumber\\
&&\phantom{\lim_{\varepsilon\rightarrow 0}\lim_{\nu\rightarrow\infty}
c_{\beta,\varepsilon}^{-1}
\exp\{\nu T/2\}\pmb{\int} \exp\{}
+\nu{\textstyle\int}\beta^{-1}q\varepsilon dt\}
\:d\mu_W^\nu\nonumber\\
&=&\lim_{\varepsilon\rightarrow 0}
c_{\beta,\varepsilon}^{-1}\langle p^{\prime\prime}q^{\prime\prime}|
p^\prime q^\prime\rangle_\varepsilon=\langle p^{\prime\prime}q^{\prime\prime}|
p^\prime q^\prime\rangle
\end{eqnarray}

\noindent The stochastic processes involved are still Brownian bridges, and, 
when the stochastic integrals are interpreted in the Stratonovich sense,
canonical (coordinate) transformations can be made in the same way as before. 
Thus, the geometric nature of the quantization is
untouched. It is just the mathematical structure which is not as elegant
as in the case $\beta>1/2$.\\

{\bf Case ${\mathbf 0{\boldsymbol{<}}{\boldsymbol{\beta}}
{\boldsymbol{\leq}}1{\boldsymbol{/}}4}$}\\

For a parameter $\beta\leq1/4$, a regularization for large $q$ is not enough. It 
turns out that an additional $p$-regularization will even make a regularization
for small
$q$ necessary (otherwise the overlap would be square integrable, but not in the
domain of $A_\varepsilon$). 

In the present case, let

\begin{equation}
\langle pq|rs\rangle_\varepsilon:=N_\varepsilon
\langle pq|rs\rangle e^{-(q+s)\varepsilon
-(q^{-1}+s^{-1})\varepsilon-(p^2+r^2)\varepsilon}
\end{equation}

\noindent where $\langle pq|rs\rangle=(qs)^{-\beta}2^{-2\beta}[(q^{-1}+s^{-1})
+i\beta^{-1}(p-r)]^{-2\beta}$ is the weak coherent state overlap, which is
analytic in the complex variable $z:=q^{-1}+i\beta^{-1}p$ 
apart from the factor $(qs)^{-\beta}$. One can write the analytic part
(previously called $Y$ - see above) as $e^{(q+s)\varepsilon+(q^{-1}+s^{-1})
\varepsilon+(p^2+q^2)\varepsilon}(qs)^\beta\langle pq|rs\rangle_\varepsilon$, 
and let the differential operator $\partial_{q^{-1}-i\beta^{-1}p}
=\frac{1}{2}(-q^2\partial_q+i\beta\partial_p)$ act on this expression.
Using $\partial_{z^*}f=0$ for an
analytic function, this results in the new operator

\begin{equation*}
B_\varepsilon=\beta^{-1}q\partial_q+1+\beta^{-1}q\varepsilon
+\beta^{-1}q^{-1}\varepsilon-2ipq^{-1}\varepsilon-iq^{-1}\partial_p
\end{equation*}

\noindent for which $B_\varepsilon\langle pq|rs\rangle_\varepsilon=0$. 
As before, define
$A_\varepsilon:=\frac{1}{2}\beta B^\dagger_\varepsilon B_\varepsilon$, then

\begin{eqnarray}
A_\varepsilon&=&{\textstyle\frac{1}{2}}
\{\beta(-iq^{-1}\partial_p+1+\beta^{-1}q\varepsilon
\beta^{-1}q^{-1}\varepsilon)^2-2\beta q^{-2}\varepsilon\nonumber\\
&&+4ip\partial_q\varepsilon+4\beta p^2q^{-2}
\varepsilon^2-\beta^{-1}\partial_1q^2\partial_q-1-2\beta^{-1}q\varepsilon\}
\end{eqnarray}

The Feynman-Kac-Stratonovich representation of the kernel of the 
operator
$\exp\{-\nu TA_\varepsilon\}$ is derived in much the same way as before and with
the same conventions for notation it reads

{\allowdisplaybreaks
\begin{eqnarray*}
&&\exp\{-\nu TA_\varepsilon\}\delta(p-p^\prime)
\delta(q-q^\prime)|_{p=p^{\prime\prime},q=q^{\prime\prime}}\\
&=&\exp\{-{\textstyle\frac{\nu T}{2}}
[\beta(-iq^{-1}\partial_p+1+\beta^{-1}q\varepsilon
\beta^{-1}q^{-1}\varepsilon)^2-2\beta q^{-2}\varepsilon\\*
&&+4ip\partial_q\varepsilon+4\beta p^2q^{-2}
\varepsilon^2-\beta^{-1}\partial_1q^2\partial_q-1-2\beta^{-1}q\varepsilon]\}\\*
&&\times\int \exp\{ix(p-p^\prime)-ik(q-q^\prime)\}
\frac{dxdk}{(2\pi)^2}|_{p=p^{\prime\prime},q=q^{\prime\prime}}\\
&=&e^{\nu T/2}\lim_{N\rightarrow\infty}[\exp\{-\nu\delta/2\\*
&&[\beta(-i q^{-1}
\partial_p+1+\beta^{-1}q\varepsilon+\beta^{-1}q^{-1}\varepsilon)^2
-2\beta q^{-2}\varepsilon-2\beta^{-1}q\varepsilon]\}\\*
&&\times \exp\{-\nu\delta/2(-\beta^{-1}\partial_qq^2\partial_q)\}
\exp\{-\nu\delta/2\cdot4ip\partial_q\varepsilon\}\\*
&&\exp\{-\nu\delta/2\cdot4\beta p^2q^{-2}\varepsilon^2\}]^N\\*
&&\int \exp\{ix(p-p^\prime)-ik(q-q^\prime)\}
\frac{dxdk}{(2\pi)^2}|_{p=p^{\prime\prime},q=q^{\prime\prime}}\\
&=&\lim_{N\rightarrow\infty}e^{\nu T/2}
\int \exp\{i\sum x_{l+1/2}(p_{l+1}-p_l)-ik_{l+1/2}(q_{l+1}-q_l)\}\\*
&&\times\exp\{-\nu\delta/2\sum
[\beta(q_l^{-1}x_{l+1/2}+1+\beta^{-1}q_l\varepsilon
+\beta^{-1}q_l^{-1}\varepsilon)^2\\*
&&-2\beta q_l^{-2}\varepsilon-2\beta^{-1}q_l
\varepsilon]\}\\*
&&\times \exp\{-\nu\delta/2\sum\beta^{-1}k_{l+1/2}^2q_l^2\}
\exp\{-\nu\delta/2\sum 4ip_l(-ik_{l+1/2})\varepsilon\}\\*
&&\exp\{-\nu\delta/2\sum 4\beta p_l^2q_l^{-2}\varepsilon^2\}
\prod_{l=0}^N
\frac{dk_{l+1/2}dx_{l+1/2}}{(2\pi)^2}\prod_{l=1}^Ndp_ldq_l\\
&=:&e^{\nu T/2}{\cal N}\int \exp\{i\int(x\dot{p}-k\dot{q})dt\}\\*
&&\times\exp\{-\nu/2\int
\{\beta(q^{-1}x+1+\beta^{-1}q\varepsilon+\beta^{-1}q^{-1}\varepsilon)^2
-2\beta q^{-2}\varepsilon-2\beta^{-1}q\varepsilon\\*
&&+\beta^{-1}k^2q^2 +4pk\varepsilon+4\beta p^2q^{-2}\varepsilon^2\}dt\}
{\cal D}x{\cal D}k{\cal D}p{\cal D}q\\
&=&e^{\nu T/2}{\cal N}\int\exp\{i\int[(x-q-\beta^{-1}q^2\varepsilon
-\beta^{-1}\varepsilon)\dot{p}
-k\dot{q}]dt\}\\*
&&\times\exp\{-\nu/2\int(\beta q^{-2}x^2
-2\beta q^{-2}\varepsilon-2\beta^{-1}q\varepsilon
+\beta^{-1}k^2q^2
+4pk\varepsilon\\*
&&+4\beta p^2q^{-2}\varepsilon^2)dt\}
{\cal D}x{\cal D}k{\cal D}p{\cal D}q\\
&=&e^{\nu T/2}{\cal N}\int\exp\{-i\int(q+\beta^{-1}\varepsilon
+\beta^{-1}q^2\varepsilon)\dot{p}dt+i\int2\beta pq^{-2}\dot{q}dt\}\\*
&&\times\exp\{\nu/2\int(2\beta^{-1}q\varepsilon
+2\beta q^{-2}\varepsilon-4\beta p^2q^{-2}\varepsilon^2) dt\}\\*
&&\exp\{-1/(2\nu)\int[\beta^{-1}q^2\dot{p}^2+\beta q^{-2}\dot{q}^2]dt\}
{\cal D}p{\cal D}q
\end{eqnarray*}
}

Introducing the Wiener measure, as above [see Eq. (\ref{WienerMeasure2})],
and performing a partial integration on $2\beta\varepsilon pq^{-2}dq
=-2\beta\varepsilon pd(q^{-1})=-2\beta\varepsilon pq^{-1}|^{(p^{\prime\prime},
q^{\prime\prime})}_{(p^\prime,q^\prime)}+2\beta\varepsilon q^{-1}dp$, the 
final result is

\begin{eqnarray}\label{Kzero2}
&&K^0_{\nu,\varepsilon}(p^{\prime\prime},p^\prime, q^{\prime\prime},q^\prime)
\nonumber\\
&:=&
\exp\{\nu T/2\}\exp\{-i\beta^{-1}\varepsilon(p^{\prime\prime}-p^\prime)
-2i\beta\varepsilon(p^{\prime\prime}q^{\prime\prime-1}-p^\prime q^{\prime-1})\}
\nonumber\\
&&\times\pmb{\int} \exp\{-i{\textstyle\int}
(q+\beta^{-1}q^2\varepsilon-2\beta q^{-1}\varepsilon)dp\nonumber\\
&&+\nu{\textstyle\int}[\beta q^{-2}\varepsilon+\beta^{-1}q\varepsilon
-2\beta p^2q^{-2}\varepsilon^2] dt\}\:d\mu_W^\nu
\end{eqnarray}

As before, the aim is to establish $\lim_{\varepsilon\rightarrow0}
\lim_{\nu\rightarrow\infty}c_{\beta,\varepsilon}^{-1}
K^0_{\nu,\varepsilon}(p^{\prime\prime},p^\prime, q^{\prime\prime},q^\prime)
=\langle p^{\prime\prime}q^{\prime\prime}|p^\prime q^\prime\rangle$. The 
questions about ``gap/no gap" and about pointwise convergence can be 
answered in the same way as in the previous section. 
The self-adjointness could previously be
proved since the ``old" $A_\varepsilon$ did not contain $p$ explicitly. 
Instead of trying to solve
the deficiency index equation for the ``new" $A_\varepsilon$ one can avoid this
question altogether.

If $A_\varepsilon$ is self-adjoint there is nothing 
to show. Assume $A_\varepsilon$ is not self-adjoint. The (sesquilinear) form
$s_\varepsilon(x,y):=\langle x|A_\varepsilon y\rangle$, generated by 
$A_\varepsilon$, is closable since $A_\varepsilon$ is symmetric and 
bounded below
\cite{Blank}. There is a bijection between the set of all (densely defined) 
closed, below-bounded forms and the set of all self-adjoint, below-bounded 
operators. Let $\bar{s}_\varepsilon$ be the closure of the form generated by 
$A_\varepsilon$ and $A_{\bar{s}_\varepsilon}$ be the self-adjoint operator
associated with $\bar{s}_\varepsilon$. Then $A_{\bar{s}_\varepsilon}$ 
preserves the lower
bound and is called the Friedrichs' extension of the operator $A_\varepsilon$. 
It is the unique extension fulfilling $D(A_{\bar{s}_\varepsilon})\subset 
D(\bar{s}_\varepsilon)$ \cite{Blank}.

In a slight abuse of notation, $A_{\bar{s}_\varepsilon}$ will be written as
$A_\varepsilon$. So from now on $A_\varepsilon$ denotes 
the Friedrichs' extension (which is
trivial in the case that $A_\varepsilon$ is already self-adjoint). Then
it is clear that $A_\varepsilon$ is non negative.

With this in mind, one can establish the desired result without further work. 
The phase factors in Eq. (\ref{Kzero2}) are $\nu$-independent, so
they come outside of the $\nu$-limit and become unity.

\begin{eqnarray}\label{Kzero3}
&&\lim_{\varepsilon\rightarrow0}\lim_{\nu\rightarrow\infty}
c_{\beta,\varepsilon}^{-1}
K^0_{\nu,\varepsilon}(p^{\prime\prime},p^\prime, q^{\prime\prime},q^\prime)
\nonumber\\
&:=&\lim_{\varepsilon\rightarrow0}\lim_{\nu\rightarrow\infty}
c_{\beta,\varepsilon}^{-1}
\exp\{\nu T/2\}\pmb{\int} \exp\{-i{\textstyle\int}
(q+\beta^{-1}q^2\varepsilon-2\beta q^{-1}\varepsilon)dp\nonumber\\
&&\phantom{\lim_{\varepsilon\rightarrow0}\lim_{\nu\rightarrow\infty}
c_{\beta,\varepsilon}^{-1}
\exp\{\nu T/2\}}
+\nu{\textstyle\int}[\beta q^{-2}\varepsilon+\beta^{-1}q\varepsilon
-2\beta p^2q^{-2}\varepsilon^2] dt\}\:d\mu_W^\nu \nonumber\\
&=&\langle p^{\prime\prime}q^{\prime\prime}|p^\prime q^\prime\rangle
\end{eqnarray}

\noindent This is a valid path integral representation for all
$0<\beta\leq1/4$.

\subsubsection{Introducing dynamics}\label{WCSRegularizingDynamics}

Dynamics are introduced in the same way as for the spectral approach. For a 
linear Hamiltonian ${\cal H}=RQ+SD$, the problem is already solved since
it reduces to an overlap with modified ending points. What remains to do is
to write down the path integral. This is straightforward since everything
stated previously concerning the measure, etc., remains valid, and the formula
for $1/4<\beta\leq1/2$ is

\begin{eqnarray}
&& \langle p^{\prime\prime}q^{\prime\prime}|e^{-i(RQ+SD)T}|p^\prime q^\prime
\rangle
=\langle p^{\prime\prime}e^{ST}+{\textstyle\frac{R}{S}}
(e^{ST}-1),q^{\prime\prime}e^{-ST}|p^\prime q^\prime\rangle\nonumber\\
&=&\lim_{\varepsilon\rightarrow0}\lim_{\nu\rightarrow\infty}
c_{\beta,\varepsilon}^{-1} e^{\nu T/2}
{\pmb \int}_{p^\prime,q^\prime}^{p^{\prime\prime}e^{ST}+
{\textstyle\frac{R}{S}}(e^{ST}-1),q^{\prime\prime}e^{-ST}}\!\!\!\!\!\exp\{-i
{\textstyle\int}(q+\beta^{-1}q^2\varepsilon)dp\nonumber\\
&&\hspace{7.5cm}
+\nu{\textstyle\int}
\beta^{-1}q\varepsilon dt\}\:d\mu_W^\nu\nonumber\\
&=&\lim_{\varepsilon\rightarrow0}\lim_{\nu\rightarrow\infty}
c_{\beta,\varepsilon}^{-1} e^{\nu T/2}
{\pmb \int}_{p^\prime,q^\prime}^{p^{\prime\prime},q^{\prime\prime}}
\exp\{-i
{\textstyle\int}(qe^{-St}+\beta^{-1}q^2e^{-2St}\varepsilon)\nonumber\\
&&\phantom{\lim_{\varepsilon\rightarrow0}\lim_{\nu\rightarrow\infty}
c_{\beta,\varepsilon}^{-1} e^{\nu T/2}}
\times d[pe^{St}+
{\textstyle\frac{R}{S}}(e^{St}-1)]
+\nu{\textstyle\int}
\beta^{-1}qe^{-St}\varepsilon dt\}\:d\mu_W^\nu\nonumber\\
&=&\lim_{\varepsilon\rightarrow0}\lim_{\nu\rightarrow\infty}
c_{\beta,\varepsilon}^{-1} e^{\nu T/2}
{\pmb \int}_{p^\prime,q^\prime}^{p^{\prime\prime},q^{\prime\prime}}
\exp\{-i[{\textstyle\int}(q+\beta^{-1}q^2e^{-St}\varepsilon)dp\nonumber\\
&&\hspace{0.5cm}+{\textstyle\int}(q+\beta^{-1}q^2e^{-St}\varepsilon)(Sp+R)dt]
+\nu{\textstyle\int}
\beta^{-1}qe^{-St}\varepsilon dt\}\:d\mu_W^\nu
\end{eqnarray}

\noindent The complexity of this expression can be hidden by introducing
the variable $q_\varepsilon:=q+\beta^{-1}q^2e^{-St}\varepsilon$ 
and the measure
$d\mu_W^{\nu,\varepsilon}:=\exp\{\nu{\textstyle\int}
\beta^{-1}qe^{-St}\varepsilon dt\}\:d\mu_W^\nu$. The new measure is 
equivalent to the
old Wiener measure since the factor $\exp\{\nu{\textstyle\int}
\beta^{-1}qe^{-St}\varepsilon dt\}$ is a Radon-Nykodym derivative. 
Then, the 
formula resembles the path integral for coherent states:

\begin{equation}
\lim_{\varepsilon\rightarrow0}\lim_{\nu\rightarrow\infty}
c_{\beta,\varepsilon}^{-1} e^{\nu T/2}
{\pmb \int}_{p^\prime,q^\prime}^{p^{\prime\prime},q^{\prime\prime}}
\exp\{-i[{\textstyle\int}q_\varepsilon dp
+(Spq_\varepsilon+Rq_\varepsilon) dt]
\:d\mu_W^{\nu,\varepsilon}
\end{equation}

\noindent The $\varepsilon$-modified Hamiltonian is given by the weak modified 
symbol
$h_{w,\varepsilon}:=Rq_\varepsilon+Spq_\varepsilon$.

The same procedure for $0<\beta\leq1/4$ leads to:

\begin{eqnarray}
&& \langle p^{\prime\prime}q^{\prime\prime}|e^{-i(RQ+SD)T}|p^\prime q^\prime
\rangle\nonumber\\
&=&\lim_{\varepsilon\rightarrow0}\lim_{\nu\rightarrow\infty}
c_{\beta,\varepsilon}^{-1} e^{\nu T/2}
{\pmb \int}_{p^\prime,q^\prime}^{p^{\prime\prime},q^{\prime\prime}}
\exp\{-i[{\textstyle\int}(q+\beta^{-1}q^2e^{-St}\varepsilon
-2\beta q^{-1}e^{2St})dp\nonumber\\
&&\phantom{\lim_{\varepsilon\rightarrow0}\lim_{\nu\rightarrow\infty}
c_{\beta,\varepsilon}^{-1}}
+{\textstyle\int}(q+\beta^{-1}q^2e^{-St}\varepsilon-2\beta q^{-1}e^{2St})
(Sp+R)dt]\nonumber\\
&&\phantom{\lim_{\varepsilon\rightarrow0}\lim_{\nu\rightarrow\infty}
c_{\beta,\varepsilon}^{-1}}
+\nu{\textstyle\int}
(\beta^{-1}qe^{-St}\varepsilon+\beta q^{-2}e^{2St}\varepsilon
-2\beta p^2q^{-2}e^{4St}\varepsilon^2) dt\}
\:d\mu_W^\nu\nonumber\\
&=&\lim_{\varepsilon\rightarrow0}\lim_{\nu\rightarrow\infty}
c_{\beta,\varepsilon}^{-1} e^{\nu T/2}
{\pmb \int}_{p^\prime,q^\prime}^{p^{\prime\prime},q^{\prime\prime}}
\exp\{-i[{\textstyle\int}\tilde{q}_\varepsilon dp+{\textstyle\int}
(Sp\tilde{q}_\varepsilon+R\tilde{q}_\varepsilon) dt]
d\tilde{\mu}_W^{\nu,\varepsilon}
\end{eqnarray}

\noindent The variable $\tilde{q}_\varepsilon:=
(q+\beta^{-1}q^2e^{-St}\varepsilon-2\beta q^{-1}e^{2St})$ and
the Ra\-don-Ny\-ko\-dym measure 
$d\tilde{\mu}_W^{\nu,\varepsilon}:=
\exp\{\nu{\textstyle\int}
(\beta^{-1}qe^{-St}\varepsilon+\beta q^{-2}e^{2St}\varepsilon
-2\beta p^2q^{-2}e^{4St}\varepsilon^2) dt\}
\:d\mu_W^\nu$ were used. The weak modified
symbol is now $h_{w,\varepsilon}=R\tilde{q}_\varepsilon
+Sp\tilde{q}_\varepsilon$.

The question, how this can be extended to, say, all polynomial Hamiltonians, was
already discussed in subsection (\ref{WCSIsolatingDynamics}). 
The remaining questions are the same.

\newpage
\section{Final Remarks}\label{FinalRemarks}

In this work, a path integral with Wiener measure for weak coherent states was 
constructed, but some questions still remain.

The path integral with Wiener measure for the coherent states 
basically contained two ``ingredients", both of which were of geometric nature. 
The canonical one-form, $\theta=p\,dq$, (where $p\,dq$ is locally 
valid if $\theta$ is 
expressed in canonical coordinates) as a phase in the integrand, and a 
(Riemannian) metric, $d\sigma^2$, in a regularization factor, by virtue of which
the Wiener measure emerged. 

These appealing features are
maintained in the spectral approach to a weak coherent state path integral, but,
unfortunately, it turned out that this approach could be applied to one special
case only. The regularized approach on the other hand introduced many new
terms, and spoils the beauty of the original idea. So the question is
whether one can avoid this. Naturally, it is possible to try different 
regularizations. One way would be a cut-off method. Instead of the 
normalization condition (which was restored by the $\varepsilon$-modification),
one could confine the problem to the square $[-L,L]^2$ 
and impose appropriate boundary conditions\footnote{The non-negative,
affine variable is supposed to be mapped to the whole real line first.}. 
The functions are periodically
continued outside of the square $[-L,L]^2$, while the boundary conditions are
enforced by sums of $\delta$-functions. The limit $L\rightarrow\infty$ is
now the final step of the whole procedure. This has not been worked out yet, 
but conceptual studies on one-dimensional problems indicate that the final
path integral seems to be closer to the old coherent state path integral 
with Wiener measure (at least under esthetic aspects). 

Another question, already raised, is the maximal class
of Hamiltonians for which the affine weak coherent state path integral is valid. 
It was rigorously established only for linear Hamiltonians. Its validity
for Hamiltonians polynomial in the kinematical operators $Q$ and $D$ was only
conjectured. However, for the regularized approach there could be another way to
prove this. This proof was used by Daubechies and Klauder \cite{DK} and by
DKP \cite{DKP} in the coherent state case for $\beta>1/2$ and basically
involves the same steps as in the case of zero Hamiltonian. The operator 
$\nu A_\varepsilon$ has to be replaced by
$\nu A_\varepsilon +ih_varepsilon$. The symbol $h_varepsilon$, representing
the Hamiltonian, could be different from the weak modified symbol introduced 
in the last section. But, a rigorous treatment of this problem is yet to be
done.

Finally, the question can be addressed as to 
whether the weak coherent state path 
integral with Wiener measure is amenable for computational pursuits. 
Calculations
using Monte Carlo methods have been done for the coherent state path integral
with Wiener measure \cite{Shabanov}, but the results were not too promising
due to the slow convergence encountered. The same can be assumed for 
the weak coherent state path integral, which is already more complicated in 
its analytic form. So it might turn out that the merits of this path integral
are rather confined to the theoretical aspect.

\newpage

\begin{appendix}

\section{Appendix}\label{AppendixA}
\subsection{The affine group}\label{AppendixA1}

Since the affine group is central to the whole present work more of its
properties are listed here for the interested reader. 

The affine group is a subgroup of $SU(1,1)$ \cite{KlauderQisG}. 
To see this, recall first that the
generators $Q$ (with $Q>0$) and $D$ fulfill the affine commutation relation
$[Q,D]=iQ$. Define the operator 

\begin{equation*}
R:=DQ^{-1}D+[(\beta-{\textstyle\frac{1}{2}})^2-{\textstyle\frac{1}{4}}]Q^{-1}
\end{equation*}

\noindent and the operators

\begin{eqnarray*}
K_0&:=&{\textstyle\frac{1}{2}}(\beta Q+\beta^{-1}R)\\
K_1&:=&{\textstyle\frac{1}{2}}(\beta Q-\beta^{-1}R)\\
K_2&:=&-D
\end{eqnarray*}

Using $[Q^\alpha,D]=i\alpha Q^\alpha$, it follows that $[R,D]=-iR$, and further

\begin{eqnarray*}
&&[K_1,K_2]=-iK_0\\
&&[K_0,K_1]=iK_2 \\
&&[K_2,K_0]=iK_1
\end{eqnarray*}

\noindent This is the Lie algebra of $SU(1,1)$. Hence, $Q=\beta^{-1}(K_0
+K_1)$ and $D=-K_2$ form a subalgebra, and the affine group is a 
subgroup of $SU(1,1)$. 

For the family $|\eta_\beta\rangle$ of minimum uncertainty states introduced in 
section (\ref{CSGroupCS}), one has 

\begin{eqnarray*}
K_0|\eta_\beta\rangle&=&\beta|\eta_\beta\rangle\\
K_+|\eta\beta\rangle&=&0
\end{eqnarray*}

\noindent where $K_\pm:=K_2\pm K_1$. These properties can most easily be 
checked in $x$-representation in which $R=-\partial_x x\partial_x
+(\beta-1/2)^2x^{-1}$.\\ 
 
For the resolution of unity, the left-invariant group measure is important. And
the fact that the affine group can generate weak coherent states (for 
certain choices of the fiducial vector) is due to the non-equality of the 
left- and right-invariant group measure. 

In general, the left-invariant group measure can be determined as follows: Let
$U(g)$ be a group representation, $X_a$, $a\in\{1,...,A\}$ 
the group generators, and
$g_0$ some group element, then $U^{-1}(g)dU(g)=X_a\omega^a(g)
=X_a\omega^a(g_0g)$, where the $\omega^a$ are one-forms. The left invariant
group measure is (proportional to) 
$d\mu_L(g)=\omega^1(g)\wedge...\wedge\omega^A(g)=d\mu_L(g_0g)$.

For the right-invariant group measure, one has (with the same notation):
$[dU(g)]U^{-1}(g)=X_a\Omega^a(g)=X_a\Omega^a(gg_0)$, where the $\Omega^a$ are
again one-forms, and the right-invariant group measure is (proportional to) 
$d\mu_R(g)=\Omega^1(g)\wedge...\wedge\Omega^A(g)=d\mu_r(gg_0)$.

Using the group representation $U(p,q)=e^{ipQ}e^{-i\ln qD}$ of the affine 
group [which is familiar from section (\ref{CSGroupCS})], one finds for
the left-invariant group measure $U^{-1}dU=i(Qqdp-Dq^{-1}dq)$, and, hence,

\begin{equation*}
d\mu_L(p,q)=qdp\wedge q^{-1}dq=dpdq
\end{equation*}

For the right-invariant group measure, one has $(dU)U^{-1}=i[Q(dp+pq^{-1}dq)
-Dq^{-1}dq]$, and, hence,

\begin{equation*}
d\mu_R(p,q)=(dp+pq^{-1}dq)\wedge q^{-1}dq=q^{-1}dpdq
\end{equation*}

\noindent This is not equal to the left-invariant group measure.

\subsection{Parameters in DKP and the present work}\label{AppendixA2}

As mentioned before, the parameter $\beta$ is different in its physical meaning
in the present work compared to the DKP-article \cite{DKP}, 
but mathematically it turns out to be 
equivalent. To make comparision easier the connection will 
be studied in the following.

Start with the original 2-parameter family of minimum uncertainty states for
the kinematical variables $D$ and $Q$

\begin{equation*}
\eta_{\alpha\beta^\prime}(x)=N_{\alpha\beta^
\prime}x^\alpha e^{-\beta^\prime x}
\end{equation*}

\noindent The norm of these states can be computed:

\begin{eqnarray*}
1 & \stackrel{!}{=} & \langle\eta|\eta\rangle = \int_0^\infty dx N_{\alpha\beta^
\prime}^2 x^{2\alpha}e^{-2\beta^\prime x} \\
& = & N_{\alpha\beta^\prime}^2(2\beta^\prime)^{-2\alpha-1}\int_0^\infty 
dx x^{2\alpha}e^{x}= N_{\alpha\beta^\prime}^2(2\beta^\prime)^{-2\alpha-1}
\Gamma(2\alpha+1)
\end{eqnarray*}

\noindent  W.l.o.g., choose $N_{\alpha\beta^\prime}$ real.
So $N_{\alpha\beta^\prime}^2=(2\beta^\prime)^{2\alpha+1}\Gamma^{-1}(2\alpha+1)$, 
and a similar calculation shows $\langle Q\rangle=\langle\eta|Q|\eta\rangle
=N_{\alpha\beta^\prime}^2 (2\beta^\prime)^{2\alpha-2}\Gamma(2\alpha+2)
=(\beta^\prime)^{-1}(\alpha+1/2)$. Setting $\langle Q\rangle=k$ for the mean of
$Q$, this means
$\alpha=\beta^\prime k-1/2$ and one is left with a family of minimum uncertainty
states: 

\begin{equation*}
\eta_{\beta^\prime k}(x)=N_{\beta^\prime k}x^{\beta^\prime k-1/2}e^{-\beta^
\prime x}
\end{equation*}

\noindent The choice in this work is $k=1$ and $\beta:=\beta^\prime$, 
whereas DKP \cite{DKP} chose
$k=\beta$ and $\beta^\prime=1$.

Let $U(p,q)=e^{ipQ}e^{-i\ln qD}$ and $|pq\rangle=U(p,q)|\eta\rangle$ (DKP:
$p\rightarrow b$, $q\rightarrow a^{-1}$), 
then the coherent state overlap becomes:

\begin{eqnarray*}
\langle pq|rs\rangle & = & \langle\eta|e^{i\ln qD}e^{-ipQ}|e^{irQ}e^{-i\ln sD}|
\eta\rangle \\
&=& \int_0^\infty dx\, [e^{ipx}e^{-i\ln q(-ix\partial_x-i/2)}\eta]^*[e^{irx}
e^{-i\ln s(-ix\partial_x-i/2}\eta(x)]\\
&=& (qs)^{-1/2}\int_0^\infty dx\, \eta^*(x/q)e^{-ix(p-r)}\eta(x/s)\\
&=& (qs)^{-1/2} N_{\beta^\prime k}^2 q^{1/2-\beta^\prime k}
s^{1/2-\beta^\prime k}\\
&&\times\int_0^\infty dx\, x^{2\beta^\prime k-1}
e^{-\beta^\prime x[(q^{-1}+s^{-1})+i\beta^{\prime-1}(p-r)]} \\
&=& (qs)^{-\beta^\prime k}(2\beta^\prime)^{2\beta^\prime k}\Gamma^{-1}(2\beta
^\prime k)(\beta^\prime)^{-2\beta^\prime k}\\
&&\times[(q^{-1}+s^{-1})+i\beta^{\prime-1}
(p-r)]^{-2\beta^\prime k}
\int_0^\infty dx\, x^{2\beta^\prime k-1}e^{-x} \\
&=& (qs)^{-\beta^\prime k}2^{2\beta^\prime k}[(q^{-1}+s^{-1})+i\beta^{\prime-1}
(p-r)]^{-2\beta^\prime k}
\end{eqnarray*}

\noindent [Here: $\langle pq|rs\rangle=(qs)^{-\beta}2^{2\beta}[(q^{-1}+s^{-1})
+i\beta^{-1}(p-r)]^{-2\beta}$\\
DKP: $\langle ab|a^\prime b^\prime\rangle=(aa^\prime)^\beta 2^{2\beta}
[(a+a^\prime)+i(b-b^\prime)]^{-2\beta}$]

Define the operator 

\begin{equation*}
B:=-ik^{-1}q^{-1}\partial_p+1+(k\beta^\prime)^{-1}q
\partial_q
\end{equation*}

\noindent [Here: $B=-iq^{-1}\partial_p+1+\beta^{-1}q\partial_q$\\
DKP: Note: $\partial_q\rightarrow -a^2\partial_a=\partial_{a^{-1}}$, so
$B=-i\beta^{-1}a\partial_b+1-\beta^{-1}a\partial_a$]

Direct computation shows $B\langle pq|rs\rangle=0$. 
By direct calculation [e.g. in $x$-representation, where $Q=x$ and
$D=-i(x\partial_x+\frac{1}{2})$] one can easily verify  

\begin{equation*}
(\frac{Q}{k}-1+\frac{iD}{k\beta^\prime})|\eta\rangle =0
\end{equation*}

\noindent and so one gets even for arbitrary
$\psi$ (with the use of $e^{i\ln qD}Qe^{-i\ln qD}=qQ$):

\begin{equation*}
B\langle pq|\psi\rangle=B\langle\eta|e^{i\ln qD}e^{-ipQ}|\psi\rangle
=\langle\eta|[{\textstyle\frac{i}{\beta^\prime k}}D+1
-{\textstyle\frac{1}{k}}Q]e^{i\ln qD}e^{-ipQ}|\psi\rangle
=0
\end{equation*}

\noindent DKP chose a different factor $c=(k\beta^\prime)^2
=\beta^2$ for the operator 
$A=c\cdot B^\dagger B$ 
than was selected for this work ($c=\frac{1}{2}k\beta^\prime 
=\frac{1}{2}\beta$).

So
here [compare Eq. (\ref{OperatorA})]: 

\begin{equation*}
A=\frac{1}{2}\{-\beta^{-1}\partial_q q^2\partial_q-\beta q^{-2}\partial_p^2
-2i\beta q^{-1}
\partial_p+\beta-1\}
\end{equation*}

\noindent and DKP has (with their $a$-$b$ notation): 

\begin{equation*}
A=-a^2(\partial_a^2+\partial_b^2)-2i\beta a\partial_b+\beta^2-\beta
\end{equation*}

\subsection{Spectrum of A}\label{AppendixA3}

{\bf Case ${\mathbf 1{\boldsymbol{/}}4{\boldsymbol{<}}
{\boldsymbol{\beta}}{\boldsymbol{\leq}}1{\boldsymbol{/}}2}$}\\

To show that there is no sequence $\psi_n$ which fulfills the necessary 
requirements, one starts with the general function

\begin{equation*}
\psi_n(q)=N_n[1+\sum_{m=1}^Ma_mq^{\sigma_m}
/n^{\rho_m}]e^{-\sum_{k=1}^Kc_kq^{\mu_k}/n^{\nu_k}}\psi(q)
\end{equation*}

\noindent In principle $(\rho_m)_n$ etc. could be sequences themselves (which have to 
fulfill some requirements, e.g. $n^{\rho_m(n)}$ must go to infinity), but since
it does not matter for the later calculations it is enough to think about
them as constant sequences (or constants). Then one must have 
$\rho_m$,$\nu_k$,$c_k\in\mathbb{ R}^+$, and
$\mu_k$, $a_m\in\mathbb{ R}$ for all $m$ or $k$ respectively. 
If there is a $k$ such that $\mu_k<0$ 
(regularization for small $q$), then $\sigma_m\in\mathbb{ R}^\mathbb{ N}$. 
Else, there is a constant $c(\beta)$, depending on $\beta$, which must ensure
$\psi_n, A\psi_n\in L^2$ when setting $\sigma_m\geq c(\beta)$ 

The function $\psi_n$ is square integrable and normalized with 
normalization constant
$N_n$, and $\psi_n/N_n$ converges to $\psi(q)$ in the limit 
$n\rightarrow\infty$. If one can show that $\parallel\!\! A\psi_n\!\!\parallel
\not\rightarrow 0$ for $n\rightarrow\infty$, then one can
conclude by a density argument that there is no function which will
do the job.

Define
$1+\sum_m:=1+\sum_{m=1}^Ma_mq^{\sigma_m}/n^{\rho_m}$, 
$\sum_m^\prime:=\sum_{m=1}^Ma_m\sigma_mq^{\sigma_m-1}/n^{\rho_m}$,
$\sum_m^{\prime\prime}:=\sum_{m=1}^Ma_m\sigma_m(\sigma_m-1)q^{\sigma_m-2}
/n^{\rho_m}$,
$\sum_k:=\sum_{k=1}^Kc_kq^{\mu_k}/n^{\nu_k}$,
$\sum_k^\prime:=\sum_{k=1}^Kc_k\mu_kq^{\mu_k-1}/n^{\nu_k}$ and
$\sum_k^{\prime\prime}:=\sum_{k=1}^Kc_k\mu_k(\mu_k-1)q^{\mu_k-2}/n^{\nu_k}$
then:

\begin{eqnarray*}
&& A\psi_n\\
&=&\{-q^2(\frac{\sum_m^\prime}{1+\sum_m})^2-q^2({\textstyle\sum_k^\prime})^2
+2\beta q(\frac{\sum_m^\prime}{1+\sum_m})-2\beta q({\textstyle\sum_k^\prime})\\
&& -4\beta [+]^{-1}(\frac{\sum_m^\prime}{1+\sum_m})
+4\beta[+]^{-1}({\textstyle\sum_k^\prime})+2q^2(\frac{\sum_m^\prime}{1+\sum_m})
({\textstyle\sum_k^\prime})\\
&& -q^2\frac{(1+\sum_m)(\sum_m^{\prime\prime})-(\sum_M^\prime)^2}
{(1+\sum_m)^2}+q^2({\textstyle\sum_k^{\prime\prime}})
-2q(\frac{\sum_m^\prime}{1+\sum_m})+2q({\textstyle\sum_k^\prime})\}\psi_n
\end{eqnarray*}

Since the norm $\parallel\!\! A\psi_n\!\!\parallel$ has to be calculated one 
needs:

\begin{eqnarray*}
&&(A\psi_n)^*(A\psi_n)\\
&=& [\{-q^2(\frac{\sum_m^\prime}{1+\sum_m})^2-q^2({\textstyle\sum_k^\prime})^2
+2\beta q(\frac{\sum_m^\prime}{1+\sum_m})-2\beta q({\textstyle\sum_k^\prime})\\
&&+2q^2(\frac{\sum_m^\prime}{1+\sum_m})({\textstyle\sum_k^\prime})
-q^2\frac{(1+\sum_m)(\sum_m^{\prime\prime})-(\sum_M^\prime)^2}
{(1+\sum_m)^2}\\
&& +q^2({\textstyle\sum_k^{\prime\prime}})-2q(\frac{\sum_m^\prime}{1+\sum_m})
+2q({\textstyle\sum_k^\prime})\}^2\\
&&-\{...\}4\beta(\frac{\sum_m^\prime}{1+\sum_m})\frac{[+]+[-]}{[\ \ ]}
+\{...\}4\beta({\textstyle\sum_k^\prime})\frac{[+]+[-]}{[\ \ ]}\\
&& 16\beta^2[\ \ ]^{-1}(\frac{\sum_m^\prime}{1+\sum_m})^2
+16\beta^2[\ \ ]^{-1}({\textstyle\sum_k^\prime})^2
-32\beta^2[\ \ ]^{-1}(\frac{\sum_m^\prime}{1+\sum_m})
({\textstyle\sum_k^\prime})]\\
&&\times \psi_n^*\psi_n
\end{eqnarray*}

\noindent where $\{...\}$ was used as an abbreviation for the terms in 
braces in 
the lines 2 to 4.

Next, proceed by calculating the norm of $\psi_n$:

\begin{eqnarray*}
&&\parallel\!\!\psi_n\!\!\parallel^2\\
&=& N_n^*N_n\cdot const\int_0^\infty dq(1+{\textstyle\sum_m})^2
e^{-2(\sum_k)}\,q^{-2\beta}
(q^{-1}+s^{-1})^{1-4\beta}\\
&&\phantom{N_n^*N_n\cdot const}
\times\int_{-\infty}^{\infty}dp[1+\beta^{-2}p^2]^{-2\beta}\\
&=& const\cdot N_n^*N_n[\int_0^c dq(1+{\textstyle\sum_m})^2e^{-2(\sum_k)}
\,q^{-2\beta}
(q^{-1}+s^{-1})^{1-4\beta}\\
&&\phantom{const\cdot N_n^*N_n[}
+\int_c^\infty dq(1+{\textstyle\sum_m})^2e^{-2(\sum_k)}\,q^{-2\beta}
(q^{-1}+s^{-1})^{1-4\beta}]
\end{eqnarray*}

\noindent In the third line distinguish two cases: 
i) There is no $\mu_k<0$, i.e., there
is no small $q$ regularization and restrictions to $(\sigma_m)$ apply. Then set
$c=0$, i.e., the first integral vanishes. ii) If there exists a $\mu_k<0$, then 
choose the biggest $\mu_k/\nu_k$ which is smaller than zero and call it 
$\mu_{max}/n_{max}$. In the fourth line choose the smallest $\mu_k/\nu_k$ 
which is bigger than zero and call it $\mu_{min}/n_{min}$. Then the equation 
continues:

\begin{eqnarray*}
&&...\\
&=& const\cdot N_n^*N_n\{\int_0^c dq\,(1+{\textstyle\sum_m})^2
\exp[-{\textstyle\frac{2c_{max}}{q^{|\mu_{max}|}
n^{\nu_{max}}}}]e^{-2(\sum_{k\neq k_{max}})}\\
&&\phantom{const\cdot N_n^*N_n\{\times\times}
\times q^{-2\beta}(q^{-1}+s^{-1})^{1-4\beta}\\
&&\phantom{const\cdot N_n^*N_n}
+\int_c^\infty dq(1+{\textstyle\sum_m})^2
\exp[-{\textstyle\frac{2c_{min}q^{\mu_{min}}}
{n^{\nu_{min}}}}]e^{-2(\sum_{k\neq k_{min}})}\\
&&\phantom{const\cdot N_n^*N_n\{\times\times}
\times q^{-2\beta}(q^{-1}+s^{-1})^{1-4\beta}\}\\
&=& const\cdot N_n^*N_n\{\int_0^c dq\,(1+\sum_{m=1}^Ma_mq^{\sigma_m}
n^{-\nu_{max}/|\mu_{max}|\cdot\,\sigma_m-\rho_m})^2 \\
&&\phantom{const\cdot N_n^*N_n\{}
\times \exp[-{\textstyle\frac{2c_{max}}{q^{|\mu_{max}|}}}]
\exp[-2\sum_{k\neq k_{max}}c_kq^{\mu_k}n^{-\nu_{max}/|\mu_{max}|\cdot\,
\mu_k-\nu_k}]\\
&&\phantom{const\cdot N_n^*N_n\{}
\times q^{-2\beta}n^{(\nu_{max}/|\mu_{max}|)(2\beta-1)}
(n^{\nu_{max}/|\mu_{max}|}q^{-1}+s^{-1})^{1-4\beta}\\
&&\phantom{const\cdot N_n^*N_n}
+\int_c^\infty dq\,(1+\sum_{m=1}^Ma_mq^{\sigma_m}
n^{\nu_{min}/\mu_{min}\cdot\,\sigma_m-\rho_m})^2 \\
&&\phantom{const\cdot N_n^*N_n\{}
\times \exp[-2c_{min}q^{\mu_{min}}]
\exp[-2\!\!\sum_{k\neq k_{min}}\!\!c_kq^{\mu_k}n^{\nu_{min}/\mu_{min}\cdot\,
\mu_k-\nu_k}]\\
&&\phantom{const\cdot N_n^*N_n\{}
\times q^{-2\beta}n^{-(\nu_{min}/\mu_{min})(2\beta-1)}
[(n^{\nu_{min}/\mu_{min}}q)^{-1}+s^{-1}]^{1-4\beta}\}\\
&\leq& const\cdot N_n^*N_n\{const\cdot n^{-\nu_{max}/|\mu_{max}|\cdot 2\beta}\\
&&\times\int_0^c dq\,
(1+\sum_{m=1}^Ma_mq^{\sigma_m}
n^{-\nu_{max}/|\mu_{max}|\cdot\,\sigma_m-\rho_m})^2 
\exp[-{\textstyle\frac{2c_{max}}{q^{|\mu_{max}|}}}]q^{2\beta-1}\\
&&\phantom{const\cdot N_n^*N_n}
+\,const\cdot n^{(\nu_{min}/\mu_{min})(1-2\beta)}\\
&&\times\int_c^\infty dq\,
(1+\sum_{m=1}^Ma_mq^{\sigma_m}
n^{\nu_{min}/\mu_{min}\cdot\,\sigma_m-\rho_m})^2 
\exp[-2c_{min}q^{\mu_{min}}]q^{-2\beta}\}
\end{eqnarray*}

\noindent By choosing $\mu_{max},\nu_{max},\mu_{min}$ and $\nu_{min}$ 
as above, one 
ensures that after a substitution ($q\rightarrow qn^{-\nu_{max}/|\mu_{max}|}$
for small $q$ and $q\rightarrow qn^{\nu_{min}/\mu_{min}}$ for large $q$),
the remaining summands of the $k$-sum quickly go to zero, i.e., their 
exponentials go to $1$ from below. Hence, they can be approximated by one. 
Observe that the different $q$-dependence ($q^{-2\beta}$ in the last and 
$q^{2\beta-1}$ in the second to last line) in the integrals stems from the 
factor $(q^{-1}+s^{-1})^{1-4\beta}$. After the different substitutions 
in the two regions, this leads to 
different integrals and to different $n$-dependences in front of the integrals 
as well. 

So,

\begin{eqnarray*}
&&N_n^*N_n\\
&\geq& \{const\cdot n^{-\nu_{max}/|\mu_{max}|\cdot 2\beta}\\
&&\times\int_0^c dq\,
(1+\sum_{m=1}^Ma_mq^{\sigma_m}
n^{-\nu_{max}/|\mu_{max}|\cdot\,\sigma_m-\rho_m})^2 
\exp[-{\textstyle\frac{2c_{max}}{q^{|\mu_{max}|}}}]q^{2\beta-1}\\
&&+const\cdot n^{(\nu_{min}/\mu_{min})(1-2\beta)}\\
&&\times\int_c^\infty\!\! dq\,
(1+\sum_{m=1}^Ma_mq^{\sigma_m}
n^{\nu_{min}/\mu_{min}\cdot\,\sigma_m-\rho_m})^2 
\exp[-2c_{min}q^{\mu_{min}}]q^{-2\beta}\}^{-1}
\end{eqnarray*}

\noindent and this is the correct asymptotic behavior.

Consider $(A\psi_n)^*(A\psi_n)$ again. This expression naturally is $\geq 0$. 
Certainly,
$A\psi_n$ can be $0$ for some $n$ (even infinitely many, as long as there are 
infinitely many for which this is not true), and one chooses w.l.o.g. the 
subsequence for which $A\psi_n\neq0$ and refers to it by the same name (i.e.  
the sequence will still be labeled by $n$).
Then $\parallel\!\! A\psi_n\!\!\parallel>0$ for all $n$ and the goal is to show that
$\lim_{n\rightarrow\infty}\parallel\!\! A\psi_n\!\!\parallel\neq0$.

Pick the summand $q^2(\sum_k^\prime)^2$ of $(A\psi_n)^*(A\psi_n)$. The 
$p$-integration will give $const\cdot (q^{-1}+s^{-1})^{1-4\beta}$. 
So compute the expression:

{\allowdisplaybreaks
\begin{eqnarray*}
&& N_n^*N_n\int_0^\infty dq\,q^{-2\beta} (q^{-1}+s^{-1})^{1-4\beta}
(1+{\textstyle\sum_m})^2
e^{-2(\sum_k)}\,q^2({\textstyle\sum_k^\prime})^2\\
&\geq& \{\int_0^cdq\,n^{-\nu_{max}/|\mu_{max}|\cdot(2\beta-1)}q^{-2\beta}
(n^{\nu_{max}/|\mu_{max}|}q^{-1}+s^{-1})^{1-4\beta}\\*
&&\phantom{\{}
\times\exp[-{\textstyle\frac{2c_{max}}{q^{|\mu_{max}|}}}]
e^{-2(\sum_{k\neq k_{max}})}(\sum_kc_k\mu_kq^{\mu_k}
n^{-\nu_{max}/|\mu_{max}|\cdot\mu_k-\nu_k})^2\\*
&&\phantom{\{}
\times(1+\sum_ma_mq^{\sigma_m}
n^{-\nu_{max}/|\mu_{max}|\cdot\sigma_m-\rho_m})^2\\*
&&+\int_c^\infty dq\,n^{\nu_{min}/\mu_{min}\cdot\,(1-2\beta)}
[(qn^{\nu_{min}/\mu_{min}})^{-1}+s^{-1}]^{1-4\beta}\\*
&&\phantom{\{}
\times\exp[-2c_{min}q^{\mu_{min}}]
e^{-2(\sum_{k\neq k_{min}})}(\sum_kc_k\mu_kq^{\mu_k}
n^{\nu_{min}/\mu_{min}\cdot\mu_k-\nu_k})^2\\*
&& \phantom{\{}
\times(1+\sum_ma_mq^{\sigma_m}
n^{\nu_{min}/\mu_{min}\cdot\,\sigma_m-\rho_m})^2\}\\*
&&\times\{const\cdot n^{-\nu_{max}/|\mu_{max}|\cdot 2\beta}\\*
&&\phantom{\times}
\times\!\int_0^c dq\,
(1+\sum_{m=1}^Ma_mq^{\sigma_m}
n^{-\nu_{max}/|\mu_{max}|\cdot\,\sigma_m-\rho_m})^2 
\exp[-{\textstyle\frac{2c_{max}}{q^{|\mu_{max}|}}}]q^{2\beta-1}\\*
&&\phantom{\times}
+const\cdot n^{(\nu_{min}/\mu_{min})(1-2\beta)}\\*
&&\phantom{\times}
\times\!\!\int_c^\infty \!\!dq\,
(1+\!\!\sum_{m=1}^M\!a_mq^{\sigma_m}
n^{\nu_{min}/\mu_{min}\cdot\,\sigma_m-\rho_m})^2 
\exp[-2c_{min}q^{\mu_{min}}]q^{-2\beta}\}^{-1}
\end{eqnarray*}
}

Define $\rho_{min},\sigma_{min}$ by:
$-\nu_{max}/|\mu_{max}|\cdot\,\sigma_{min}-\rho_{min}\geq
-\nu_{max}/|\mu_{max}|\cdot\,\sigma_m-\rho_m$ for all m and
define $\rho_{max},\sigma_{max}$ by:
$\nu_{min}/\mu_{min}\cdot\,\sigma_{max}-\rho_{max}\geq
\nu_{min}/\mu_{min}\cdot\,\sigma_m-\rho_m$ for all m.

The $n$-dependence is discussed with the help of these definitions:

1.term [``$\int_0^c/(\int_0^c+\int_c^\infty)$"] contains:

Numerator: \\
a) $n^{-\nu_{max}/|\mu_{max}|\cdot 2\beta}$\\
b) the ``worst" term of $(\sum_kc_k\mu_kq^{\mu_k}
n^{-\nu_{max}/|\mu_{max}|\cdot\mu_k-\nu_k})^2$ is the one where $k=k_{max}$ and 
it is $O(n^0)=const$.\\
c) the ``worst" term of $(1+\sum_ma_mq^{\sigma_m}
n^{-\nu_{max}/|\mu_{max}|\cdot\sigma_m-\rho_m})^2$ is the one with
$n^{-2\nu_{max}/|\mu_{max}|\cdot\sigma_{min}-\rho_{min}}$ (or the 1)

Denominator:\\
a.i) $n^{-\nu_{max}/|\mu_{max}|\cdot 2\beta}$\\
b.i) the dominant term of $(1+\sum_m(max))^2$ is 
$n^{2(-\nu_{max}/|\mu_{max}|\cdot\sigma_{min}-\rho_{min})}$ (or the 1)\\
a.ii) $n^{\nu_{min}/\mu_{min}\cdot(1-2\beta)}$\\
b.ii) the dominant term of $(1+\sum_m(min))^2$ is 
$n^{2(\nu_{min}/\mu_{min}\cdot\sigma_{max}-\rho_{max})}$ (or the 1)

If a.i) and b.i) dominate [i.e. 
$-\nu_{max}/|\mu_{max}|\cdot 2\beta+2(-\nu_{max}/|\mu_{max}|\cdot\sigma_{min}
-\rho_{min})\geq \nu_{min}/\mu_{min}\cdot(1-2\beta)+
2(\nu_{min}/\mu_{min}\cdot\sigma_{max}-\rho_{max})$] then the expression
will go to a constant, else it will go to 0.

2.term [``$\int_c^\infty/(\int_0^c+\int_c^\infty)$"] contains:

Numerator: \\
a) $n^{\nu_{min}/\mu_{min}\cdot (1-2\beta)}$\\
b) the ``worst" term of $(\sum_kc_k\mu_kq^{\mu_k}
n^{\nu_{min}/\mu_{min}\cdot\mu_k-\nu_k})^2$ is the one where $k=k_{min}$ and 
it is $O(n^0)=const$.\\
c) the ``worst" term of $(1+\sum_ma_mq^{\sigma_m}
n^{\nu_{min}/\mu_{min}\cdot\sigma_m-\rho_m})^2$ is the one with
$n^{2(\nu_{min}/\mu_{min}\cdot\sigma_{max}-\rho_{max})}$ (or the 1)

Denominator: same as above

If a.ii) and b.ii) dominate [i.e. 
$-\nu_{max}/|\mu_{max}|\cdot 2\beta+2(-\nu_{max}/|\mu_{max}|\cdot\sigma_{min}
-\rho_{min})\leq \nu_{min}/\mu_{min}\cdot(1-2\beta)+
2(\nu_{min}/\mu_{min}\cdot\sigma_{max}-\rho_{max})$], then the expression
will go to a constant, or else it will go to 0.

Since the two conditions for the two terms 
mutually exclude each other, except for equality, when both terms yield 
constants, there will always be one term going to zero and the other going to a 
constant. Therfore, the 
whole expression must contain at least one constant term.

To complete the proof one must show that this constant can not be cancelled
exactly in the limit by negative constant terms. Unfortunately, there
are 36 positive ``pseudo-terms" (each containing many terms in the 
sums over k and m) and 30 negative ones. It would be very difficult
to study these terms in detail to make sure that no total
cancelation can occur. 
So, strictly speaking, the proof is not complete. From a practical point of
view, however, there is enough evidence to safely assume this. Hence,
$0\not\in spec(A)$ for $1/4<\beta<1/2$.\\

{\bf Case ${\mathbf 0{\boldsymbol{<}}{\boldsymbol{\beta}}
{\boldsymbol{\leq}}1{\boldsymbol{/}}4}$}\\

Start with an example and choose:

\begin{equation*}
\psi_n:=N_n({\textstyle\frac{1}{2}})^{-2\beta}
(qs)^{-\beta}[(q^{-1}+s^{-1})+i\beta^{-1}(p-r)]
^{-2\beta} e^{-q/n-1/(qn)}e^{-(p-r)^2/n^2}
\end{equation*}

\noindent Without the regularization for 
small $q$ (the $e^{-1/(qn)}$ factor), $\psi_n$
would be square integrable. But, 
$A\psi_n=\{-q^2/n^2+2\beta[+]^{-1}1/n+(2-2\beta)q/n
-\beta^2q^{-2}(4ip/n^2[+]^{-1}+4p^2/n^4-2/n^2)+\beta^2q^{-1}8ip/n^2\}\psi_n$ 
would not be square integrable 
because of the terms with the factors $q^{-2}$ and $q^{-1}$. 
One must have $\psi_n\in D(A)$, and this is why the regularization
for small $q$ was additionally introduced. 

This complicates things, not only 
because there will be 
even more terms, but it will be hard to extract the correct $n$-dependence 
as well.
(A regularization term like $e^{-1/(qn)}$ can not at the same time with 
$e^{-q/n}$ get rid of its $n$-dependence by a substitution like 
$q\rightarrow qn$, 
and that was the trick used above). The problem must be split in regions 
($\int_0^c$ and $\int_c^\infty$).

\begin{eqnarray*}
\parallel\!\!\psi_n\!\!\parallel^2
& \propto &
N_n^*N_n\int_0^\infty dq\, e^{-2q/n-2/(qn)}q^{-2\beta}\\
& &\phantom{N_n^*N_n}
\times\int_{-\infty}^\infty dp\, [(q^{-1}+s^{-1})^2+\beta^{-2}p^2]^{-2\beta}
e^{-2p^2/n^2} \\
& = & N_n^*N_n\,n^{1-4\beta}\int_0^\infty dq\, e^{-2q/n-2/(qn)}q^{-2\beta}\\
& &\phantom{N_n^*N_n\,n^{1-4\beta}}
\times\int_{-\infty}^\infty dp [(q^{-1}+s^{-1})^2n^{-2}
+\beta^{-2}p^2]^{-2\beta}
e^{-2p^2} \\
& = & N_n^*N_n\,n^{1-4\beta}\{\int_0^c dq\, e^{-2q/n-2/(qn)}q^{-2\beta}\times
I_{(0,0)}(n)\\
& &\phantom{N_n^*N_n\,n^{1-4\beta}}
+\int_c^\infty dq\, e^{-2q/n-2/(qn)}q^{-2\beta}\times I_{(0,0)}(n)\}\\
& \leq & N_n^*N_n\,n^{1-4\beta}\{n^{2\beta-1}\int_0^c dq\, 
e^{-2/q}q^{-2\beta}\cdot
I^{small}_{(0,0)}(n,q)\\
& &\phantom{N_n^*N_n\,n^{1-4\beta}}
+n^{1-2\beta}\int_c^\infty dq\, e^{-2q}q^{-2\beta}\cdot e^{-2/(cn^2)}
I^{large}_{(0,0)}(n,q)\}\\
&\leq& N_n^*N_n\{n^{-2\beta}\cdot c_1+n^{2-6\beta}\cdot c_2\}
\end{eqnarray*}

\noindent where (for reasons that become obvious later)

\begin{equation*}
I_{(0,0)}(n,q):=\int_{-\infty}^\infty dp\, 
[(q^{-1}+s^{-1})^2n^{-2}+\beta^{-2}p^2]^{-2\beta}
e^{-2p^2}
\end{equation*}

\begin{equation*}
I^{small}_{(0,0)}(n,q):=\int_{-\infty}^\infty dp\, 
[(nq^{-1}+s^{-1})^2n^{-2}+\beta^{-2}p^2]^{-2\beta}
e^{-2p^2}\stackrel{n\rightarrow\infty}{\longrightarrow}I^{small}_{(0,0)}(q)
\end{equation*}

\begin{equation*}
I^{large}_{(0,0)}(n,q):=\int_{-\infty}^\infty \!\!dp\, 
[((qn)^{-1}+s^{-1})^2n^{-2}+\beta^{-2}p^2]^{-2\beta}
e^{-2p^2}\stackrel{n\rightarrow\infty}{\longrightarrow}I^{large}_{(0,0)}=
const
\end{equation*}

\begin{equation*}
c_1:=\int_0^c dq\, e^{-2/q}q^{-2\beta}\cdot I^{small}(q)
\end{equation*}

\begin{equation*}
c_2:=\int_c^\infty dq\, e^{-2/q}q^{-2\beta}\cdot I^{large}
\end{equation*}

So $N_n^*N_n\geq const\cdot [c_1n^{-2\beta}+c_2n^{6\beta-2}]^{-1}
\stackrel{n\rightarrow\infty}{\longrightarrow}const\cdot n^{-2+6\beta}$.

Compute $A\psi_n$ and write ``$p$" instead of ``$(p-r)$" since this will
make no difference upon integration:

\begin{eqnarray*}
A\psi_n&=&\{-q^2(1-q^{-2})n^{-2}+2q[1-\beta(1-q^{-2})]n^{-1}\\
&&\phantom{\{}
-4\beta^2q^{-2}
p^2n^{-4}+2\beta^2q^{-2}n^{-2}
+4i\beta^2q^{-1}pn^{-2}\\
&&\phantom{\{}
+2\beta[+]^{-1}(1-q^{-2})n^{-1}
+4i\beta^2q^{-2}[+]^{-1}pn^{-2}\}\psi_n
\end{eqnarray*}

The norm is:

\begin{eqnarray*}
\parallel\!\! A\psi_n\!\!\parallel^2 &=& \int_0^\infty dq
\int_{-\infty}^\infty dp\,
(A\psi_n)^*(A\psi_n)\\
&=& \int_0^\infty dq\int_{-\infty}^\infty dp\, 
[\{-q^2(1-q^{-2})n^{-2}+2q[1-\beta(1-q^{-2})]n^{-1}\\
&&-4\beta^2q^{-2}p^2n^{-4}+2\beta^2q^{-2}n^{-2}\}^2
+16\beta^4q^{-2}p^2n^{-4}\\
&&+\{...\}2\beta\frac{[+]+[-]}{[\ \ ]}(1-q^{-2})n^{-1}
+\{...\}4\beta^2q^{-2}pi\frac{[-]-[+]}{[\ \ ]}n^{-2}\\
&&+8\beta^3q^{-1}(1-q^{-2})pi\frac{[+]-[-]}{[\ \ ]}n^{-3}
+16\beta^4q^{-3}p^2\frac{[+]+[-]}{[\ \ ]}n^{-4}\\
&&+4\beta^2(1-q^{-2})^2[\ \ ]^{-1}n^{-2}+16\beta^4q^{-4}p^2[\ \ ]^{-1}n^{-4}]
\psi_n^*\psi_n
\end{eqnarray*}

\noindent where the definitions of $[+],[-],[\ \ ]$ from 
above were used. $[+]+[-]=q^{-1}
s^{-1}$, $\pm([-]-[+])=\pm(2\beta^{-1}p)$. Here, $\{...\}$ are the ``$q$-terms"
in the braces above.

Some of the terms will be explicitly calculated, the others will be discussed
in words.
 
The first term is proportional to:

\begin{eqnarray*}
&& N_n^*N_n\int_0^\infty dq\,e^{-2q/n-2/(qn)}q^{-2\beta}(q^2-1)^2n^{-4}
\int_{-\infty}^\infty dp\,e^{-p^2/n^2}[\ \ ]^{-2\beta}\\
&=& N_n^*N_n\,n^{1-4\beta}\int_0^\infty dq\,e^{-2q/n-2/(qn)}q^{-2\beta}
(q^4-2q^2+1)n^{-4}I_{(0,0)}(n,q)\\
&=& N_n^*N_n\,n^{1-4\beta}[\int_0^c dq\,e^{-2q/n-2/(qn)}q^{-2\beta}
(q^4-2q^2+1)n^{-4}I_{(0,0)}(n,q)\\
&&\phantom{N_n^*N_n\,n^{1-4\beta}}
+\int_c^\infty dq\,e^{-2q/n-2/(qn)}q^{-2\beta}
(q^4-2q^2+1)n^{-4}I_{(0,0)}(n,q)] \\
&\geq& N_n^*N_n\,n^{1-4\beta}[n^{2\beta-1}\int_0^c dq\,e^{-2c/n^2}e^{-2/q}
q^{-2\beta}\\
&&\phantom{N_n^*N_n\,n^{1-4\beta}[n^{2\beta-1}}
\times(n^{-4}q^4-2n^{-2}q^2+1)n^{-4}I^{small}_{(0,0)}(n,q)\\
&&\phantom{N_n^*N_n\,n^{1-4\beta}}
+n^{1-2\beta}\int_c^\infty dq\,e^{-2q}e^{-2/(cn^2)}
q^{-2\beta}\\
&&\phantom{N_n^*N_n\,n^{1-4\beta}+n^{1-2\beta}}
\times(n^4q^4-2n^2q^2+1)n^{-4}I^{large}_{(0,0)}(n,q)]
\end{eqnarray*}

\noindent Notice that the first part (small $q$) goes rapidly to zero. 
In the second part 
everything goes to zero, except the $q^4$-term, which will go to a constant.

One can easily generalize this: Terms with a higher power of $1/n$ than of $q$ 
will all go to zero (e.g. $q/n^2$ etc.) and will be called ``more-$n$" terms. 
Terms with the same power will go to 
a constant (e.g. $q/n$, $q^2/n^2$ etc.). Terms with $O(q^0)=O(1)$ will vanish
in the limit since they come with some power of $1/n$. 
All these terms can be called
``large $q$ dominated" - since the positive powers of $q$ help for small $q$.

Terms with a negative power of $q$ (e.g. $1/(qn), 1/(qn^2), 1/(q^2n^2)$ etc.) 
can be called ``small $q$ dominated". 
It is for them that one needs 
the regularization
for small $q$. (They help for large $q$, so the $\int_c^\infty$-part 
will quickly go to zero). Here is an example:

\begin{eqnarray*}
&& N_n^*N_n\int_0^\infty dq\,e^{-2q/n-2/(qn)}q^{-2\beta}q^{-1}n^{-1}
\int_{-\infty}^\infty dp\,e^{-p^2/n^2}[\ \ ]^{-2\beta}\\
&=& N_n^*N_n\,n^{1-4\beta}\int_0^\infty dq\,e^{-2q/n-2/(qn)}q^{-2\beta}
q^{-1}n^{-1}I_{(0,0)}(n,q)\\
&=& N_n^*N_n\,n^{1-4\beta}[\int_0^c dq\,e^{-2q/n-2/(qn)}q^{-2\beta}
q^{-1}n^{-1}I_{(0,0)}(n,q)\\
&&\phantom{N_n^*N_n\,n^{1-4\beta}}
+\int_c^\infty dq\,e^{-2q/n-2/(qn)}q^{-2\beta}
q^{-1}n^{-1}I_{(0,0)}(n,q)] \\
&\geq& N_n^*N_n\,n^{1-4\beta}[n^{2\beta-1}\int_0^c dq\,e^{-2c/n^2}e^{-2/q}
q^{-2\beta}q^{-1}I^{small}_{(0,0)}(n,q)\\
&&\phantom{N_n^*N_n\,n^{1-4\beta}}
+n^{1-2\beta-2}\int_c^\infty dq\,e^{-2q}e^{-2/(cn^2)}
q^{-2\beta}q^{-1}I^{large}_{(0,0)}(n,q)]
\end{eqnarray*}

\noindent Because $N_n^*N_n\geq const\cdot 
[c_1n^{-2\beta}+c_2n^{6\beta-2}]^{-1}$ 
everything goes to zero. 
This can be generalized to all ``small $q$ dominated" terms.

The ``$p$"- and ``$[\ \ ]$"-terms:

There are two kinds of terms involving $p$ and the $[\ \ ]$-bracket. The first 
kind involves just $[\ \ ]^{-1}$, and so one sets:

\begin{eqnarray*}
\int_{-\infty}^\infty dp\,e^{-p^2/n^2}[\ \ ]^{-2\beta}\cdot [\ \ ]^{-1}
&=& n^{1-4\beta}n^{-2}\int_{-\infty}^\infty dp\,e^{-p^2}[\ \ ]^{-2\beta-1}\\
&=:& n^{1-4\beta}n^{-2}I_{(0,-2)}(n,q)
\end{eqnarray*}

\noindent Now the meaning of the subscripts of the integrals $I$ becomes 
clear. The $0$ 
means no extra power of $p$, the $-2$ means one extra $[\ \ ]^{-1}$ 
(and the
$2$ rather than a $1$ was chosen just because the $[\ \ ]$-bracket is 
quadratic in $p$ and so
one gets an extra $n^{-2}$ from a simple $[\ \ ]^{-1}$). In the different 
regions, the $I_{(0,-2)}$ will go to $I^{small}_{(0,-2)}(n,q)
\stackrel{n\rightarrow\infty}{\longrightarrow}I^{small}_{(0,-2)}(q)$, and 
$I^{large}_{(0,-2)}(n,q)
\stackrel{n\rightarrow\infty}{\longrightarrow}I^{large}_{(0,-2)}=const$.

The second term involves $p^2$ and $[\ \ ]^{-1}$ and one sets:

\begin{eqnarray*}
\int_{-\infty}^\infty dp\,e^{-p^2/n^2}p^2[\ \ ]^{-2\beta}\cdot [\ \ ]^{-1}
&=&n^{1-4\beta}\int_{-\infty}^\infty dp\,e^{-p^2}p^2[\ \ ]^{-2\beta-1}\\
&=:& n^{1-4\beta}I_{(2,-2)}(n,q)
\end{eqnarray*}

\noindent Compared to the original term without the extra factors,
there is no different $n$-dependence, only the integral changed 
[$I_{(2,-2)}(n,q)$ instead of $I_{(0,0)}(n,q)$].

So all terms involving the extra $p$- and $[\ \ ]$-factors are of the 
``small $q$ dominated" or ``more-$n$" type and will vanish in the limit. 
So the only
terms going to constants are the ``large $q$ dominated", non-``more-$n$" 
terms which already appeared
in the easier case $1<4\beta<2$ as well.

The conclusion is again that 
the norm of $A\psi_n$ will not vanish for $n\rightarrow\infty$ (the same
comment concerning cancelations as in the case $1/4<\beta<1/2$ applies).

Terms coming from a $p$-regularization used to vanish
in the limit, so one does not expect new things to happen for a general
$\psi_n$, and the previous discussion of the general case (for $1/4<\beta<1/2$) 
is enough to convince for $0<4\beta\leq 1$ as well. 

Note that the case $4\beta=1$ as a limiting case is well behaved, namely the
norm-squares of $\psi_n$ go to $n^{-1/2}$ \{for $4\beta\searrow1$: $N_n^*N_n
\propto n^{2\beta-1}$ and for $4\beta\nearrow1$: $N_n^*N_n
\geq const\cdot [c_1n^{-2\beta}+c_2n^{6\beta-2}]^{-1}
\stackrel{n\rightarrow\infty}{\longrightarrow}const\cdot n^{-2+6\beta}$\}

\subsection{Alternative computation of the denominator of 
Eq. (\ref{limitkernel})}\label{AppendixA4}

Another way to find the $\nu$-dependence of the denominator of
Eq. (\ref{limitkernel}) is the following. 
Use 
the definition of the Whittaker function [Eq. (\ref{Whittaker})]
$W_{\mu,\nu}(z)=e^{-z/2}z^{1/2+\nu} U(1/2-\mu+\nu,1+2\nu,z)$
and of the confluent hypergeometric function
$U(a,b,z)=\Gamma^{-1}(a)\int_0^\infty e^{-zt}t^{a-1}(1+t)^{b-a-1}dt$ applied
to $\mu=1/2$, $\nu=i\lambda$
and extract the the different orders of $\lambda$ directly. 
Attention is required because the integral definition demands a real part 
for $a$, which
is introduced as $\epsilon$, and all expressions are computed with the complex
variable $\xi:=\epsilon+i\lambda$. The limit $\epsilon\rightarrow 0$ is taken in 
the end (or when not critical). Although this approach to
the problem is different from the one before, it does 
not come as a surprise that there is a need for
an auxiliary parameter $\epsilon$. In the first approach, the critical
point was the evaluation of the hypergeometric series after the $x$-integration.
Here, it is the definition of the confluent hypergeometric function $U$ before
the $x$-integration. 

The wave function becomes:

\begin{eqnarray*}
&&x^{1/2}\psi_\lambda^\prime(x)=x^{1/2}x^{-1/2}\psi_\lambda(\ln (x))\\
&=& ({\textstyle\frac{\lambda\sinh(2\pi\lambda)}
{\pi^2}})^{1/2}\Gamma(i\lambda)x^{-1/2}e^{-x/2}x^{1/2+i\lambda}
\Gamma^{-1}(i\lambda)\int_0^\infty e^{-xt}t^{-1+\xi}(1+t)^{\xi}dt
\end{eqnarray*}

\noindent Notice that the small real part has been added 
only in the integrand, leading to a somewhat mixed notation.

This comes together with its complex conjugate. 
So, by omitting the pure $\lambda$-dependent factors for a while, 
one gets an integral

\begin{eqnarray*}
&& \lim_{\epsilon\rightarrow0}\int_0^\infty dx\, e^{-x}
\{\int_0^\infty e^{-xt}t^{-1+\xi}(1+t)^\xi dt\}
\{\int_0^\infty e^{-xt}t^{-1+\xi^*}(1+t)^{\xi^*} dt\}\\
&=& \lim_{\epsilon\rightarrow0}\int_0^\infty dx\, e^{-x}
\{\int_0^\infty e^{-xt}t^{-1+\xi}[1+\ln(1+t)\xi
+{\textstyle\frac{1}{2}}\ln^2(1+t)\xi^2+...]dt\}\\
&&\times\{\int_0^\infty e^{-xt}t^{-1+\xi^*}
[1+\ln(1+t)\xi^*+{\textstyle\frac{1}{2}}\ln^2(1+t)\xi^{*2}+...]dt\}
\end{eqnarray*}

The next step is to interchange the sum, from the power series expansion, and
the integral. This is only justified if
one can make sure
the existence of all integrals (to all orders) and the existence of the limit
$\epsilon\rightarrow 0$. The latter is trivial, except for the summands with
a ``1".
Higher orders come at least with one factor of $\ln(1+t)$, which goes
linearly to 0 for $t\rightarrow0$, thus cancelling the otherwise critical
$t^{-1}$.  

The lowest order (take the 1's from both braces) is:

\begin{eqnarray*}
&&\lim_{\epsilon\rightarrow0}\int_0^\infty dx\, e^{-x}
\{\int_0^\infty e^{-xt}t^{-1+\xi}dt\}
\{\int_0^\infty e^{-xt}t^{-1+\xi^*}dt\}\\
&=&\lim_{\epsilon\rightarrow0}\int_0^\infty dx\, e^{-x}x^{-2\epsilon}
\{\int_0^\infty e^{-t}t^{-1+\xi}dt\}
\{\int_0^\infty e^{-t}t^{-1+\xi^*}dt\}\\
&=&\lim_{\epsilon\rightarrow0}\Gamma(1-2\epsilon)
\Gamma(\xi)\Gamma(\xi^*)
=\Gamma(i\lambda)\Gamma(-i\lambda)=O(1/\lambda^2)
\end{eqnarray*}

Notice that the terminology ``lowest order" refers to the expansion
of $(1+t)^\xi$ only. Naturally, this ``lowest order" term contains all 
possible
powers of $\lambda$, but is the
only term to really contain the lowest order which is $1/\lambda^2$.
For the ``higher orders", the non-critical limit can be taken first 
and the expression is now

\begin{eqnarray*}
&&\int_0^\infty \!\!dx\, e^{-x}\biggl[
\{\int_0^\infty \!\!e^{-xt}t^{-1+i\lambda}[\mathbf{1}+\ln(1+t)i\lambda
+{\textstyle\frac{1}{2}}\ln^2(1+t)
(i\lambda)^2+...]dt\}\\
&&\phantom{\times}
\times\{\int_0^\infty \!e^{-xt}t^{-1-i\lambda}
[\mathbf{1}+\ln(1+t)(-i\lambda)
+{\textstyle\frac{1}{2}}\ln^2(1+t)(-i\lambda)^2+...]dt\}\\
&&\phantom{\int_0^\infty \!\!dx\, e^{-x}}
-{\mbox``\mathbf{1},\mathbf{1}-term"}\biggr]\\
&=&\int_0^\infty dx\, e^{-x}\biggl[
\{\int_0^\infty e^{-xt}t^{-1}
[1+i\ln(t)\lambda+{\textstyle\frac{1}{2}}(i\ln(t))^2\lambda^2+...]\\
&&\hspace{3cm}
\times[\mathbf{1}+\ln(1+t)i\lambda
+{\textstyle\frac{1}{2}}\ln^2(1+t)(i\lambda)^2+...]dt\}\\
&&\phantom{\int_0^\infty dx\, e^{-x}}
\times\{\int_0^\infty e^{-xt}t^{-1}
[1-i\ln(t)\lambda+{\textstyle\frac{1}{2}}(-i\ln(t))^2\lambda^2+...]\\
&&\hspace{3cm}
\times[\mathbf{1}+\ln(1+t)(-i\lambda)
+{\textstyle\frac{1}{2}}\ln^2(1+t)(-i\lambda)^2+...]dt\}\\
&&\phantom{\int_0^\infty dx\, e^{-x}}
-{\mbox``\mathbf{1},\mathbf{1}-term"}\biggr]
\end{eqnarray*}

\noindent where $-{\mbox``\mathbf{1},\mathbf{1}-term"}$ 
indicates the term already taken care of. This is just a notation to 
avoid having to write out everything. The only thing  
to prove is the existence of all integrals. They are of the type

\begin{eqnarray}\label{xtIntegrals}
&&\int_0^\infty dx\, e^{-x}\biggl[
\{\int_0^\infty e^{-xt}t^{-1}
[\ln(t)]^n[\ln(1+t)]^m dt\}\nonumber\\
&&\phantom{\int_0^\infty dx\, e^{-x}}
\times\{\int_0^\infty e^{-xs}
[\ln(s)]^{\tilde{n}}[\ln(1+s)]^{\tilde{m}}ds\}\biggr]
\end{eqnarray}

\noindent where $m,\tilde{m}\in\mathbb{N}$, $n,\tilde{n}\in\mathbb{N}_0$.

Look at the integral $\int_0^\infty e^{-xt}[\ln(t)]^n[\ln(1+t)]^m dt$ first. 
Rewrite it as $\int_0^\infty=\int_0^c+\int_c^\infty$
where $c>e-1$; $e$ is the Euler number.

\begin{equation}\label{definitionf}
\int_0^c e^{-xt}t^{-1}[\ln(1+t)]^m[\ln(t)]^n dt=f(x)
\end{equation}

The ``worst case" is $m=1$, but $[\ln(t)]^n$ is integrable at 0 since $\ln(1+t)
\approx t$ for small $t$, which takes care of the $t^{-1}$. Therefore, there is 
no problem for any $x$. Now,
$f(x)$ is a constant for $x=0$ and goes to $0$ for $x\rightarrow\infty$. Since
it is continuous, it is thus bounded, and so $f(x)\leq c_1=const$.

\begin{equation}\label{definitiong}
\int_c^\infty e^{-xt}t^{-1}[\ln(1+t)]^m[\ln(t)]^n dt=g(x)
\end{equation}

Since $c>e-1>1$ the function $g(x)$ is certainly positive and
$g(x)$ goes to 0 for large $x$. For $0<\delta<1$, the value of the function 
$g(x=\delta)$ is finite. Because
$g(x)$ is continuous, it is bounded on the interval $[\delta,\infty)$ with
bound $d_1$.
The integral does not exist for $x=0$, so one has to extract the $x$-
dependence for small $x$ to determine $x$-integrability at $0$ 
(in the following assume at least $x<1$ so that $-\ln(x)>0$). Since $c>e-1$,
there is a $k\in\mathbb{N}$ such that $[\ln(1+t)]^m\leq[\ln(t)]^{m+k}$. So
Eq. (\ref{definitiong}) becomes

\begin{eqnarray}\label{xBigtIntegrals}
&&\int_c^\infty e^{-xt}t^{-1}[\ln(1+t)]^m[\ln(t)]^n dt\nonumber\\
&\leq&\int_c^\infty e^{-xt}t^{-1}[\ln(t)]^{n+m+k}dt=
\int_{cx}^\infty e^{-t}t^{-1}[\ln(t/x)]^{n+m+k}dt\nonumber\\
&=&\int_{cx}^\infty e^{-t}t^{-1}[-\ln(x)+\ln(t)]^{n+m+k}dt\nonumber\\
&=&\int_{cx}^\infty e^{-t}t^{-1}\{\sum_{l=0}^{n+m+k}\binom{n+m+k}{l}
[-\ln(x)]^l[\ln(t)]^{n+m+k-l}\}dt\nonumber\\
&\leq&\int_{cx}^\infty e^{-t}t^{-1}\{\sum_{l=0}^{n+m+k}\binom{n+m+k}{l}
[-\ln(x)]^l|\ln(t)|^{n+m+k-l}\}dt\nonumber\\
&\leq&\sum_{l=0}^{n+m+k}\binom{n+m+k}{l}[-\ln(x)]^l
\int_{x}^\infty e^{-t}t^{-1}
|\ln(t)|^{n+m+k-l}dt\nonumber\\
&=:&\sum_{l=0}^{n+m+k}\binom{n+m+k}{l}[-\ln(x)]^l\cdot h_l(x)
\end{eqnarray}

Examine $h_l$ alone for a moment:

\begin{eqnarray*}
&&h_l(x)=\int_{x}^\infty e^{-t}t^{-1}
|\ln(t)|^{n+m+k-l}dt\\
&=&\int_x^1(...)+\int_1^\infty(...)\\
&\leq&\int_x^1 t^{-1}|\ln(t)|^{n+m+k-l}dt+const\\
&=&\int_{\ln(x)}^0|u|^{n+m+k-l}du+const=\int_{\ln(x)}^0(-u)^{n+m+k-l}du+const\\
&=&(n+m+k-l+1)^{-1}(-1)^{n+m+k-l+1}[\ln(x)]^{n+m+k-l+1}+const\\
&=&(n+m+k-l+1)^{-1}[-\ln(x)]^{n+m+k-l+1}+const
\end{eqnarray*}

Putting this in the last line of Eq. (\ref{xBigtIntegrals}), one sees that,
for small $x$, the function
$g(x)$ is dominated by a polynomial in $-\ln(x)$ of degree $(n+m+k+1)$.
This is named $\mbox{Poly}^{n+m+k+1}[-\ln(x)]$. 

The same arguments hold for the $s$-integral in Eq.
(\ref{xtIntegrals}). Denote the new constants in an obvious way 
by $c_2$, $d_2$ and $\tilde{k}$ and get  
functions $\tilde{f}$, $\tilde{g}$, $\tilde{h}_l$
at an intermediate step, and 
$\mbox{Poly}^{\tilde{n}+\tilde{m}+\tilde{k}+1}[-\ln(x)]$.
The integral (\ref{xtIntegrals}) is bounded by

\begin{eqnarray*}
&&\int_0^\infty e^{-x}[f(x)+g(x)][\tilde{f}(x)+\tilde{g}(x)]dx\\
&\leq&\int_0^\infty e^{-x}[c_1+g(x)][c_2+\tilde{g}(x)]dx\\
&=&\int_0^\epsilon e^{-x}[c_1+g(x)][c_2+\tilde{g}(x)]dx+
\int_\epsilon^\infty e^{-x}[c_1+g(x)][c_2+\tilde{g}(x)]dx\\
&\leq&\int_0^\epsilon e^{-x}\{c_1+\mbox{Poly}^{n+m+k+1}[-\ln(x)]\}
\{c_2+\mbox{Poly}^{\tilde{n}+\tilde{m}+\tilde{k}+1}[-\ln(x)]\}dx\\
&&+\int_\epsilon^\infty e^{-x}[c_1+d_1][c_2+d_2]dx
\end{eqnarray*}

This expression is smaller than infinity since $[\ln(x)]^r$ is integrable 
at 0 for any $r$, and $e^{-x}$ takes 
care of the integrability at infinity. 

One could go on and calculate the numerical factors for the different 
orders of $\lambda$, i.e. develop the prefactor $\lambda\sinh(2\pi
\lambda)/\pi^2$ and the Gamma functions $\Gamma(i\lambda)\Gamma(-i\lambda)$
in power series, compute the relevant integrals for each power, and collect
the numerical coefficients belonging to that power. 
Surprisingly, the coefficients are not the same as in
Eq. (\ref{hNuT}). But, they do not have to be, since the two approaches
to the problem used different regularizations to deal with only conditionally
convergent expressions. The final result does not depend on these 
coefficients since a quotient is taken.

\section{Acknowledgments}

Thanks are expressed to Prof. Dr. A. Schenzle as first referee of this 
Diplom-thesis and for his support of it, to Prof. Dr. J. von Delft for 
taking on the task of second referee, to the Deutsche Studienstiftung for
financial support and encouragement, to the University of Florida and the
members of its
physics department for hospitality and to S. Shabanov and W. Bomstad
for comments
and helpful remarks.

Very special thanks go to Prof. Dr. J.R. Klauder who suggested the topic
of this article - weak coherent states - which is based on his outstanding
lifelong work. The research for this thesis was conducted under his supervision
and could not have been done without his being a constant source of 
information as well
as of ideas and encouragement at all times. This made working with him not only
an honor but a great pleasure, too.

\end{appendix}

\end{document}